\documentclass[smallcondensed]{svjour3}
\usepackage[round]{natbib}
\usepackage[graphicx]{realboxes}
\usepackage{url}
\usepackage{soul}
\usepackage{color, xcolor}
\usepackage{makecell}
\usepackage{array}
\usepackage{booktabs,lscape}
\usepackage[T1]{fontenc}
\soulregister{\cite}7 
\soulregister{\ref}7 
\soulregister{\citep}7
\usepackage{lineno}

\begin{document}

\title{A Comprehensive Review of Image Analysis Methods for Microorganism Counting: From Classical Image Processing to Deep Learning Approaches}

\author{Jiawei Zhang \textsuperscript{1}  \and Chen Li  \textsuperscript{1,*} \and Md Mamunur Rahaman \textsuperscript{1} \and Yudong Yao \textsuperscript{1,2} \and Pingli Ma \textsuperscript{1} 
\and Jinghua Zhang \textsuperscript{1,5} \and Xin Zhao \textsuperscript{1,3} \and Tao Jiang \textsuperscript{1,4} \ Marcin Grzegorzek \textsuperscript{1,5} }

\authorrunning{C. Li et al.}
\titlerunning{A Review for Image Analysis based Microorganism Counting}

\institute{
\email{lichen201096@hotmail.com}
\at
{1} Microscopic Image and Medical Image Analysis Group, College of Medicine and Biological Information Engineering, Northeastern University, Shenyang, 110169, China.
\at
{2} Department of Electrical and Computer Engineering, Stevens Institute of Technology, Hoboken,
NJ 07030, USA.
\at
{3} School of Resources and Civil Engineering, Northeastern University, Shenyang 110004, China.
\at
{4} School of Control Engineering, Chengdu University of Information Technology, Chengdu 610225,
China.
\at
{5} Institute of Medical Informatics, University of Luebeck, Luebeck 23538, Germany.
}

%
\maketitle              

\begin{abstract}
Microorganisms such as bacteria and fungi play essential roles in many application fields, like biotechnique, medical technique and industrial domain. Microorganism counting techniques are crucial in microorganism analysis, helping biologists and related researchers quantitatively analyze the microorganisms and calculate their characteristics, such as biomass concentration and biological activity. 
However, traditional microorganism manual counting methods, such as plate counting method, hemocytometry and turbidimetry, are time-consuming, subjective and need complex operations, which are difficult to be applied in large-scale applications.
In order to improve this situation, image analysis is applied for microorganism counting since the 1980s, which consists of digital image processing, image segmentation, image classification and suchlike. 
Image analysis-based microorganism counting methods are efficient comparing with traditional plate counting methods. In this article, we have studied the development of microorganism counting methods using digital image analysis. Firstly, the microorganisms are grouped as bacteria and other microorganisms. Then, the related articles are summarized based on image segmentation methods. Each part of the article is reviewed by methodologies. Moreover, commonly used image processing methods for microorganism counting are summarized and analyzed to find common technological points. More than 144 papers are outlined in this article. In conclusion, this paper provides new ideas for the future development trend of microorganism counting, and provides systematic suggestions for implementing integrated microorganism counting systems in the future. Researchers in other fields can refer to the techniques analyzed in this paper.

\end{abstract}

\section{Introduction}

\subsection{Basic knowledge of microorganisms} 

Microorganism is a kind of tiny organism which cannot be observed by naked eyes but can be observed by light microscope or electron microscope~\citep{Madigan-1997-BBOM}. There are many different types of microorganisms, and the classification standards are various. Generally, microorganisms are composed of bacteria, viruses, fungi and some algae. 
\paragraph a Bacteria are unicellular organisms with minimal size, simple structure, lack of nuclei, cytoskeletons, and membranous organelles. It widely distributes in soil and water, and most of them are decomposers at the bottom of the biological chain, such as \emph{Escherichia coli}. Some bacteria are consumers and producers. For example, sulfur bacteria and iron bacteria are producers. They can use inorganic materials to produce organic substances they need. The rhizobia can consume organic substances produced by the photosynthesis of legumes~\citep{Doetsch-2012-IBAT}. 
\paragraph b The virus is a kind of microorganism that can spread and infect other organisms. It is small and has a simple structure. It contains only one type of nucleic acid, such as ribonucleic acid(RNA) virus and deoxyribonucleic acid(DNA) virus. It must parasitize in living cells and proliferate in the way of replication. Viruses consist of single and double-stranded RNA virus, single and double-stranded DNA virus~\citep{Cui-2019-OAEP}. For example, severe acute respiratory syndrome coronavirus 2 (SARS-CoV-2) is a single-stranded RNA virus~\citep{Andersen-2020-TPOS}. 
\paragraph c The fungus is one type of eukaryotic microorganism, including mold, yeast and mushroom, that can produce spores through asexual and sexual reproduction. Tinea pedis is a kind of foot skin disease caused by pathogenic fungi, which is widely spread globally.  There are no sebaceous glands between the soles of human feet and toes, so the environment lacking fatty acids and poor air circulation is conducive to the growth of filamentous fungi~\citep{Perea-2000-PARF}. 
\paragraph d Algae are eukaryotes of the protozoa and most of them are aquatic organisms, which can carry out photosynthesis.  Algae can be composed of one or a few cells, or many cells aggregate into tissue-like structures.  According to the color, algae can be divided into green algae, brown algae and red algae. Red tide is an abnormal phenomenon in the marine ecosystem. It is caused by the explosive proliferation of red tide algae under specific environmental conditions, which is a signal of marine pollution. During the red tide period, a large number of fish, shrimp, crabs, and shellfish die, causing significant damage to aquatic resources and human health~\citep{Kirkpatrick-2004-LRFR}.

Some microorganisms are harmful to human beings by causing food decomposition, infect humans and cause diseases, but some microorganisms are beneficial to human beings. \emph{Penicillin} is an epoch-making discovery in the medical field, which has saved countless lives. Yeast is widely used in industrial fermentation, ethanol production and food production for human beings~\citep{Brill-1981-AM}. Some microorganisms can degrade plastics, treat waste-water, gas, and have great potential in renewable resources~\citep{Rizzo-2013-UWTP}. There are also many microorganisms in the intestines of healthy people, which can help humans decompose and absorb food and toxic substances. Some microorganisms have adverse effects on the human body and industrial production. For example, the human immunodeficiency virus (HIV) can cause the loss of immune function of patients and cause infection. The disease spreads rapidly, has high mortality and cannot be cured, which has caused a significant threat to world health; SARS-CoV-2 breaks out at the end of December in 2019~\citep{Hui-2020-TCET}. More than 183,000,000 people have been infected worldwide till July 1st, 2021, which becomes a global malignant epidemic~\citep{JHU-2020-CCGC}. The SARS-CoV-2 is highly infectious and mainly transmitted through close contact and respiratory droplets. Microorganisms play an essential role in human's daily life and production. Therefore,  beneficial microorganisms should be used wildly, and harmful microorganisms should be prevented.

Microorganism counting is an essential part of microbial research, which is widely used in food and drug safety tests, biomedical tests, and environmental monitoring~\citep{Liu-2004-HTIB}. At present, there are two main methods for microorganism counting and quantification, one of the methods is manual counting, the other one is computer image analysis counting~\citep{Rajapaksha-2019-ARMT}. 
Manual counting mainly includes the plate counting method, hemocytometry and turbidimetry. 
In the plate counting method, the bacteria are placed in a suitable medium and then wait for them to grow into colonies. After that, the number of colonies is counted through the microscope. The advantage of the plate counting method is that the number of live bacteria can be estimated. However, the operation is complicated, and it takes a period to culture the microorganisms and gets the results. In general, the number of colonies obtained is lower than the actual number of living bacteria because when more than two living bacteria cells stick together, the observed number is still one colony~\citep{Balestra-1997-ITET}.
In the hemocytometry method, the bacteria are diluted and dropped on a blood cell counting plate, which is then observed under a microscope to calculate the average number of bacteria in each compartment. Finally, the total number of bacteria is estimated. However, the hemocytometry method cannot distinguish the dead bacteria from the live bacteria and can only estimate the total number of bacteria by the average value, which carries out the low accuracy~\citep{Sambrook-2006-ECNH}.
In the turbidimetry method, a spectrophotometer is applied to measure the optical density of bacterial suspension at a particular wavelength. The cell concentration in bacterial suspension is proportional to the turbidity of bacterial suspension within a specific range, that is, the cell concentration is proportional to the optical density. So, the number of bacteria can be expressed in terms of optical density. However, the turbidimetry method has specific requirements for the wavelength of light in the experimental environment, which should be controlled within the line limit range where the bacterial concentration is proportional to the optical density. Otherwise, the measurement result will have a large error~\citep{Dalgaard-1994-EBGR}.

\begin{figure}[ht]
\centering
\includegraphics[trim={0cm 0cm 0cm 0cm},clip,width=1.0\textwidth]{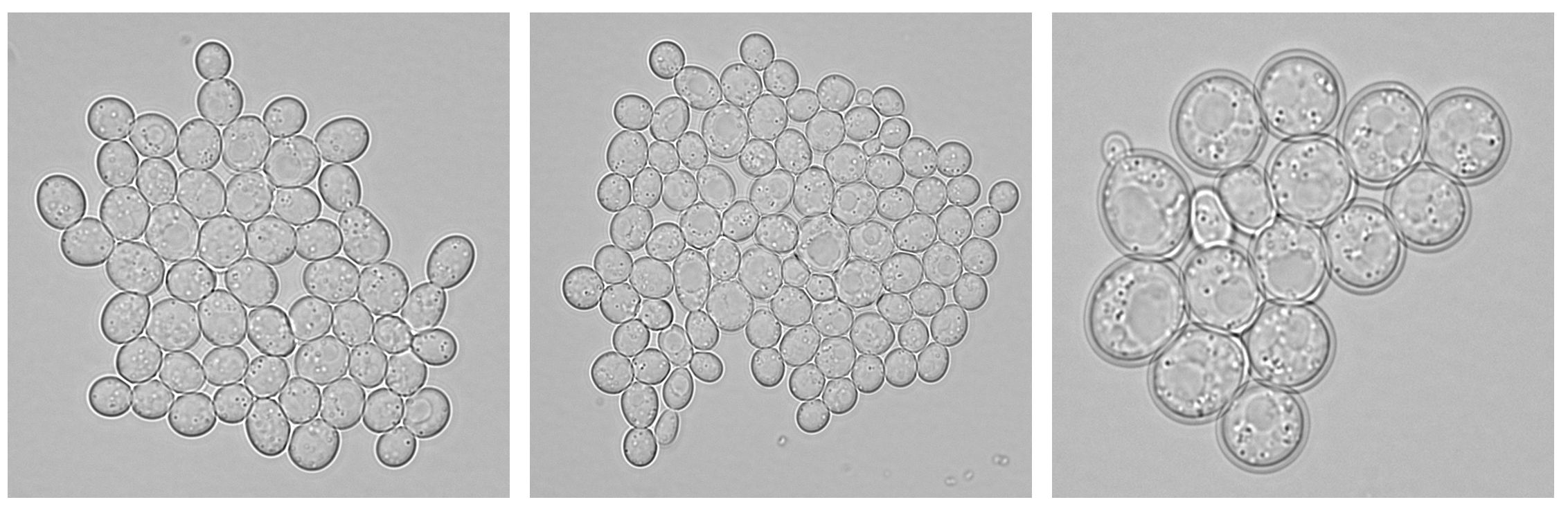}
\caption{An example of yeast cells image (in \citep{Dietler-2020-ACNN} proposed dataset).}
\label{fig:Dietler-2020-ACNN-Example}
\end{figure}

It can be seen that the traditional methods can obtain satisfactory counting results under certain conditions, such as when the number of samples is small and the imaging effect is good under the microscope. However, when the sample becomes larger, it is often encountered that the colony is small, the contrast between the colony and the culture medium is not clear, and it is not easy to detect and count with naked eyes. The detection results have the problems like large errors and poor reliability. The sample image contains many particles, and the workload is heavy and dull, which is easy to cause misjudgment. Moreover, the subjectivity of manual counting is common. Even if the same staff member observes the same sample in different periods, different observation results will be obtained~\citep{Chien-2007-USEA}. 
With the development of computer image analysis technology, automatic particle image analysis systems based on image processing and visual analysis can automatically, quickly and objectively count the number of particles contained in the image and extract various characteristic parameters of particles, which significantly reduces the workload and improves the analysis accuracy, so it has been widely used. The image analysis system for microorganism counting can improve counting performance if the quantity of sample is large~\citep{Thiran-1994-ARCC}. An example image of yeast cells is shown in Fig.~\ref{fig:Dietler-2020-ACNN-Example}. The precise boundaries of microorganisms make it possible to separate and count the number of colonies by image analysis. 

\subsection{Motivation of this review} 

\begin{figure}[ht]
\centering
\includegraphics[trim={0cm 0cm 0cm 0cm},clip,width=1.0\textwidth]{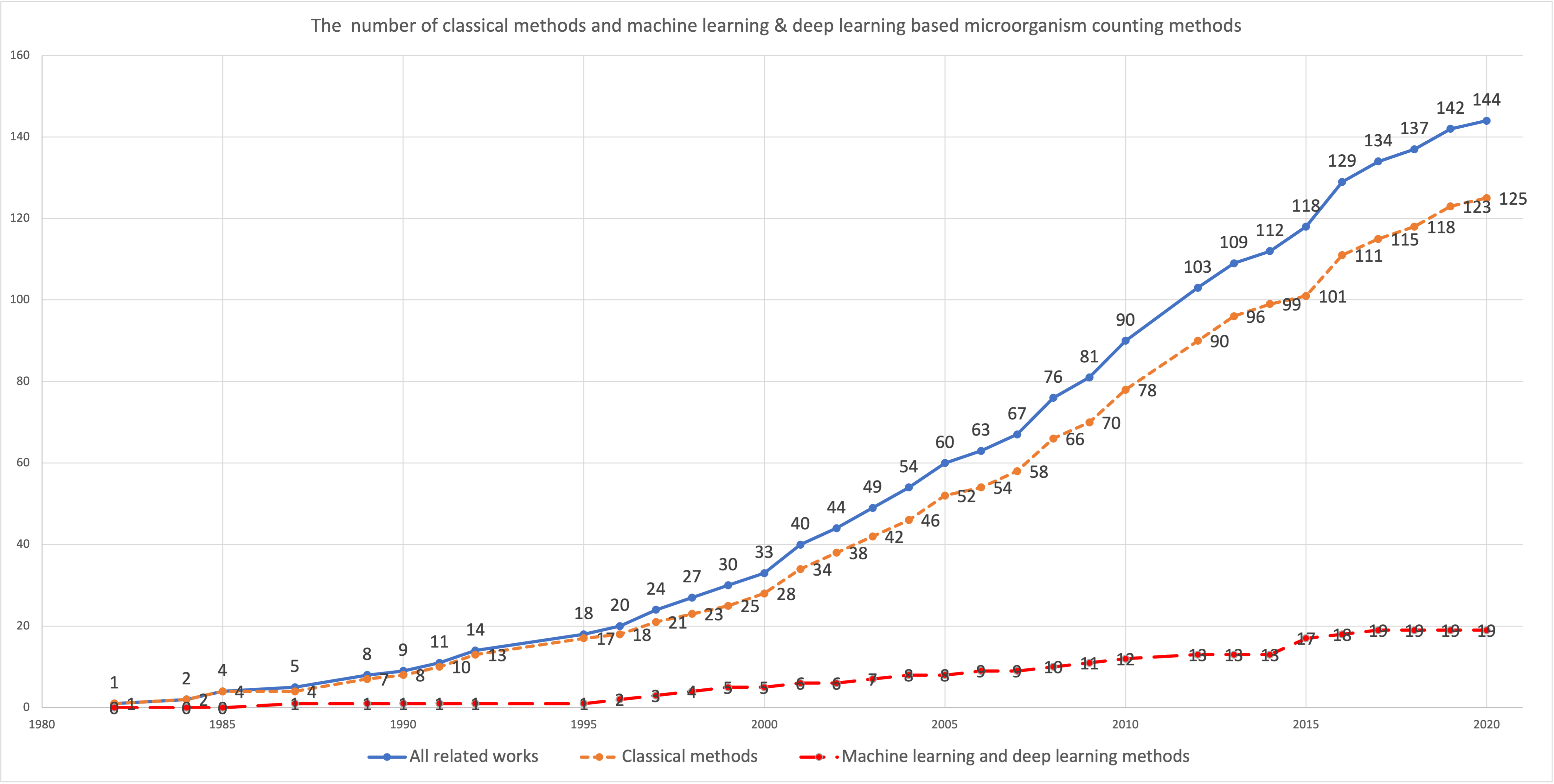}
\caption{The total number of related works on microorganism counting approaches.}
\label{fig:totalnum}
\end{figure}

Digital image processing (DIP), also known as computer image processing, refers to converting an image signal into a digital signal and processing it by computer. DIP first appeared in the 1950s, when the electronic computer has developed to a certain level. People can use the computer to process images and improve image quality~\citep{Gonzalez-2004-DIPU}. The commonly used digital image processing methods include image enhancement, denoising, restoration, coding and compression. 
DIP has been widely used in many fields. Agricultural and forestry departments understand the growth of plants through remote sensing images, estimate the yield, and monitor the development and management of diseases and insect pests~\citep{Amrita-2016-IPTA}. Through remote sensing image analysis, the water conservancy department can obtain the change of water disaster~\citep{Dodi-2012-APIP}. The meteorological department is used to analyze the meteorological cloud chart and improve the accuracy of the forecast~\citep{Chatterjee-2019-VISU}. The department of national defense, surveys, and mapping use aerial surveys or satellites to obtain regional landform and ground facilities~\citep{Xie-2017-BTAA}. The mechanical department can use image processing technology to analyze and identify the metallographic diagram automatically~\citep{Denis-2019-ATMM}. Medical departments use various digital image analysis technologies to diagnose various diseases automatically~\citep{Salvi-2020-TIPP, Madabhushi-2016-IAAM, Chen-2020-ARCH}. Because of the flexibility and universality of DIP, there are no complex measurement steps involved, which means it has low learning cost. In the field of microorganism analysis, expensive equipment is usually needed to ensure the accuracy of the measurement. DIP can save this part of this cost~\citep{Ekstrom-2012-DIPT}. Therefore, DIP has been widely used in microbial counting in many types of research. Its development trend is shown in Fig.~\ref{fig:totalnum}, which has shown a good development trend so far.

As shown in Fig.~\ref{fig:totalnum}, the application of digital image processing in the field of microorganism counting has been explored.  Since the 1980s, DIP has been applied to microorganism counting. From 1980 to 1995, the application and development of this field is relatively slow, but it is rapidly developed from 1995 to 2010. After 2010, the number of research for microorganism counting increases faster. 
As for machine learning and deep learning based microorganism counting methods, machine learning is firstly applied in this field in 1987, and then it is slowly developed from 1990 to 1995. Since 1990, microorganism counting methods based on machine learning and deep learning is increasing steadily, and it has a tremendous development since 2015. 
According to the content of the papers, a possible reason is summed up, that is, the development of the deep learning algorithm can lead to more accurate image segmentation. For example, the segmentation of adherent colonies can lead to more precise microorganism counting.

\subsection{Related reviews} 
The microorganism counting is an essential topic in microbial research, and the relevant works are also relatively abundant. Many researchers have written relevant reviews, which are summarized as follows:

Review~\cite{Gray-2002-CIAS} outlines several image analysis methods for algal cell estimating, and several image segmentation methods based on thresholding, edge tracking and template matching are compared. There are 32 papers summarized, and only three are about the algal counting method.
Review~\cite{Qiu-2004-AMTA} describes the development course of bacteria counting and cell size measurement, which contains the classical methods and automated flow analysis technology.  More than 33 papers are summarized and 7 of them are about bacteria counting.
Review~\cite{Gracias-2004-ARCD} describes the use of fluorogenic or chromogenic to classify different species of bacteria and impedance technology for enumeration. There are more than 25 papers are about traditional food bacteria counting methods in total 103 papers.
Review~\cite{Daims-2007-QUMF} indicates that the difference between microorganism counting and biovolume measurement is whether or not to identify individual objects (cell or cell clusters) in the biomass.  There are six papers about automatic cell counting in total 92 papers. 
Review~\cite{Barbedo-2012-ARMA} describes the object counting methods using digital image processing.  The methods are composed of morphological operation, filtering operation, contrast enhancement, transformation, edge detection and image segmentation.  They summarized over 130 papers, among them, 29 papers are used for cell counting and 13 papers are about bacteria counting.
Review~\cite{Dazzo-2015-UCIA} describes the use of CMEIAS for both microorganism counting and biovolume measurement based on image processing. The hierarchical tree classifier and $k$-Nearest Neighbour classifier are applied for classification. There are 65 papers in total, and more than 30 papers are used for cell counting.
Review~\cite{Li-2019-ASTA} describes computer-based microorganism image analysis development and introduces different methods for different microorganism classification. This review is a comprehensive microorganism classification paper. It uses plenty of works of literature for quoting, but there is no significant description for microorganism counting in more than 300 papers in total.
Review~\cite{Puchkov-2019-QMSC} describes the main quantitative analysis methods of single bacterial and yeast cells at the cellular and subcellular levels. More than 150 papers are summarized. This review mainly introduces several techniques for scanning, but there is no straightforward application of DIP in microorganism quantification.

\begin{figure}[ht]
\centering
\includegraphics[trim={0cm 0cm 0cm 0cm},clip,width=1.0\textwidth]{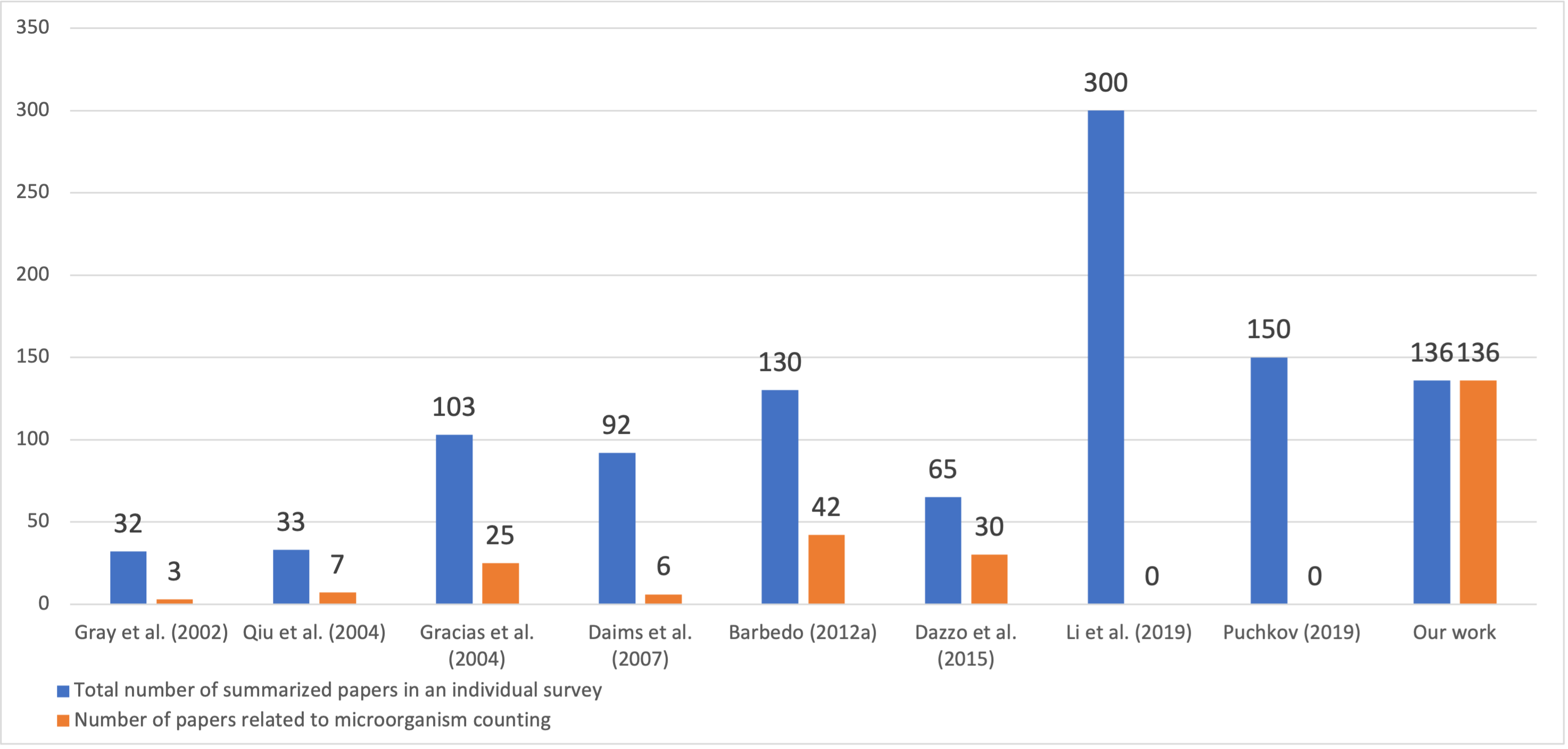}
\caption{A comparison among recent survey papers. Number of summarized papers in the existing review in comparison to their contribution to image analysis based microorganism counting methods.}
\label{fig:reviewimg}
\end{figure}

Although the reviews above are excellent enough and the descriptions about the current situation of microbial research are objective and detailed. However, there is no targeted research about image analysis based microorganism counting, so it is necessary to do additional research on this aspect. 
For a clear overview, a histogram (Fig. ~\ref{fig:reviewimg}) is used to show each of the related survey papers and their contribution to microorganism counting with our proposed studies.
Because of the vital role microorganism quantification plays in microbial research, this review focuses on the application of microorganism counting and summarizes each method's development and prospects. This review has great reference value for microbiological researchers and computer vision researchers.  There are more than 136 papers are used for microorganism counting.

\subsection{Microorganism counting methods} 
\begin{figure}[ht]
\centering
\includegraphics[trim={0cm 0cm 0cm 0cm},clip,width=1.0\textwidth]{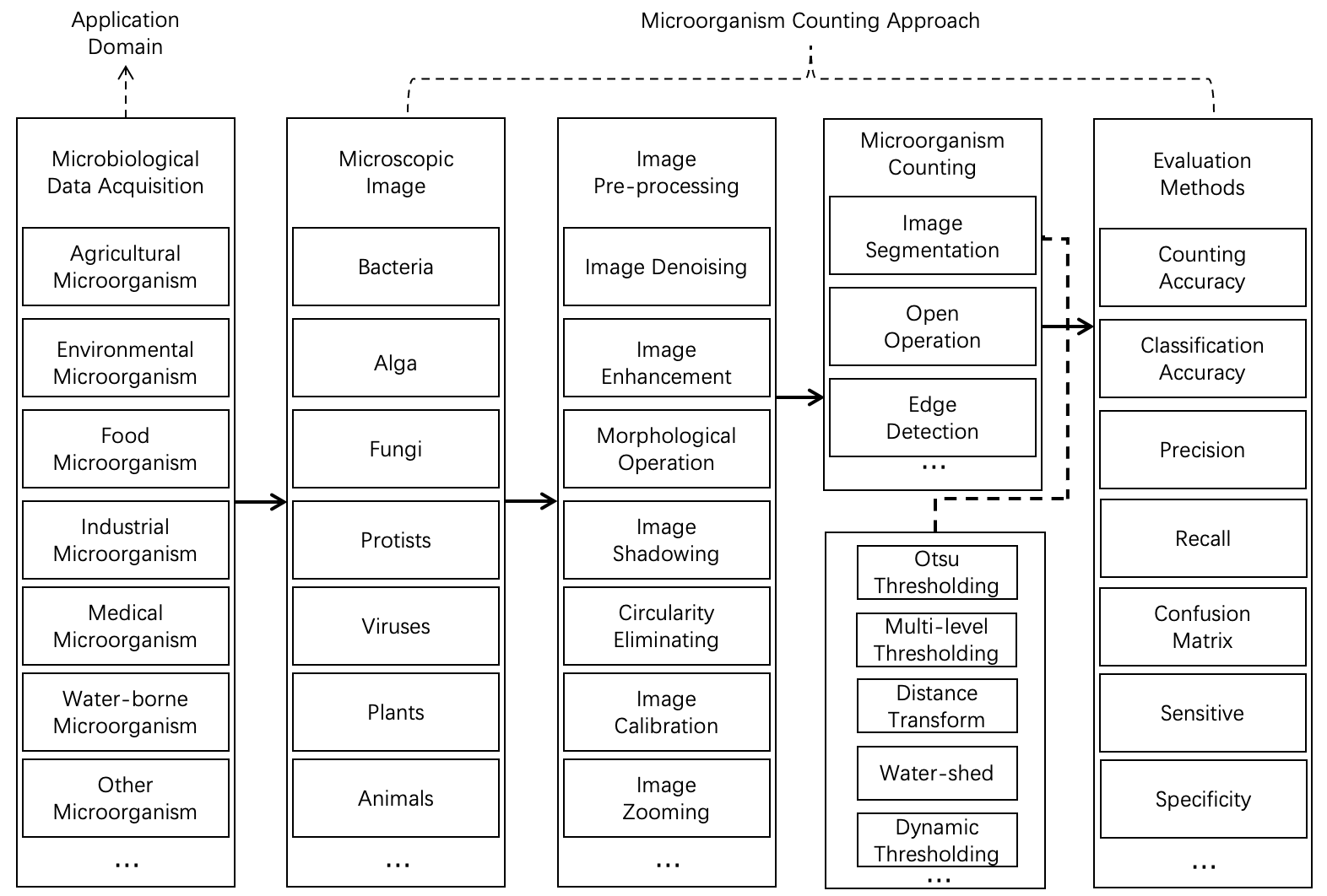}
\caption{The organisation chart of microorganism counting approaches in this paper.}
\label{fig:flowchart}
\end{figure}

In order to expound the approach of microorganism counting, the organization chart of this review is shown in Fig.~\ref{fig:flowchart}. The approach contains five steps: microbiological data acquisition, microscopic image, image pre-processing, microorganism counting methods and evaluation methods.  

Firstly, according to the different application domains, microorganisms are composed of the following seven categories: agricultural microorganism, environmental microorganism, food microorganism, industrial microorganism, medical microorganism, water-borne microorganism and other microorganisms. 
Then, the samples are stained and sliced. After that, the microscopic images are captured by the imaging equipment, such as a charge-coupled device (CCD) camera~\citep{Gmur-2000-AIES}.  

Next, in the pre-processing part, the images are denoised and enhanced to improve the contrast between the object particles and the background.
In the process of microorganism microscopic image scanning, the loss of information in the process of electronic transmission and the pretreatment process, such as staining may bring some noise to the microscopic image. In order to improve image quality, image processing can be used to reduce or remove these noises. The main methods to remove noise are wire filter (such as Gaussian filter and Mean filter), median filter and so on~\citep{Li-2020-ARCM}. 

The next step is microorganism counting~\citep{Li-2020-ARCM}. The objects of the microorganism counting method are separating the adherent colonies and counting. Image segmentation is an essential part of this task, which contains three broad categories: threshold segmentation, edge detection and region extraction. 
The initial image segmentation method is threshold segmentation, whose core algorithm is the selection of the threshold.  At present, there are two main methods, one is based on the iterative method and the other one is Otsu thresholding. For the image with prominent double peaks and deep valley bottom, the iterative method can get satisfactory results quickly, but for images with significant differences in the ratio of target and background, the iterative method cannot segment the target well~\citep{Perez-1987-AITA}. Another standard method is the maximum inter class variance based Otsu method which can achieve good segmentation results for most images~\citep{Otsu-1979-ATSM}. The advantages of Otsu segmentation are fast and straightforward calculation, not affected by brightness and contrast of images, and most of the segmentation results are satisfactory.  Nevertheless, it has limitations such as the sensitivity to noise and cannot support semantic segmentation functions~\citep{Xu-2011-CAOT}. 
Edge detection mainly includes gradient and second-order differential operator based methods,  Laplace of Gaussian function (LoG) edge detection method and Canny edge detection method.  The gradient detection method is the most widely used method among them that usually contains Roberts, Sobel, Prewitt, Kirsch and Robinson~\citep{Gonzales-2002-DIP}.
The Watershed method is one of the most popular methods in the region extraction domain, a closed region signature method based on region growth~\citep{Levner-2007-CDWS}.  The image segmentation is considered according to the composition of the watershed. The calculation process of the watershed is an iterative labeling process, which has an excellent response to weak edges. However, the watershed algorithm may lead to over-segmentation because of the noise or slight gray-level change of object surface~\citep{Strahler-1957-QAWG}.

Another critical part of microorganism counting is morphological operations which contain erosion, dilation, open and close. The erosion operation uses structural elements to erode the input image, eliminating the image's boundary points. It can reduce the size of the object, filter the image interior and eliminate the isolated noise points effectively~\citep{Jackway-1996-SSPT}.
Dilation operation is the dual operation of erosion operation. The dilation operation can merge all the background points contacted by the target object into the object, which can increase the target and the shrink holes~\citep{Jackway-1996-SSPT}.
The open operation is using the erosion operation firstly and then use the dilation operation. The open operation can eliminate the isolated points in the image, eliminate the burr and connect the two domains so that the outer boundary of the image can be polished by the open operation~\citep{Chudasama-2015-ISMO}.
The close operation is the opposite of the open operation, which means the image is dilated first and then eroded. The close operation can fill small holes, close small cracks, and polish the inner boundary of the image~\citep{Chudasama-2015-ISMO}.

After image segmentation, the microorganisms need to be classified and counted respectively. Machine learning is widely applied in image classification, which has been developed rapidly.
Principle component analysis (PCA) is an unsupervised machine learning algorithm, which is always applied for exploration and dimension reduction of higher dimensional data~\citep{Roweis-1998-EMAP}.
More comprehensible features can be extracted, and valuable information of the sample can be processed faster by using dimension reduction. In addition, dimension reduction can also be applied to visualization and denoising.
The primary process of PCA is to map $n$-dimensional features to $k$-dimensional features, which are new orthogonal features, also known as principal components. Then the $k$-dimensional features are re-constructed based on the original $n$-dimensional features.
PCA can increase the sampling density by dropping part of the information, which is helpful for the curse of dimensionality. 
However, PCA retains the primary information, which is only for the training set, but the primary information is not necessarily meaningful. So the overfitting may be exacerbated by using PCA~\citep{Karamizadeh-2013-AOPC}.
Support vector machine (SVM) is one kind of generalized linear classifier for data classification by using supervised learning~\citep{Vishwanathan-2002-SSVM}.
The object of SVM learning is to find the separation hyperplane with the most considerable geometric interval, which can divide the training data set correctly.
The learning strategy of SVM is to maximize the interval, which can be formalized into a problem to solve the convex quadratic programming. 
The selection of SVM kernels can make it to be a nonlinear classifier, such as polynomial kernel, RBF kernel, Laplacian kernel and Sigmoid kernel~\citep{Han-2012-PSSV}.

\begin{figure}[ht]
\centering
\includegraphics[trim={0cm 0cm 0cm 0cm},clip,width=0.7\textwidth]{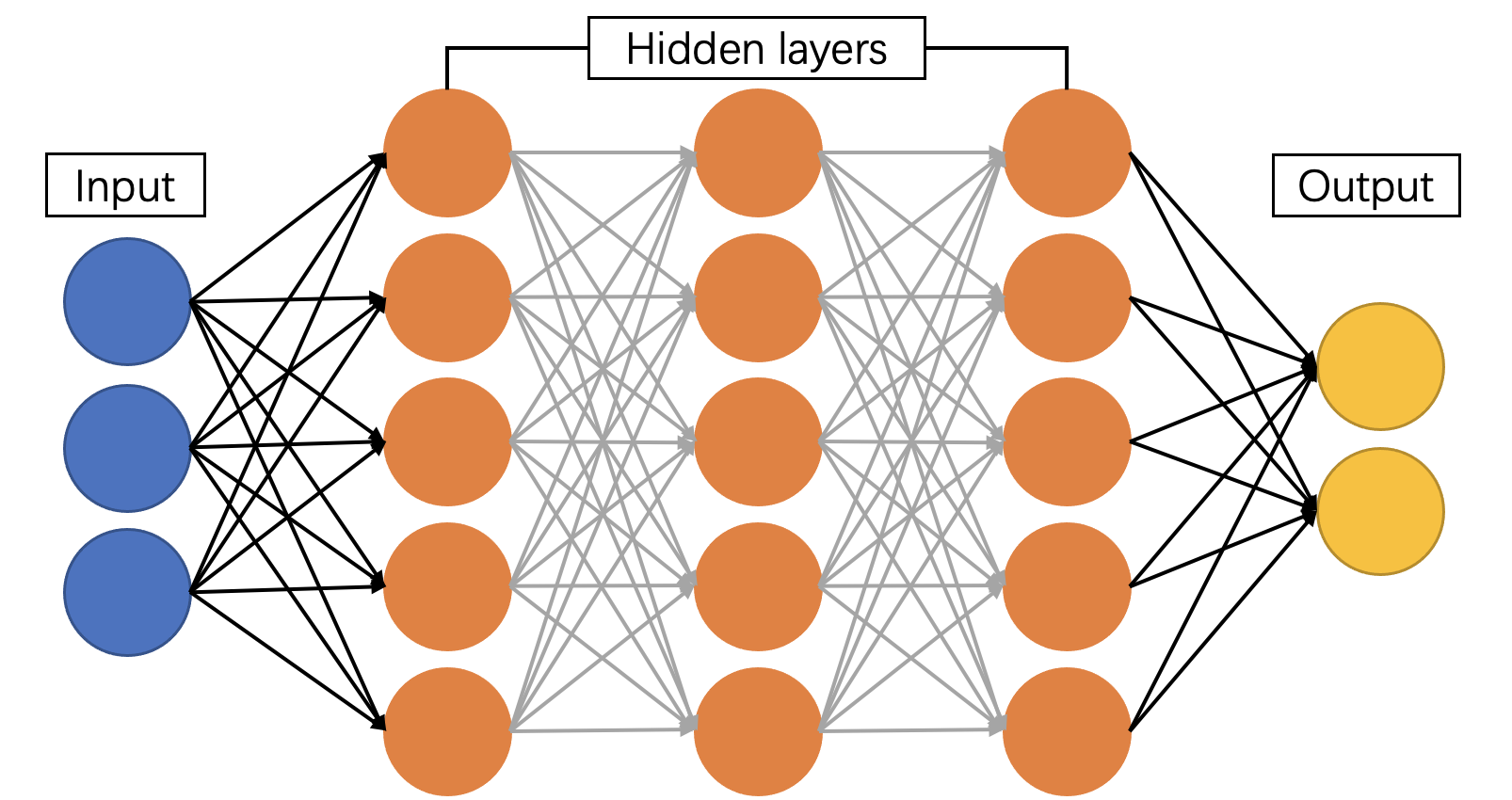}
\caption{The structure of ANN.}
\label{fig:ANN}
\end{figure}

Artificial neural network (ANN) is a mathematical model of distributed and parallel information processing that imitates animal neural networks' behavior characteristics.
The most commonly used ANN is Multilayer perceptron (MLP), which is a feedforward ANN model~\citep{Ghate-2010-OMLP}. ANN is composed of input layers, hidden layers and output layers, which are fully connected to each other. The structure of ANN is shown in Fig.~\ref{fig:ANN}. ANN is composed of many simple neurons, and each neuron receives input from other neurons. In this way, every neuron restricts and influences each other to achieve nonlinear mapping from input state space to output state space. ANN is a combination of many same simple processing units in parallel. Although the function of each unit is simple, the parallel activities and the ability of information processing are unique. ANN can realize the memory of information through its network structure, and the memory information is stored in the weights between neurons. This makes the network has good fault tolerance and can handle the pattern information processing such as clustering analysis, feature extraction and defect pattern restoration~\citep{Zupan-1994-IANN}.

Back propagation neural network (BPNN) is a supervised learning, which is developed from ANN~\citep{Larsoliya-2012-ANHL}. The loss function in BPNN is optimized based on back propagation. In forward-propagating, the data is processed from input layers to output layers. In back propagation, the loss function is transmitted from output layers to input layers, then the weights and biases are optimized based on the gradient descent method. BPNN can carry out the nonlinear mapping from input to output, and can still make the correct mapping for the new non-sample data, which has a specific generalization ability and fault tolerance ability~\citep{Dai-1997-ELPL}.

\begin{figure}[ht]
\centering
\includegraphics[trim={0cm 0cm 0cm 0cm},clip,width=0.8\textwidth]{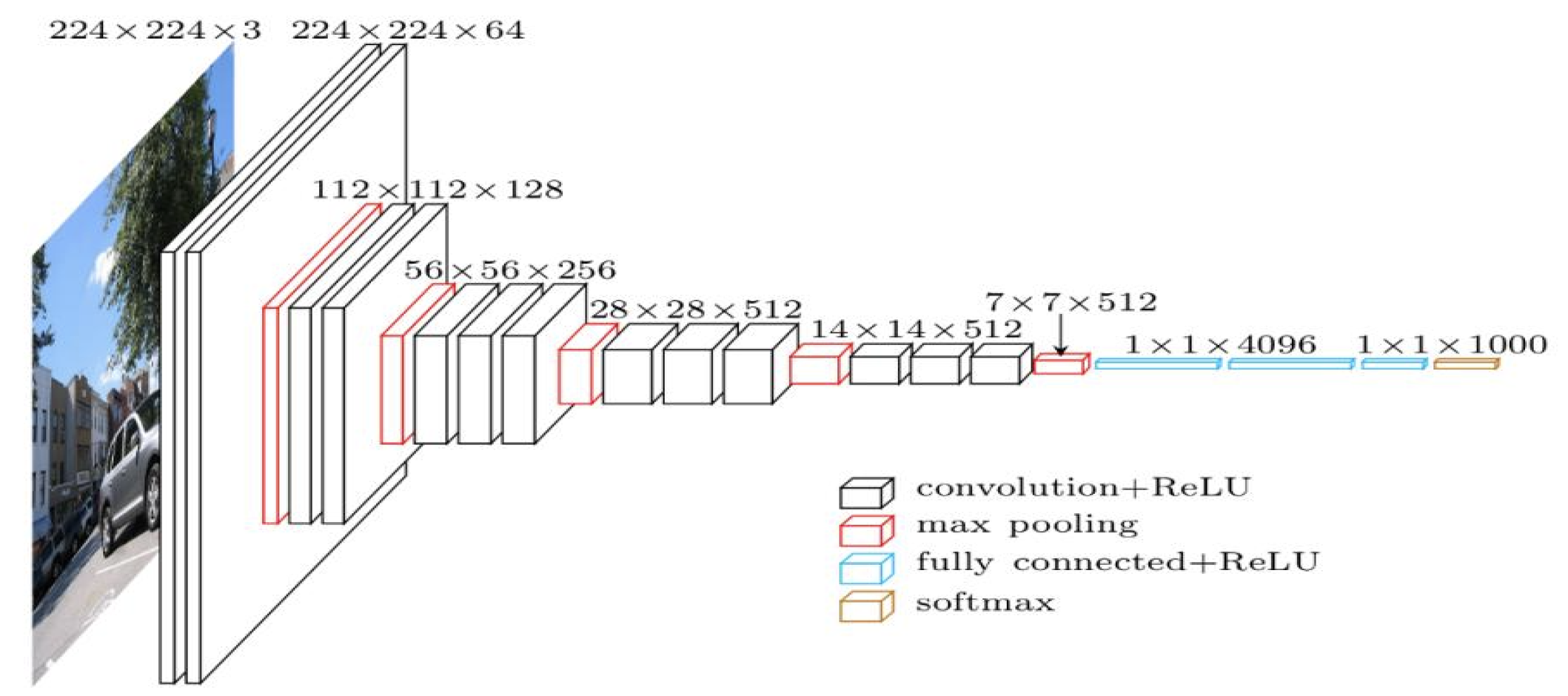}
\caption{The structure of VGG-16 (in~\cite{Nash-2018-ARDL} fig.3).}
\label{fig:VGG-16}
\end{figure}

Convolutional neural network (CNN) is one kind of feedforward ANN, that is wildly applied in DIP and computer vision~\citep{Li-2016-SCNN}. The convolutional kernels are applied to scan the whole image and the deep features are then extracted. After pooling, the image can be classified through a fully connected layer. The loss function is minimized with back propagation~\citep{Chauhan-2018-CNNI}. CNN is developed rapidly after 2010, and it is not only be applied for classification, but segmentation (Unet) and image generation (GAN). 
VGG-16 is one of the most popular CNN, which is composed of five convolution layers, three pooling layers and three fully connected layers~\cite{Simonyan-2014-VDCN}. The structure of VGG-16 is shown in Fig.~\ref{fig:VGG-16}.
Only 3 $\times$ 3 filters are applied in VGG-16 because the combination of small filters can simulate a larger filter, reducing the parameters and improving the nonlinear ability.

The last step is system evaluation, which can help researchers systematically perceive the image processing results. The results of the evaluation can help to prompt the accuracy of the system. 
Counting accuracy is a standard evaluation method in target counting, which is the ratio of the number of detected targets to the ground truth. Generally, the accuracy can only be used to evaluate the global accuracy, because it cannot show whether the detected target and the ground truth target are one-to-one corresponding~\citep{Bloem-1995-FADS}.
Evaluation of image segmentation and image classification can also reflect the performance of the counting system. True positive (TP), false negative (FN), false positive (FP) and true negative (TN) are four basic metrics in image classification~\citep{Zhang-2008-ISEA}. 
Pixel accuracy (PA) is one of the simplest evaluation methods for image segmentation, which means the ratio of the number of correctly classified pixels and the number of whole pixels. The mean pixel accuracy (MPA) is the improved method of PA, which indicates the mean PA of all classes~\citep{Zhang-2008-ISEA}.  
Mean intersection over union (MIoU) is the ratio of intersection and union of ground truth and predicted segmentation result. It can be regarded as the mean ratio of TP and the union of TP, FN and FP in the process of image segmentation~\citep{Rahman-2016-OIOU}.

\subsection{Structure of this review} 
\begin{figure}[ht]
\centering
\includegraphics[trim={0cm 0cm 0cm 0cm},clip,width=1.0\textwidth]{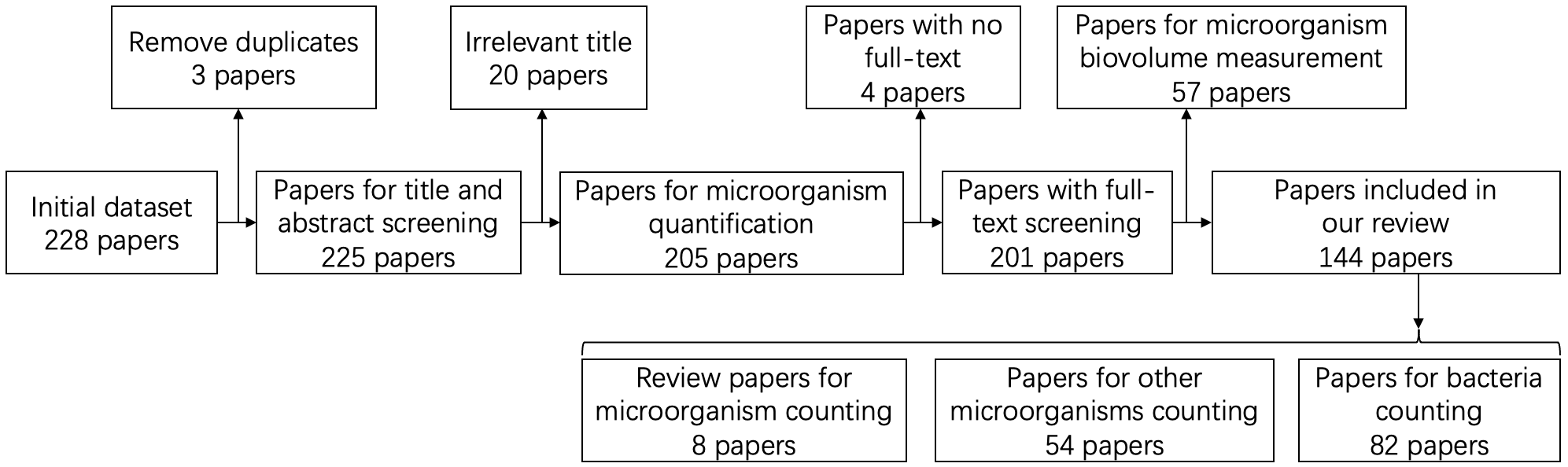}
\caption{The systematic flow chart of paper selection for our work.}
\label{fig:reviewselect}
\end{figure}

In this review, a comprehensive overview of microorganism counting using image analysis is presented. The relevant research in the microbial application has been investigated since 1980, and the applications of microorganism counting in different situations are discussed. Furthermore, this paper also summarizes the research motivation and research methods of microorganism counting in the microbial field. The review articles related to this research are also summarized, and the structures of their references are recorded.
More than 144 papers are selected from the initial paper dataset and the structure of the systematic
review is shown in Fig. ~\ref{fig:reviewselect}. 
The initial papers are searched from Google Scholar, IEEE, ACM, Nature, Science, Cell, Elsevier, Wiley, Hindawi, IOP, PloS, BMC and Springer, and the keywords contain ``microorganism counting'',  ``bacteria counting'', ``cell counting'', ``algae counting'' and ``fungus counting''. Then the duplicate and irrelevant papers are deleted. There are 57 papers about microorganism biovolume counting, which do not conform to this review.  Finally, 144 papers are about microorganism counting methods, containing 8 review papers, 82 bacteria counting methods, and 54 other microorganism counting methods.

The review is structured as follows: In Sect. 2,  the related works of image analysis based bacteria counting are introduced.  In Sect.3, the related works of other microorganisms counting based on image analysis are introduced. 
Then in Sect. 4, the commonly used microorganism counting methods are analyzed, and their different application domains are summarized.  Finally, in Sect. 5 this review is concluded by summarizing the whole paper. This review structure can help microbiological workers clearly and quickly understand the development status of this field and obtain the relevant content they need.

\section{Bacteria counting methods}
Bacteria are one crucial part of the ecosystem because of the cardinal role they play in the carbon and nitrogen cycle, which are closely related to human daily life~\citep{Madigan-1997-BBOM}. Therefore, bacteria counting has become one of the most important directions in the field of microorganism counting, including the study of bacteria number and colony size. Thus, the bacteria counting methods are summarized in this chapter. Bacteria counting is of great significance in food safety monitoring and industrial safety detection, but manual counting is tedious and redundant work, that is very subjective. Therefore, the research of computer image analysis based bacteria counting is significant. 
This chapter is structured as follows: the first part summarizes the classic bacteria counting methods. The second part is bacteria counting methods based on machine learning and deep learning. The last part is third-party tools methods.

\subsection{Classic counting methods}
\subsubsection{Bacteria counting method based on image enhancement}

\begin{figure}[ht]
\centering
\includegraphics[trim={0cm 0cm 0cm 0cm},clip,width=1.0\textwidth]{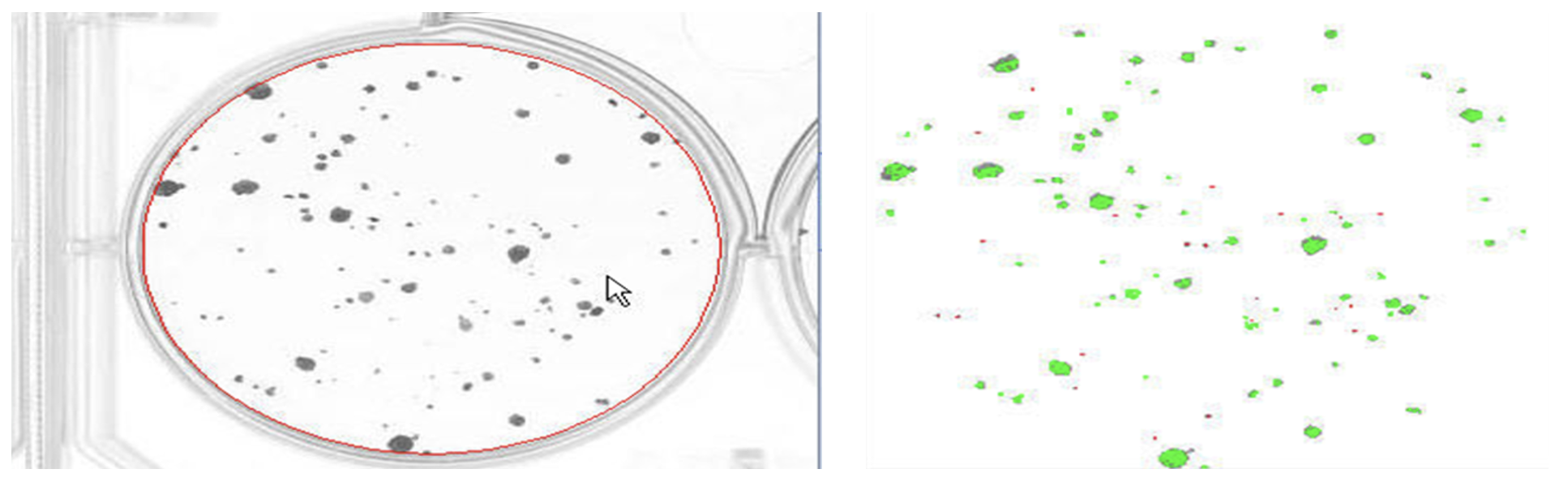}
\caption{Screenshot of the program. The left panel is the scanned flask and the right panel is the segmented result (in~\cite{Niyazi-2007-CCCA} fig.1).}
\label{fig:Niyazi-2007-CCCA}
\end{figure}

In~\cite{Pettipher-1982-SACB, Niyazi-2007-CCCA}, the gray-level contrast is applied for bacteria counting.  The contrast is used for counting bacteria and somatic cells of milk (\cite{Pettipher-1982-SACB}).
In~\cite{Niyazi-2007-CCCA}, the maximum size of one colony (defined by the area) and the distribution of the gray color within the colony are further measured beside the gray-level contrast, which has the mistake of less than 3\%. The colony counting result is shown in Fig.~\ref{fig:Niyazi-2007-CCCA}.

In~\cite{Shenglang-2005-TJBR}, histogram equalization is used to enhance the contrast of bacteria images as pre-processing. Then a convolutional filter is used to extract the curvature feature of bacteria and remove the background features. After that, a median filter is used to remove noises. The basis of judging whether there are bacteria in the image is whether the maximum connected number of non-zero gray pixels in the image exceeds a certain threshold. Finally, the number of connected domains is calculated as the number of bacteria in the image. The result shows that the accuracy is more than 95\% while counting the number of bacteria. 

In~\cite{Buzalewicz-2010-IPGA}, optical transforms are used for the determination of bacteria colony number. Because of the scale invariant of Mellin transform, the combination method of Fourier transform and Mellin transform is applied. First, the Fourier transform of the input objects is calculated and the high pass frequency filter is applied to eliminate the zero-order component of the Fourier spectrum. Then the Mellin transform followed by the log-polar transformation is performed. Moreover, the two-dimensional Fourier transform is computed in order to obtain scale and rotation invariance. Finally, the value of the Mellin spectrum is used to evaluate the number of the analyzed object. A good agreement between calculations and manual counting for twelve samples is achieved (the differences range from 1 to 3\% and the standard deviation is equal 4.51).

\subsubsection{Bacteria counting method based on thresholding}
\begin{figure}[ht]
\centering
\includegraphics[trim={0cm 0cm 0cm 0cm},clip,width=0.6\textwidth]{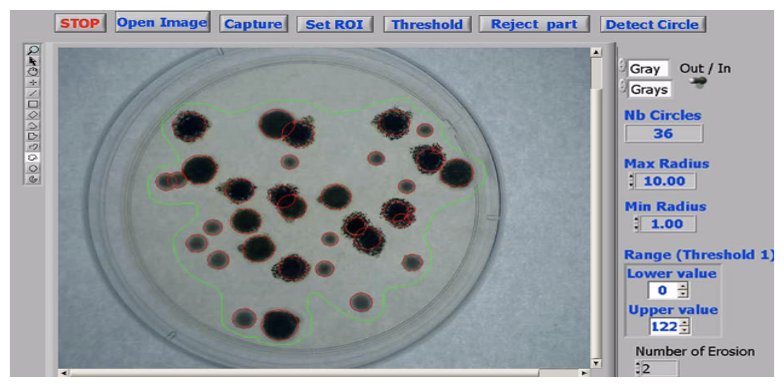}
\caption{The detection result (in~\cite{Chunhachart-2016-CAVE} fig.8).}
\label{fig:Chunhachart-2016-CAVE}
\end{figure}

\begin{figure}[ht]
\centering
\includegraphics[trim={0cm 0cm 0cm 0cm},clip,width=0.6\textwidth]{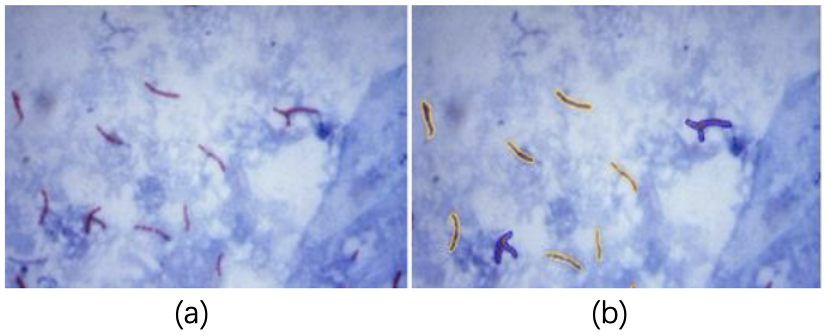}
\caption{The counting result. (a) Original image. (b) Image after shape characterization and segmentation (in~\cite{Payasi-2017-DACT} fig.8).}
\label{fig:Payasi-2017-DACT}
\end{figure}

In~\cite{Masuko-1991-ANMD,Trujillo-2001-AMVS, Chunhachart-2016-CAVE,Gupta-2012-MABC,Sethi-2012-BCCM,Kaur-2012-ANMA,Nayak-2010-ANAA,Pernthaler-1997-SCAI,Sotaquira-2009-DAQB,Maretic-2017-ACCH,Chunhachart-2016-CAVE,Kaur-2012-ANMA,Payasi-2017-DACT}, global thresholding is applied for bacteria enumeration. 
The RGB image is firstly converted to YCbCr and Lab color spaces in~\cite{Sotaquira-2009-DAQB}.
An adaptive median filter (\cite{Gupta-2012-MABC,Kaur-2012-ANMA}), a`flatten filter' (\cite{Trujillo-2001-AMVS}) or a top-hat algorithm (\cite{Pernthaler-1997-SCAI}) is applied for denoising. Besides, a combination method of Gaussian low-pass filter, simple symmetric moving average filter and median filter is applied for noise removal in~\cite{Maretic-2017-ACCH}. After thresholding, the morphological operations are applied for image enhancement (\cite{Chunhachart-2016-CAVE,Kaur-2012-ANMA}), and the detected circles are used for counting the number of colonies that appeared on the selected region. The detection result is shown in Fig.~\ref{fig:Chunhachart-2016-CAVE} and the average percentage error of 2.13\% is obtained in comparison to the counting by the expert.
In~\cite{Payasi-2017-DACT}, the RGB image is converted to HSI color space image and the image is then segmented based on thresholding. Then the noises are removed, so the labeling and counting of the bacilli in the image will be possible.  Afterward, the boundaries of bacteria are detected and stored, and the area and perimeter are calculated. If there is a clump of bacilli, the count of bacilli is increased by the integer, which is closed to the ratio of the area of the clump to the average area of bacilli. The result is shown in Fig.~\ref{fig:Payasi-2017-DACT} and accuracy of 90\% is obtained.

In~\cite{Shen-2010-ESAC,Austerjost-2017-ASDA}, the iterative local threshold is applied for bacteria colony counting. 
First, a median filter and contrast enhancement are applied to remove noises and enhance images (\cite{Shen-2010-ESAC}). Then the iterative local threshold method is used for image segmentation. After that, the petri dish edge is removed by detecting the connected region with the maxima white pixel. 
Finally, the number of bacteria is counted based on eight neighborhoods in~\cite{Shen-2010-ESAC} and the average relative error of 2.5\% is obtained. 
In~\cite{Austerjost-2017-ASDA}, after thresholding, the region of interest is examined for objects which will be divided into single colonies and colony clusters by using a classification algorithm based on the previously defined threshold. Afterward, a Hough circle transformation is applied for the segregation of colony clusters into single colonies. The last part of the algorithm is dedicated to finding colonies that could not be detected within the previous steps. For this, the sizes of previously found colonies are compared with other objects found on the plate. If these objects fit into the size range of previously found colonies and have a suitable roundness, they are recognized as a colony. Finally, the detected colonies are all counted with an average accuracy of 86.76$\pm$9.76\%.

\begin{figure}[ht]
\centering
\includegraphics[trim={0cm 0cm 0cm 0cm},clip,width=1.0\textwidth]{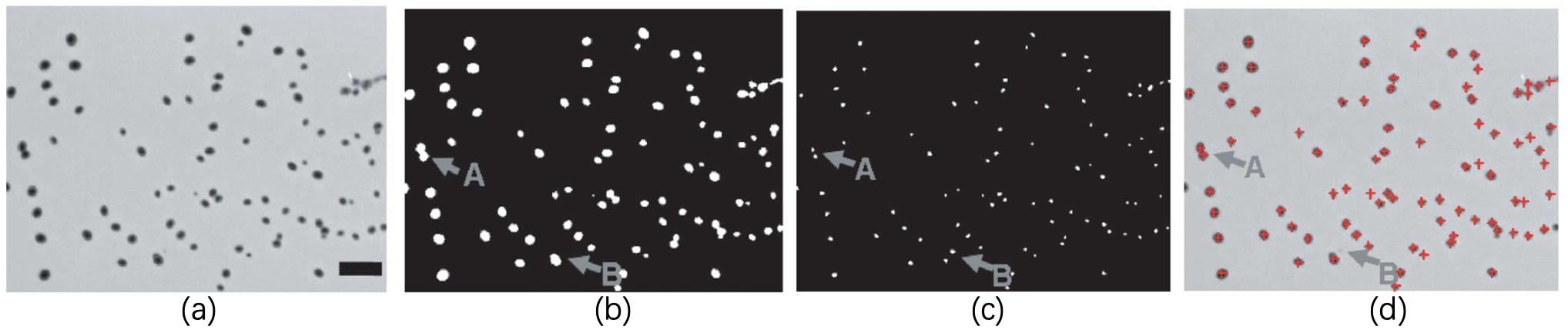}
\caption{Illustration of the colony counting procedure. (a) The initial image. (b) The image after thresholding. (c) The extended minima of the original image. (d) The counted colonies (in~\cite{Clarke-2010-LCHT} fig.4).}
\label{fig:Clarke-2010-LCHT}
\end{figure}

\begin{figure}[ht]
\centering
\includegraphics[trim={0cm 0cm 0cm 0cm},clip,width=0.6\textwidth]{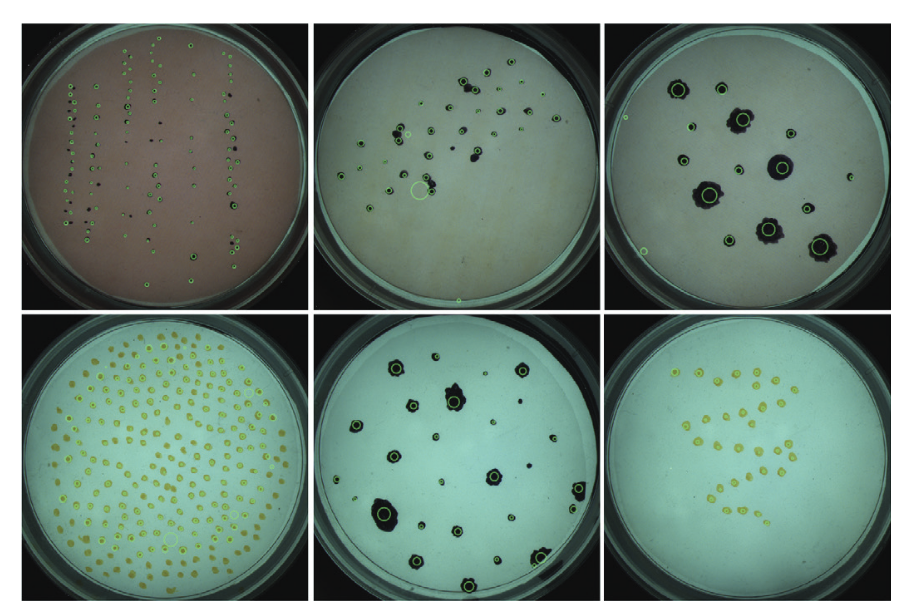}
\caption{Test result (in~\cite{Matic-2016-SAPS} fig.7).}
\label{fig:Matic-2016-SAPS}
\end{figure}

In~\cite{Jun-2010-RDRM,Clarke-2010-LCHT,Marotz-2001-EORA,Feng-2013-ACAI,Boukouvalas-2018-ACCS,Matic-2016-SAPS,Siqueira-2017-MFSA}, the adaptive threshold is used for bacteria counting. 
First, a median filter (\cite{Jun-2010-RDRM,Boukouvalas-2018-ACCS,Siqueira-2017-MFSA}),  a Gamma correction (\cite{Matic-2016-SAPS}) and a Gaussian filter (\cite{Clarke-2010-LCHT}) are used for noise removal. Then the extended minima function is used to find the center of the colonies (\cite{Clarke-2010-LCHT}). 
Finally,  a saturation based adaptive thresholding is applied for image segmentation. The morphological operations such as opening and closing are used for adherence colonies segmentation and image smoothing. The segmentation procedure is shown in Fig.~\ref{fig:Clarke-2010-LCHT}. The results correlate well with the results obtained from manual counting, with a mean difference of less than 3\%. Moreover, the distance transform and progressive erosion are applied in~\cite{Feng-2013-ACAI} to separate connected colonies into a single one. The counting result of~\cite{Matic-2016-SAPS} is shown in Fig.~\ref{fig:Matic-2016-SAPS}. 
In~\cite{Boukouvalas-2018-ACCS}, the circular area is detected through Hough transform to obtain only the inner area of the dish and a mask is created for the removal of the unwanted area. Afterward, Gaussian adaptive thresholding is performed for image segmentation because of the different lighting conditions in different areas. Then the histogram of vertical projections of the image is analyzed by varying its rotation angle to align the stripe at a 90$^\circ$ angle. Finally, cross correlation-based granulometry is applied for the determination of the amount of bacteria colonies.

In~\cite{Zhang-2008-AABC,Alves-2016-CCVA,Sanchez-2016-MAAC,Boukouvalas-2019-ASMC}, Otsu thresholding is applied for bacteria counting. 
First, a linear expansion of the histogram (\cite{Sanchez-2016-MAAC}) and a multi-directional Sobel operation (\cite{Boukouvalas-2019-ASMC}) is applied for image enhancement and edge detection.
In~\cite{Zhang-2008-AABC}, the RGB and achromatic are processed, respectively.
For RGB images, the Otsu thresholding method is firstly used for segmentation, then the color similarity in HSV (Hue-Saturation-Value) color space is adopted to assist the colony boundaries detection. For achromatic images, the sizes of all objects detected by the Otsu method from the dish/plate region are collected, and the frequency distribution with log base of those size values is generated. Colonies of similar size should occupy the high frequency segment in this distribution, and the frequencies for those massive artifacts should be very low. By this assumption, the large size objects can be removed. Then the hypothesis testing is used to remove minor artifacts which are very similar to the colonies. 
After Otsu thresholding, a Laplacian filter is applied for edge detection and circular Hough transform is used to detect circular bacteria colonies in~\cite{Alves-2016-CCVA,Boukouvalas-2019-ASMC}. The mean error between the proposed method and the manual counting method is less than 10\%. In~\cite{Sanchez-2016-MAAC},  Euler’s method is applied for colony counting. The accuracy of 98\% is obtained by comparing with the proposed method and manual counting method.

\begin{figure}[ht]
\centering
\includegraphics[trim={0cm 0cm 0cm 0cm},clip,width=1.0\textwidth]{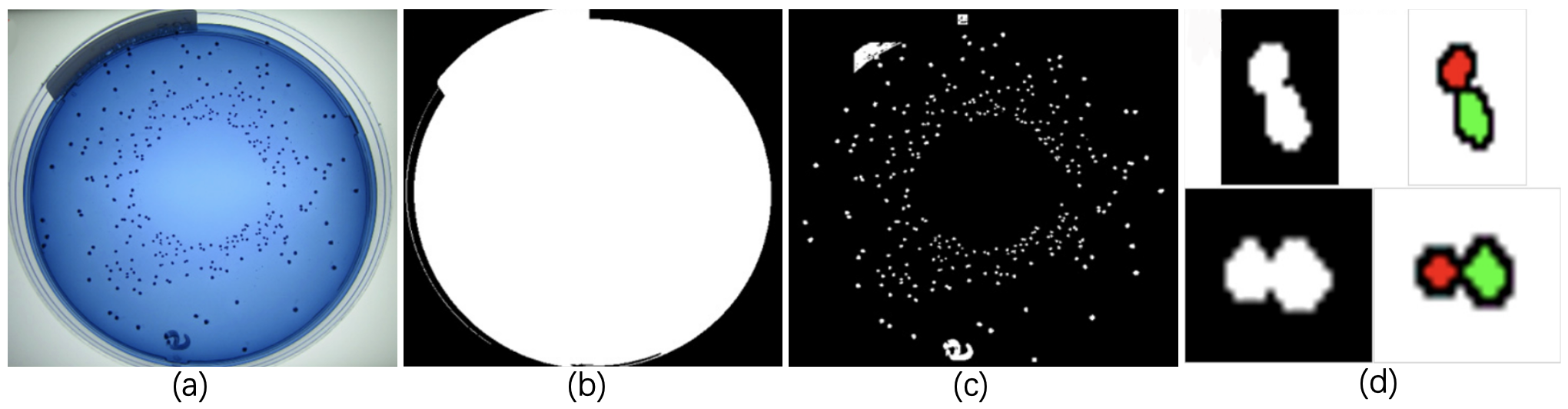}
\caption{(a) The original image. (b) The plate mask.  (c) The colony mask. (d) Colonies separated from aggregated colony clusters (in~\cite{Chen-2008-BCEA} fig.2).}
\label{fig:Chen-2008-BCEA}
\end{figure}

In~\cite{Zhang-2007-AEAR,Chen-2008-BCEA}, the Otsu and watershed are applied for automatic detection and enumeration of bacteria colonies. First, the contrast-limited adaptive histogram equalization (CLAHE) is used on the converted gray-scale images to enhance the dish/plate contour (\cite{Zhang-2007-AEAR}). 
Then, the Otsu threshold is used to detect the dish/plate region and binarize the images automatically.  After the morphological operation is used to fill holes, the color similarity values between a pixel and its eight neighbors are calculated and the minimum value is used to detect the object boundaries. Moreover, the watershed algorithm is used for clustered colony separation and the number of viable colonies is counted. The proposed counter performs very well on the blue medium dish/plate, which has average precision, recall, and F-measure values of 0.97, 0.96, and 0.96, respectively in~\cite{Zhang-2007-AEAR}. In~\cite{Chen-2008-BCEA},  The precision, recall, and F-measure values of the proposed counter are 0.61$\pm$0.29, 0.94$\pm$0.06, and 0.69$\pm$0.20, while the corresponding values of the Clono-Counter are 0.22$\pm$0.25, 1.00$\pm$0.00, 0.29$\pm$0.31, respectively. The segmentation method is shown in Fig.~\ref{fig:Chen-2008-BCEA}.

\subsubsection{Bacteria counting method based on edge detection}
\begin{figure}[ht]
\centering
\includegraphics[trim={0cm 0cm 0cm 0cm},clip,width=1.0\textwidth]{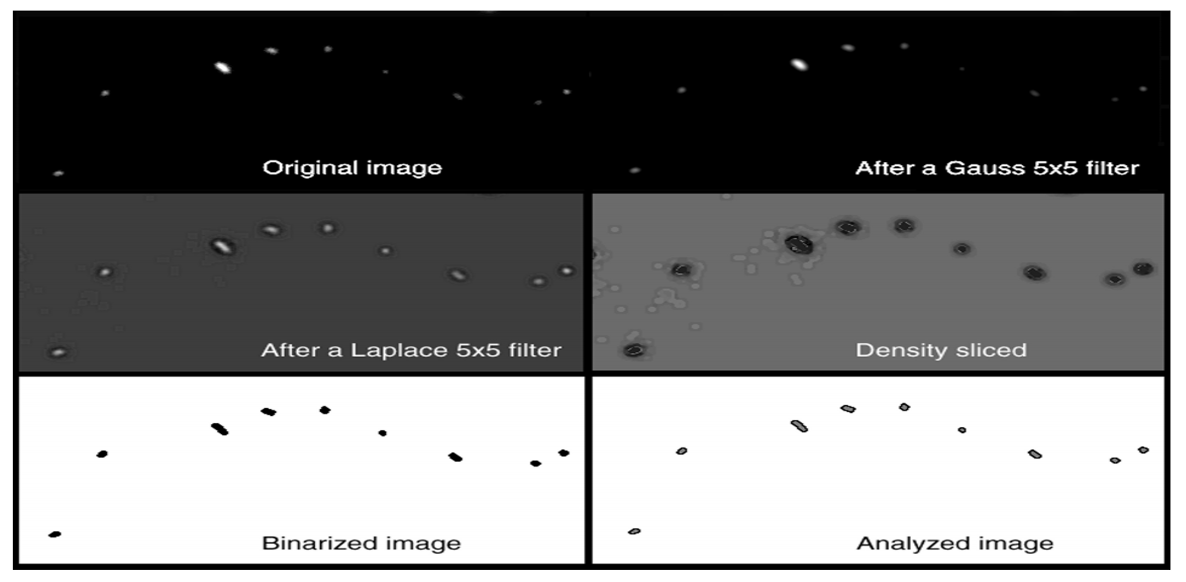}
\caption{Overview of the whole process of image processing (in~\cite{Massana-1997-MBSV} fig.2).}
\label{fig:Massana-1997-MBSV}
\end{figure}

In~\cite{Massana-1997-MBSV,Ogawa-2003-DMDI,Yamaguchi-2004-MEDC,Choudhry-2016-HTMA},edge detection is applied for bacteria counting.
A Gauss filter (kernel 5$\times$5), a Laplace filter (kernel 5$\times$5), and a median filter (rank 3) are used for edge detection before thresholding (\cite{Massana-1997-MBSV}), which is shown in Fig.~\ref{fig:Massana-1997-MBSV}. On the contrast, the Sobel and Laplacine filter are used to detect the edges after thresholding (\cite{Yamaguchi-2004-MEDC}). 
In~\cite{Choudhry-2016-HTMA}, the edge detection system has six major steps. First, the background is subtracted to enhance contrast and reduce the effects caused by uneven illumination. The radius for background subtraction is determined empirically. A starting number can be the average radius of colonies. The next step is sharpening and enhancing the image, which follows by finding the edges. Sobel filter is used in the macro. Then the image is smoothed using Gaussian blur, and converted to black and white. Alternatively, the image can be smoothed by sequential dilate and erode steps. This is followed by the closing of the edges to form a closed circle. Closed objects are filled black using holes filling resulting in images containing black colonies on a white background. To ensure that all colonies are detected, an additional step of closing and filling holes is performed. Here, the size of each pixel is increased, in order to bring the detected edges closer to each other that allows the detection of colonies whose entire edge along the perimeter fails to be otherwise detected. After filling, the size of the pixels is reduced to return the colony size to their original values. After that, denoising and segmentation are applied to remove small particles and separate clustered colonies using thresholding. Finally, the objects are filtered based on size, circularity and measured. Then a new pipeline is developed for the detection of cells and colonies from images. The background is corrected and then the colonies are detected. After that, the parameters are measured and the number of colonies is counted. 

In~\cite{Barbedo-2013-AACM}, five digital processing methods for automatic colony counting are proposed and compared. In the first method, a Gaussian Laplacian filter is applied for edge detection and the connected regions are identified and counted after holes filling. In the second method, the Gaussian Laplacian filter is replaced by the Canny filter. In the third method, three thresholding values are used for image segmentation. In the fourth method, thresholding is used for histogram equalization but not for image segmentation. In the fifth method, the region growing method is applied for segmentation. After that, the concave surface between the connected colonies can be detected to separate the colonies into a single one.  Finally, the number of colonies is counted. The accuracy of the first method performs best that obtains the accuracy of 99\%.

\subsubsection{Bacteria counting method based on watershed}
\begin{figure}[ht]
\centering
\includegraphics[trim={0cm 0cm 0cm 0cm},clip,width=0.5\textwidth]{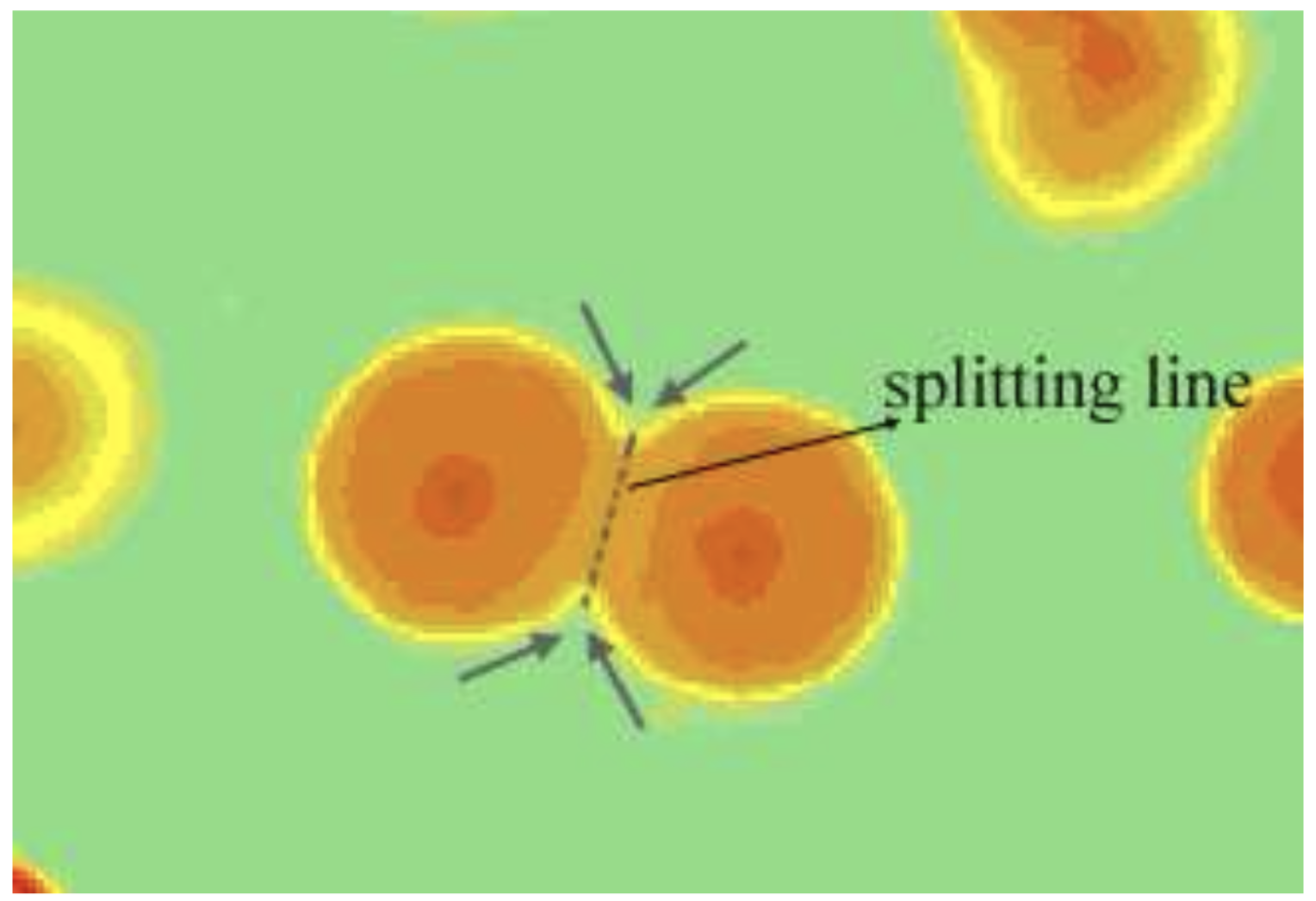}
\caption{Watershed segmentation for two merged colonies (in~\cite{Ates-2009-AIPB} fig.5).}
\label{fig:Ates-2009-AIPB}
\end{figure}

In~\cite{Ates-2009-AIPB,Selinummi-2005-SQLB}, watershed is applied for bacteria counting. 
In~\cite{Ates-2009-AIPB}, a median filter is applied first for noise removal and the petri dish boundary is detected and removed. Then the patterns are separated into two groups: colonies and clusters of colonies, based on the classification of circularity ratio. After that, the cluster colonies are segmented based on the watershed (\cite{Ates-2009-AIPB}) and marker-controlled watershed(\cite{Selinummi-2005-SQLB}). The watershed segmentation method is shown in Fig.~\ref{fig:Ates-2009-AIPB}. Finally, the number of actual colonies is estimated as the ratio of cluster area to an average colony area.

\begin{figure}[ht]
\centering
\includegraphics[trim={0cm 0cm 0cm 0cm},clip,width=0.6\textwidth]{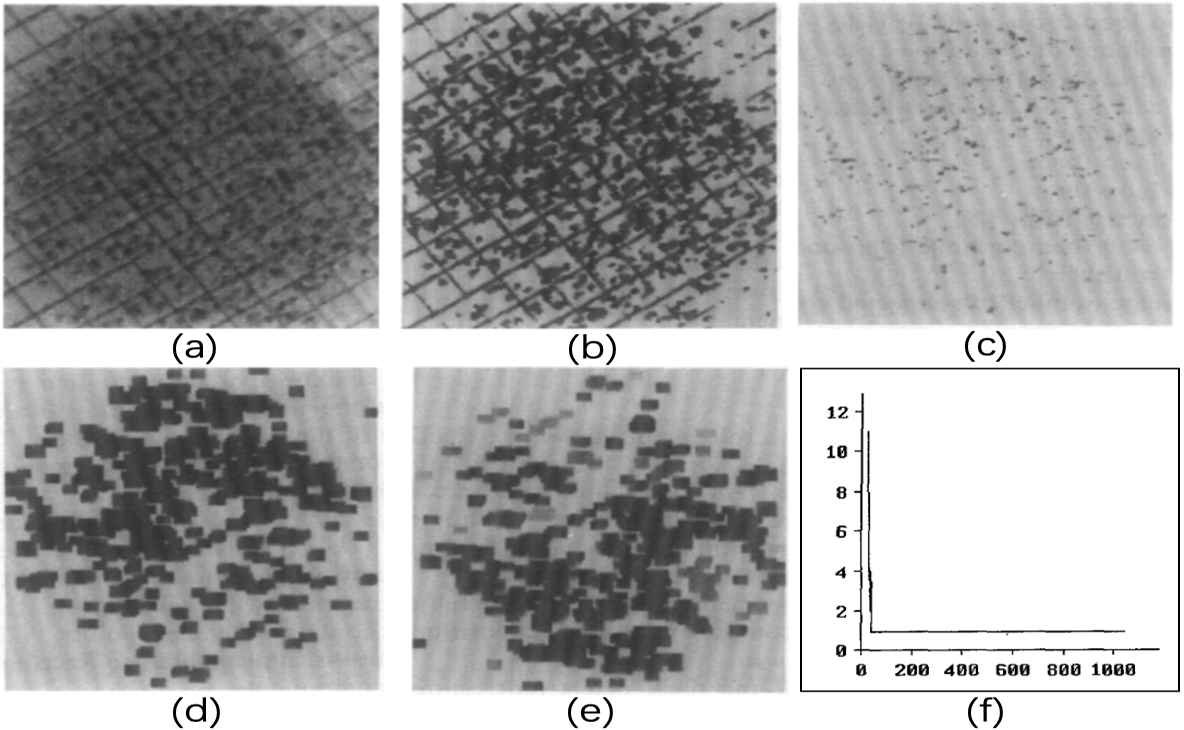}
\caption{(a) Original image. (b) Image after thresholding at gray value 125. (c) Image after distance transform. (d) Image after region growing. (e) Image after component labelling. (f) Frequency distribution (in~\cite{Mukherjee-1995-BCCD} fig.3).}
\label{fig:Mukherjee-1995-BCCD}
\end{figure}

\begin{figure}[ht]
\centering
\includegraphics[trim={0cm 0cm 0cm 0cm},clip,width=0.6\textwidth]{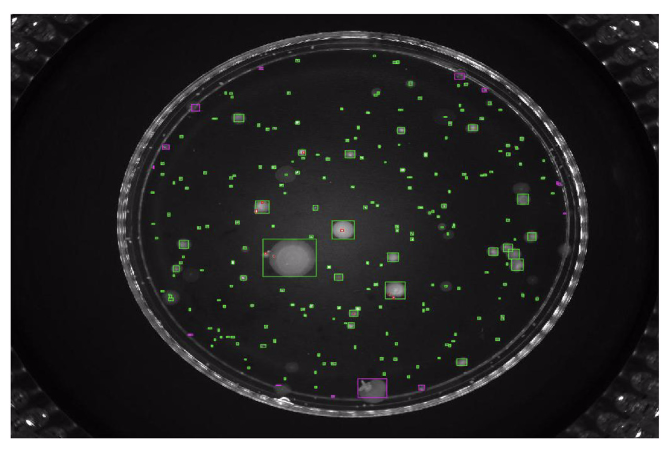}
\caption{Identified colonies displayed in different colors (in~\cite{Zhu-2018-ACBC} fig.11).}
\label{fig:Zhu-2018-ACBC2}
\end{figure}

In~\cite{Hong-2008-SHBC,Brugger-2012-ACBC,Masschelein-2012-TACC,Zhu-2018-ACBC,Wong-2016-ACCA,Minoi-2016-MVBA,Yujie-2009-DIMB,Kan-2008-QQSB,Fang-2008-ESAC,Martinez-2016-NTBC,Mukherjee-1995-BCCD}, the distance transform and watershed are applied for bacteria counting. First, a median filter is used to remove noise and determine the threshold for every single patch in~\cite{Kan-2008-QQSB,Fang-2008-ESAC}.
After that,  the iterative threshold (in~\cite{Fang-2008-ESAC}), a gray-scale weighted thresholding method (in~\cite{Hong-2008-SHBC}), a combination method of distance transform and region growing (\cite{Mukherjee-1995-BCCD}),  and Otsu thresholding (in~\cite{Brugger-2012-ACBC,Minoi-2016-MVBA}) are used to obtain the binary image.  Then the objects are detected based on eight neighbor regions in~\cite{Martinez-2016-NTBC}. Moreover, an adaptive thresholding method is applied in~\cite{Brugger-2012-ACBC} for secondary binarization to solve the challenges that come from the fact that bacterial strains from the same species may exhibit different colony phenotypes.  In~\cite{Zhu-2018-ACBC}, image subtraction is carried out to extract the candidate colonies, which are connected to the inner circle of the agar plate and a nonlinear gray transformation is used to enhance the gray-scale.  Afterward, a distance transformation is performed on the binarized image and segmentation is done with a watershed transformation. Furthermore, the sharp corners produced by the watershed transformation are removed by using the morphological opening method. After the segmentation algorithm is completed, the Bayes classifier distinguishes the remaining concatenated groups into classes of one, two, three or four containing colonies, and the final colony is counted.  Finally, in~\cite{Masschelein-2012-TACC,Yujie-2009-DIMB},  the GLCM is extracted and the SVM is applied for classification.
The total number of single and clustered colonies is counted with an average relative error of 0.2\% in~\cite{Zhu-2018-ACBC}. The counting result is shown in Fig.~\ref{fig:Zhu-2018-ACBC2}.
The processing method of~\cite{Mukherjee-1995-BCCD} is shown in Fig. ~\ref{fig:Mukherjee-1995-BCCD}.

\subsubsection{Bacteria counting method based on color segmentation}

In~\cite{Ogawa-2005-MDIA}, a distinctive multicolor segmentation algorithm is applied for the accurate and simultaneous differentiation of triple-stained bacteria. The result has the 95\% confidence intervals of the regression. 

In~\cite{Schonholzer-2002-AIAA,Peitz-2010-SCBG,Mukti-2010-DACT}, RGB images are separated into three channels. The preliminary detection of both bacteria and debris is based on the green channel, and the differentiation between bacteria and debris is based on the processing of the green and blue channels. Then the debris particles are eliminated by the combination of two output images above. Then a Gaussian filter is applied for noise removal and Otsu thresholding is used to roughly separate the data of the relatively dark electrodes from data belonging to the electrode gaps in~\cite{Peitz-2010-SCBG}. Moreover, the numbers of single and dividing cells and cell agglomerates are determined by a method based on the number of local grey value maxima in~\cite{Schonholzer-2002-AIAA}.  Finally, cell numbers and cell sizes are calculated based on area and perimeter measurements for each single or dividing cell.

\subsection{Machine learning and deep learning counting methods}
\begin{figure}[ht]
\centering
\includegraphics[trim={0cm 0cm 0cm 0cm},clip,width=1.0\textwidth]{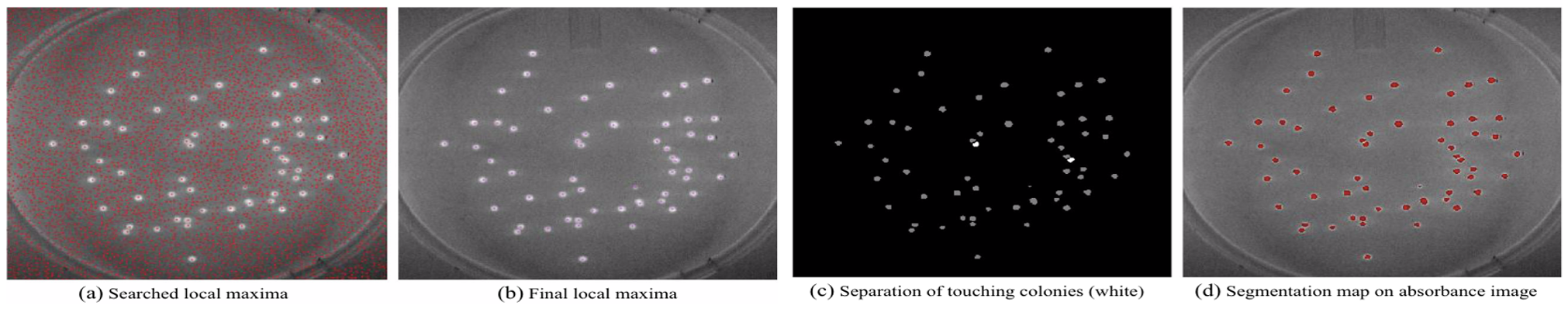}
\caption{Local absorbance maxima search and separation of touching colonies (in~\cite{Yoon-2015-ACAC} fig.14).}
\label{fig:Yoon-2015-ACAC1}
\end{figure}

In~\cite{Ishii-1987-TICA,Yoon-2015-ACAC,Chiang-2015-ACBC},  principal-component analysis (PCA) is used to separate the biological pattern with the surrounding area. The type of pattern for selection is identified and the objective biological pattern is counted.  Moreover, the nearest neighbor searching algorithm is applied to separate touching colonies after PCA in~\cite{Yoon-2015-ACAC}, which contains three main steps. First, the local maxima on an absorbance image is found, and a mask image is created in which the locations of the local maxima are marked with 255 and otherwise with 0. Then the local maximal pixels outside the binary segmentation image are masked out by a logical AND operation. Afterward, the clumped blobs are split when the number of local maxima is greater than the number of blobs. The separation result is shown in Fig.~\ref{fig:Yoon-2015-ACAC1}. Finally, the image of the bacteria colony is segmented and counted. The accuracy of the colony segmentation and counting algorithm is over 99\%. However, in~\cite{Chiang-2015-ACBC}, the Otsu thresholding is applied for segmentation after PCA. Then, the distance transform and waster-shed are applied for the division of overlapping colonies. Afterward, the bottom-hat transformation is applied to extract colonies from the rim image. Comparisons show that the proposed system is an effective method with excellent accuracy with a mean value of absolute percentage error of 3.37\%.

\begin{figure}[ht]
\centering
\includegraphics[trim={0cm 0cm 0cm 0cm},clip,width=0.6\textwidth]{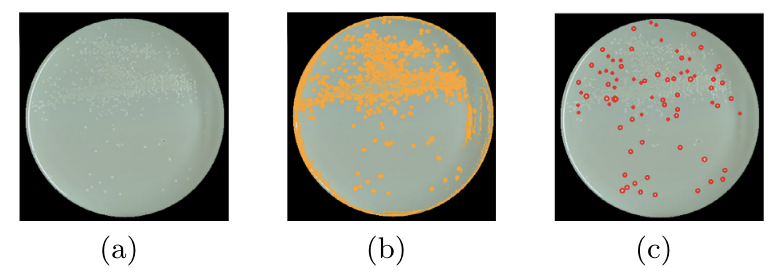}
\caption{(a) The original image. (b) The identified edges within the background. (c) Candida colonies found on the Petri dish (in~\cite{Andreini-2015-AIAA} fig.4).}
\label{fig:Andreini-2015-AIAA}
\end{figure}

\begin{figure}[ht]
\centering
\includegraphics[trim={0cm 0cm 0cm 0cm},clip,width=1.0\textwidth]{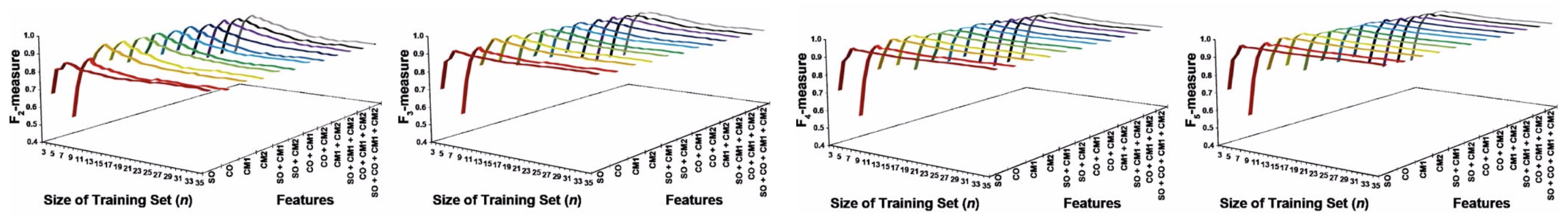}
\caption{The classification results for different sizes of the training set (in~\cite{Chen-2009-AABC} fig.15).}
\label{fig:Chen-2009-AABC}
\end{figure}

In~\cite{Andreini-2015-AIAA,Andreini-2016-AICT,Zhang-2010-ASTF,Chen-2009-AABC}, SVM is applied for bacteria counting and classification. 
In~\cite{Andreini-2015-AIAA,Andreini-2016-AICT}, the colonies are segregated from the background by a background removal process based on chromatic information about the specific chromogenic medium used in the culture. Then, a supervised training technique is adopted to obtain a chromatic description of the background and the uncertainty region is obtained as a union of the intersections of some binary masks obtained by imposing a threshold on the probability level of the background and of the infected regions. After that, a mean shift segmentation algorithm is used to associate each image pixel to the corresponding modal density value and a Sobel based edge enhancement is applied to distinguish different classes. Moreover, the considered uncertainty region is divided into two subregions based on the computed thresholding and the histograms of the two sub regions are calculated and compared to establish if a significant separation exists. Finally, SVM is applied for classification and the number of colonies is counted with an accuracy of 99.2\%.  The segmentation result is shown in Fig.~\ref{fig:Andreini-2015-AIAA}. 
In~\cite{Zhang-2010-ASTF,Chen-2009-AABC},  a subtraction operation is applied between the original image and background image to eliminate the background unevenness caused by the light source. Then a median filtering algorithm is applied to smooth the image because it reduces the hot-electron noise and the noise caused by environmental disturbance during image collection, quantify and transmission, and overcomes the blur of the image details created by linear filtering. After that,  the gray-level histogram equalization is applied for image enhancement. Then the Otsu thresholding (\cite{Zhang-2010-ASTF}) and watershed (\cite{Chen-2009-AABC}) are used to obtain the binary image. Finally, the shape features are extracted and used for SVM training to identify and count bacteria. It can be seen that the counting results of SVM have a small difference from that of human eye recognition and its relative error is less than 3\%, which means that SVM can be used for rod-shaped bacteria counting.The classification results of~\cite{Chen-2009-AABC} is shown in Fig.~\ref{fig:Chen-2009-AABC}.

\begin{figure}[ht]
\centering
\includegraphics[trim={0cm 0cm 0cm 0cm},clip,width=1.0\textwidth]{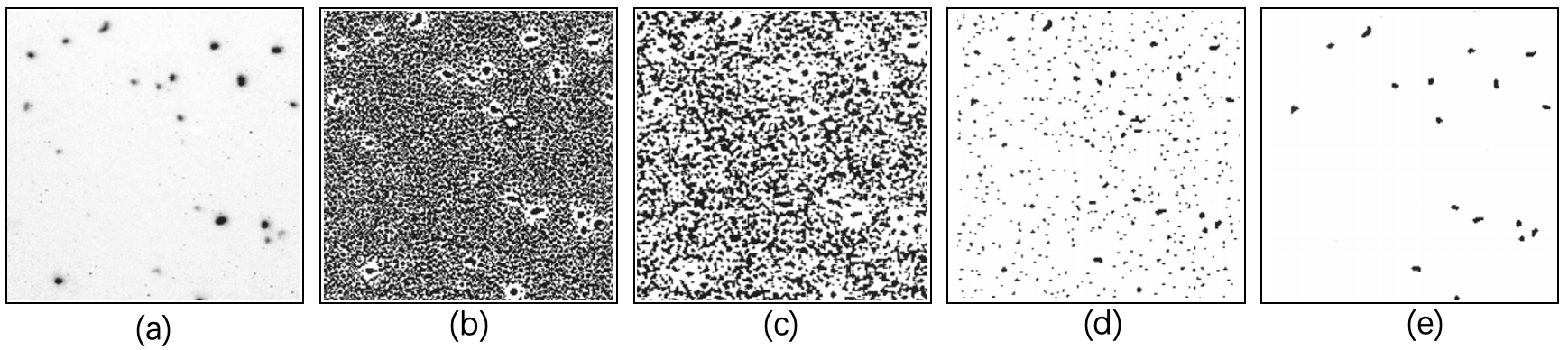}
\caption{Edge detection procedure. (a)The original image. (b) Image after application of the Marr-Hildreth operator with 3$\times$3 kernel. (c) Image after application of a rank 3 filter. (d) Binary image after thresholding. (e) Image after erosion (in~\cite{Blackburn-1998-RDBA} fig.1).}
\label{fig:Blackburn-1998-RDBA}
\end{figure}

In~\cite{Blackburn-1998-RDBA}, the Marr-Hildreth operator is used for edge detection of bacteria image and threshold is used for image binarization. Then a rank 3 filter is applied to remove pixels that have intensities equivalent to the intensities of amplified background noise. An artificial neural network (ANN) is firstly applied for the classification of bacteria. The ANN is composed of 6 input nodes, 5 intermediate nodes, and 3 output nodes, and then the ANN is activated by using a sigmoid activation function.
After training, the images can be analyzed automatically at a rate of 100 images per h.
Minimal variation in cell counts between filters is observed (5\%) with the filtering procedure used. The bacteria counting procedure is shown in Fig.~\ref{fig:Blackburn-1998-RDBA}.

In~\cite{Shenglang-2008-RDTN,Hongwei-2012-TTMI}, back propagation (BP) neural network and DIP are used for analysis and counting for the microscopic image of bacteria.  Median filtering and adaptive filtering are used for denoising and background elimination, and then the iterative algorithm is used for image segmentation in~\cite{Shenglang-2008-RDTN}. In~\cite{Hongwei-2012-TTMI}, the Otsu thresholding method and the combination of square and circle filter are used for segmentation and edge detection of other microorganisms. After that, morphological operations are applied to smooth the contour of cells and the binary images are obtained. Moreover, the morphology and colorimetry features are extracted and trained in BP neural network for identification and counting. The detection error between the proposed method and the manual counting method is no more than 5\%. 
In~\cite{Shenglang-2008-RDTN}, the perimeter, area, shape factor, rectangularity, extension length and gray-scale of the object are input to the BP neural network.  Then the neural network is activated by using Sigmoid function, which contains six hidden layers and one output layer. The counting system can analyze the sample in less than 10 minutes, whereas the classical manual counting method takes 48 hours.

\begin{figure}[ht]
\centering
\includegraphics[trim={0cm 0cm 0cm 0cm},clip,width=0.6\textwidth]{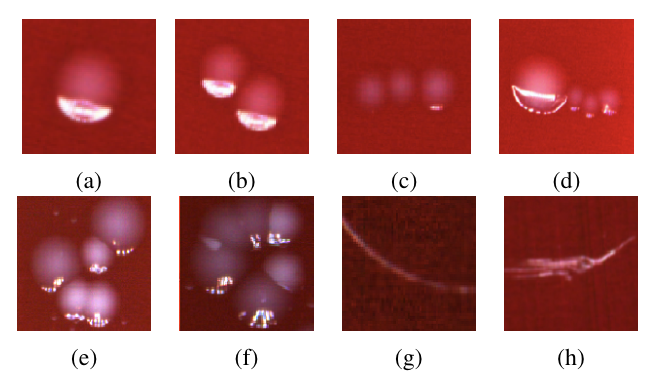}
\caption{Example of dataset images representing a certain number of colonies, from 1 (a) to 6 (f), and two example of outliers (g) and (h) (in~\cite{Ferrari-2015-BCCC} fig.2).}
\label{fig:Ferrari-2015-BCCC1}
\end{figure}

\begin{figure}[ht]
\centering
\includegraphics[trim={0cm 0cm 0cm 0cm},clip,width=0.7\textwidth]{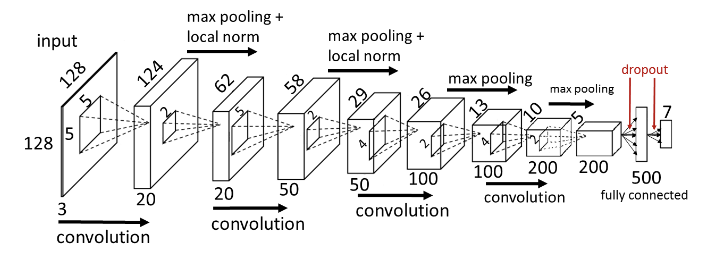}
\caption{Convolutional Neural Network topology (in~\cite{Ferrari-2015-BCCC} fig.3).}
\label{fig:Ferrari-2015-BCCC2}
\end{figure}

\begin{figure}[ht]
\centering 
\includegraphics[trim={0cm 0cm 0cm 0cm},clip,width=0.7\textwidth]{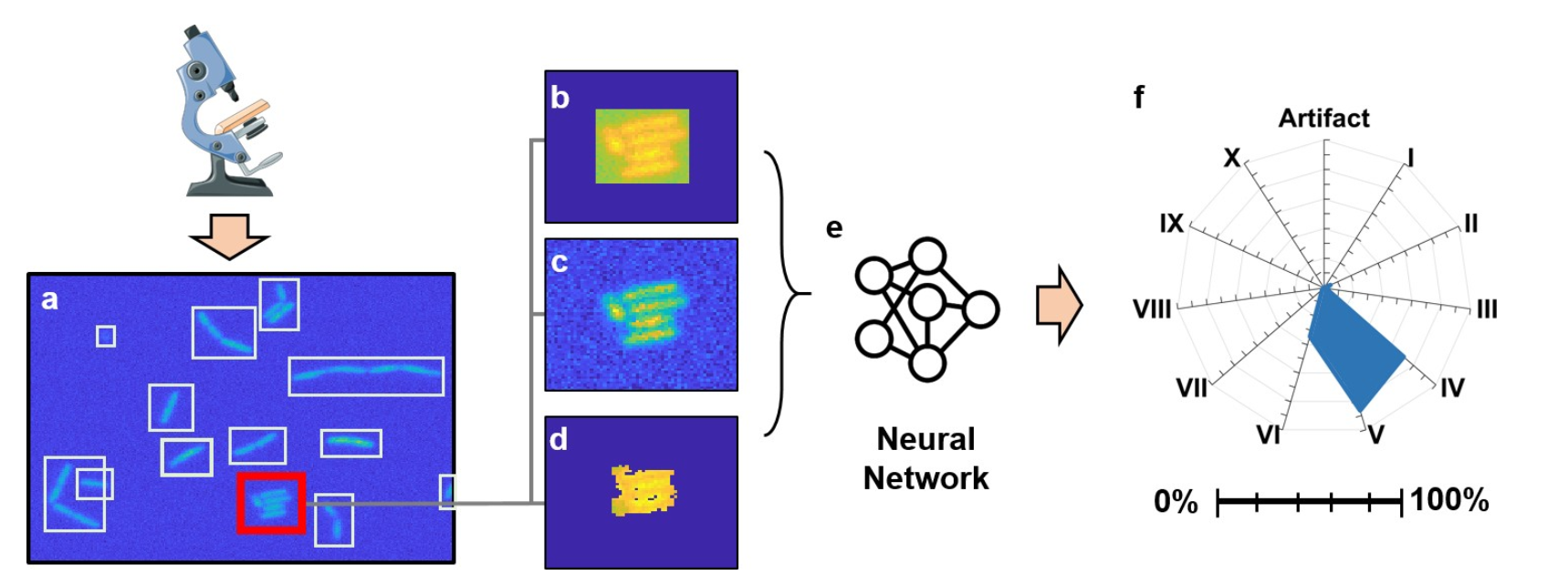}
\caption{The work flow of image processing. (a) The image is binarized  and annotated manually. (b)Null Bumper. (c) Blended. (d) Masked.  (e) Neural network training. (f) The input of neural network (in~\cite{Tamiev-2020-ACBC} fig.2).}
\label{fig:Tamiev-2020-ACBC}
\end{figure}

In~\cite{Ferrari-2015-BCCC,Ferrari-2017-BCCC,Tamiev-2020-ACBC}, the convolutional neural network (CNN) is applied for bacterial colony counting. 
The example of the dataset is shown in Fig.~\ref{fig:Ferrari-2015-BCCC1}. Then a horizontal flip is performed on the images to double the training dataset and three different artificial color distortions on RGB color space are applied. After that, another transformation is the conversion of the masked dataset in gray-scale color space and seven different values of spatial rescaling before cropping is performed. 
Then the images are enhanced through normalization concerning the segment orientation (\cite{Ferrari-2015-BCCC}) and contrast limited adaptive histogram equalization (\cite{Ferrari-2017-BCCC}). Finally, CNN is applied for classification and counting that contains five learned layers, four convolutional and one fully connected as shown in Fig.~\ref{fig:Ferrari-2015-BCCC2}. During the training, the testing accuracy flattens after 15000 iterations. 50,000 iterations have taken approximately 3 hours on an Nvidia Titan Black GPU. The accuracy of 92.8\% is obtained. 
After CNN classification, a watershed algorithm is applied for colony separation in~\cite{Ferrari-2017-BCCC}. 
The testing accuracy increases with the number of training iterations and flattens around 30,000 iterations. 50,000 iterations take approximately one hour on an Nvidia Titan X GPU. The accuracy of 92.1\% is obtained after data augmentation.  
In~\cite{Tamiev-2020-ACBC}, a classification-type convolutional neural network (cCNN) is proposed for automatic bacteria classification and counting, and an efficient method for microscope image preprocessing is presented. First, the raw images are segmented with an adaptive binary thresholding method and images with individual cells or cell clusters are cropped. Then the images are trained using cCNN. The network's output corresponds to the number of cells in given cell clusters and the individual outputs are then added to find the total cell count. The counting accuracy of 86\% is obtained. The workflow is shown in Fig.~\ref{fig:Tamiev-2020-ACBC}. The result shows a 3.8X increase in processing speed by using an NVIDIA Quadro K620 GPU.

\subsection{Third-party tools}

\begin{figure}[ht]
\centering
\includegraphics[trim={0cm 0cm 0cm 0cm},clip,width=0.7\textwidth]{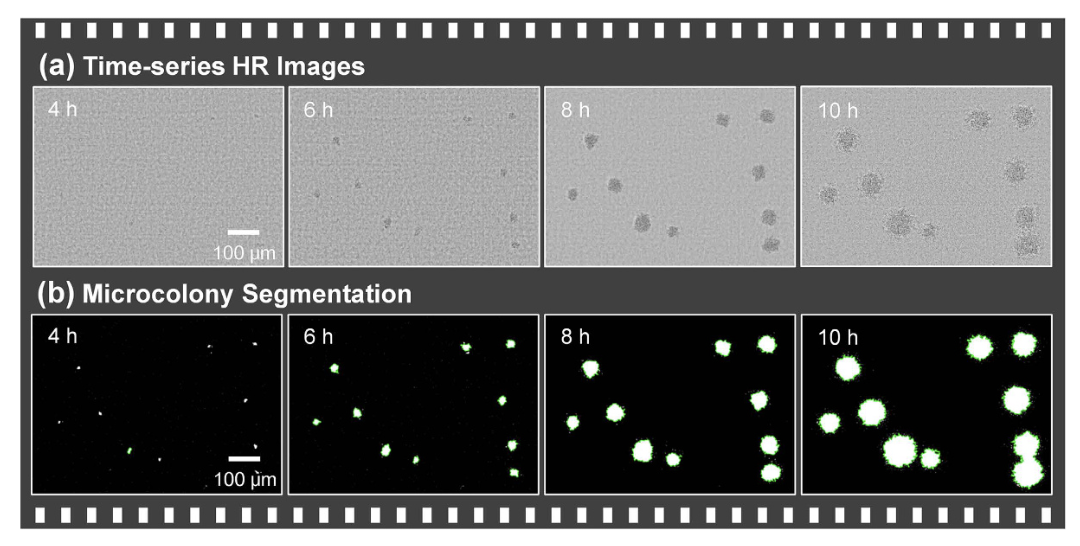}
\caption{Image processing. (a) Time-series high-resolution (HR) images. (b) Following reconstruction of the HR image (in~\cite{Jung-2016-RTBM} fig.3).}
\label{fig:Jung-2016-RTBM}
\end{figure}

In~\cite{Jung-2016-RTBM}, image analysis is used for real-time bacterial counting. First, the time-series high-resolution (HR) images of bacterial microcolonies are reconstructed using sub-pixel sweeping perspective microscopy (SPSM). Then the images are segmented, and the equivalent diameter and number of colonies in each time-lapse image are then calculated. The processed images are shown in Fig.~\ref{fig:Jung-2016-RTBM}.

\begin{figure}[ht]
\centering
\includegraphics[trim={0cm 0cm 0cm 0cm},clip,width=1.0\textwidth]{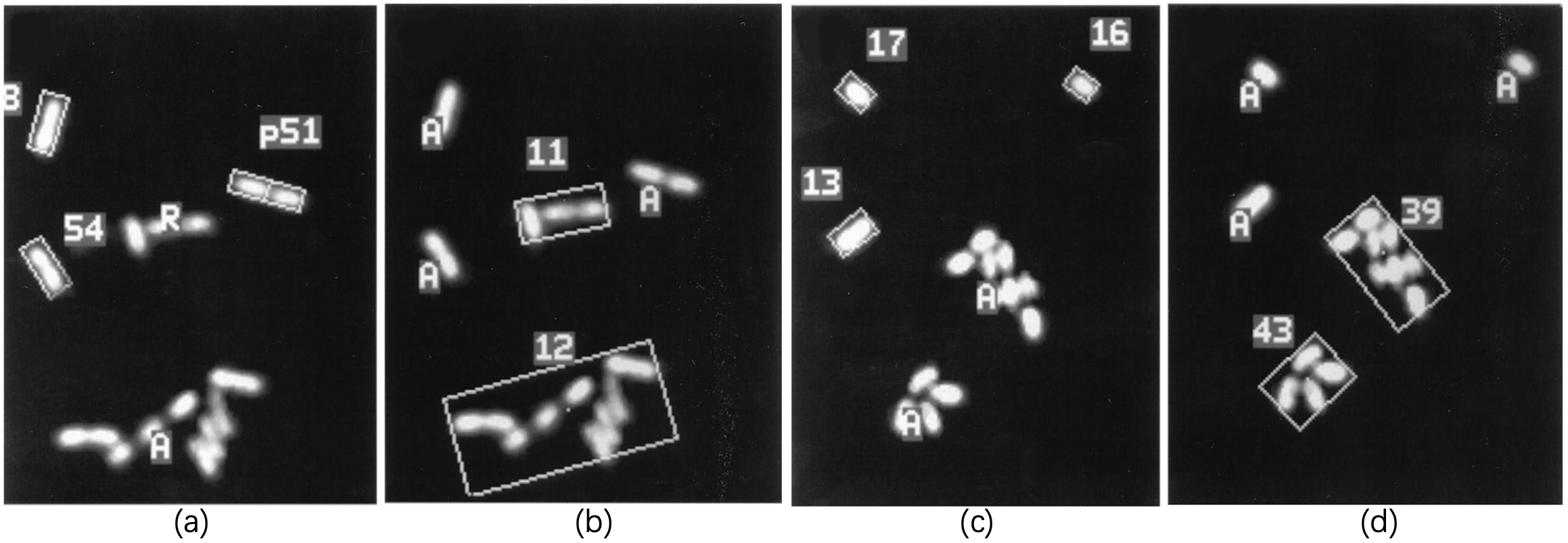}
\caption{Automatic identification of bacteria with Cellstat. (a) The cell between 75 and 700 pixels. (b)The cell  between 350 and 700 pixels. (c) Identification of single cells between 75 and 175 pixels. (d) Identification of micro colonies on the surface (in~\cite{Moller-1995-BGSA} fig.1).}
\label{fig:Moller-1995-BGSA}
\end{figure}

In~\cite{Moller-1995-BGSA}, Cellstat image analysis program is developed to determine the biovolume of bacteria.  They present a method for simultaneous quantitative staining of RNA and DNA using the metachromatic dye AO and quantify the RNA and DNA. The automated image analysis is not biased by the operator, and it allows the analysis of a number of objects, ensuring good statistics. Choosing the right parameters for cell identification makes it possible to discriminate between single cells and clumps of cells. By using a different set of parameters for object recognition, it is possible to detect and measure the intensities of surface-associated microcolonies and single cells on the surface independently. The result of automatic identification of bacteria with Cellstat is shown in Fig.~\ref{fig:Moller-1995-BGSA}.

In~\cite{David-1989-EASA},  `Model 2000' (Image Technology Corporation, Deer Park, New York) image analysis system is used for enumeration and sizing of bacteria, which can detect and enhance each individual cell at the same time.  There is no statistical difference in cell counts made manually or by the image analysis system.

In~\cite{Kildeso-1997-EAAM}, `Kontron Vidas Plus' (Kontron Elektronik GmbH, Germany) image analyser system is used for airborne microorganisms counting. Gaussian filter is used to remove noises and the edge is derived using Laplace filter.  This work has established a possibility of improving exposure assessment of airborne microorganisms through image processing instead of manual counting.

In~\cite{Shopov-2000-IIAA}, a program `Skidaway Tools' (Skidaway Institute of Oceanography, 10 Ocean Science Circle,  Savannah, USA) is developed based on Marr-Hildreth Gaussian-smoothed Laplacian edge-detection protocol that is proposed in~\cite{Viles-1992-MMPC},  with added flat-fielding and edge-strength operators.  The alpha-channel is applied in bacteria image segmentation, masking the background and providing a count of the attached bacteria cells.

In~\cite{Gmur-2000-AIES},  `IBAS 2.0' (Kontron Inc., Eching, West Germany) is used to process the images of dental bacteria. A gradient convolution filter is used to process images firstly, then, any white objects below or above an acceptable size range are excluded and the remaining spots are counted automatically after image binary.  In~\cite{Singleton-2001-AFAM}, `IBAS 2.0' is used for oral microbial quantification. Thresholdiinging is used for images binarization, and the edge-effect rule is used to eliminate the objects of the wrong size.  Then the individual bacteria are segmented, and the white spot of images are counted.  A close agreement between the automated system and the manual visual counts is observed.

In~\cite{Nunan-2001-QTSD}, `Zeiss KS300 Imaging System 3.0' is used for bacteria image processing. The RGB images are decomposed into 3 channels that can be processed separately. Sigma smoothing and top-hat transform are used for edge detection and segmentation in green channel images that can detect all features in the bacteria size range.  High pass filter and morphological opening are used to remove autofluorescent objects in red channel images. The top hat transform is used to distinguish bacteria from other objects by detecting the blue halos in blue channel images. The binary images are obtained based on the three-channel images above. The number of cells and other parameters such as area is measured. In~\cite{Stoderegger-2005-DBSP},  `Zeiss KS300' is used to quantify the natural bacterial community. The binary images are obtained by adjusting the threshold level. Then the images are corrected by excluding or adding cells originally not detected by the channel settings. The individual cell area is determined, and the total area is calculated. Therefore, the number of cells can be obtained. 

In~\cite{Pena-2002-CAVA}, `Image-Pro Plus' image analysis software (Media Cybernetics, USA) is used for the quantification of bacteria.  For aggregates, contour extraction is used for automatic image segmentation. The average equivalent diameter (AED) is used to characterize the size of aggregates, which is used for the automatic elimination of individual cells and debris.  A High-Gauss filter is used for image enhancement for individual cells, and a Gaussian filter is used for noise reduction. The roundness value is calculated to select the objects that correspond to individual cells.  The binary images based on the two methods above are used for bacteria counting and biovolume measurement. 

In~\cite{O-2003-MQPS},  `Optimas 6.5' (Media Cybernetics Inc., Silver Spring, MD) image processing system is used to quantify bacteria. Firstly, the low-frequency background noises are isolated and removed, then a combination method of binary erosions to point and dilations within image masks is used to separate cells.  The parameters such as cell count and cell volume are measured by using the `Optimas' image processing system. The result shows that the deviation of the experimentally measured density from the known density is 3.2\%.

In~\cite{Putman-2005-SMAC}, `ProtoCOL' (Version 4.04 from Synoptics Ltd., Cambridge, UK) is used to count the bacteria colonies. When the `ProtoCOL' software is used to process the digital camera image, the count result is highly correlated with the true count but slightly less than the true count. 

In~\cite{Thiel-2005-AIMT}, `KS400' (Carl Zeiss Vision, Hallbergmoos, Germany) is used for automated enumeration of fluorescently labeled bacteria. First, the DAPI images are analyzed to detect single signals at a high spatial resolution. Then, a second analysis system is developed to detect signals with a low signal-to-noise ratio. After, the binary images are obtained by merging the resulting images above.  Finally, the third step processes the Cy3 image and thus provides information on the signals that derive from target organisms. The logical `AND' operation of the processed DAPI and Cy3 images ensures that only those signals are counted in both channels. The calculated correlation coefficient of 0.984 indicates that the manual and the automatic counting result are in agreement.

\begin{figure}[ht]
\centering
\includegraphics[trim={0cm 0cm 0cm 0cm},clip,width=1.0\textwidth]{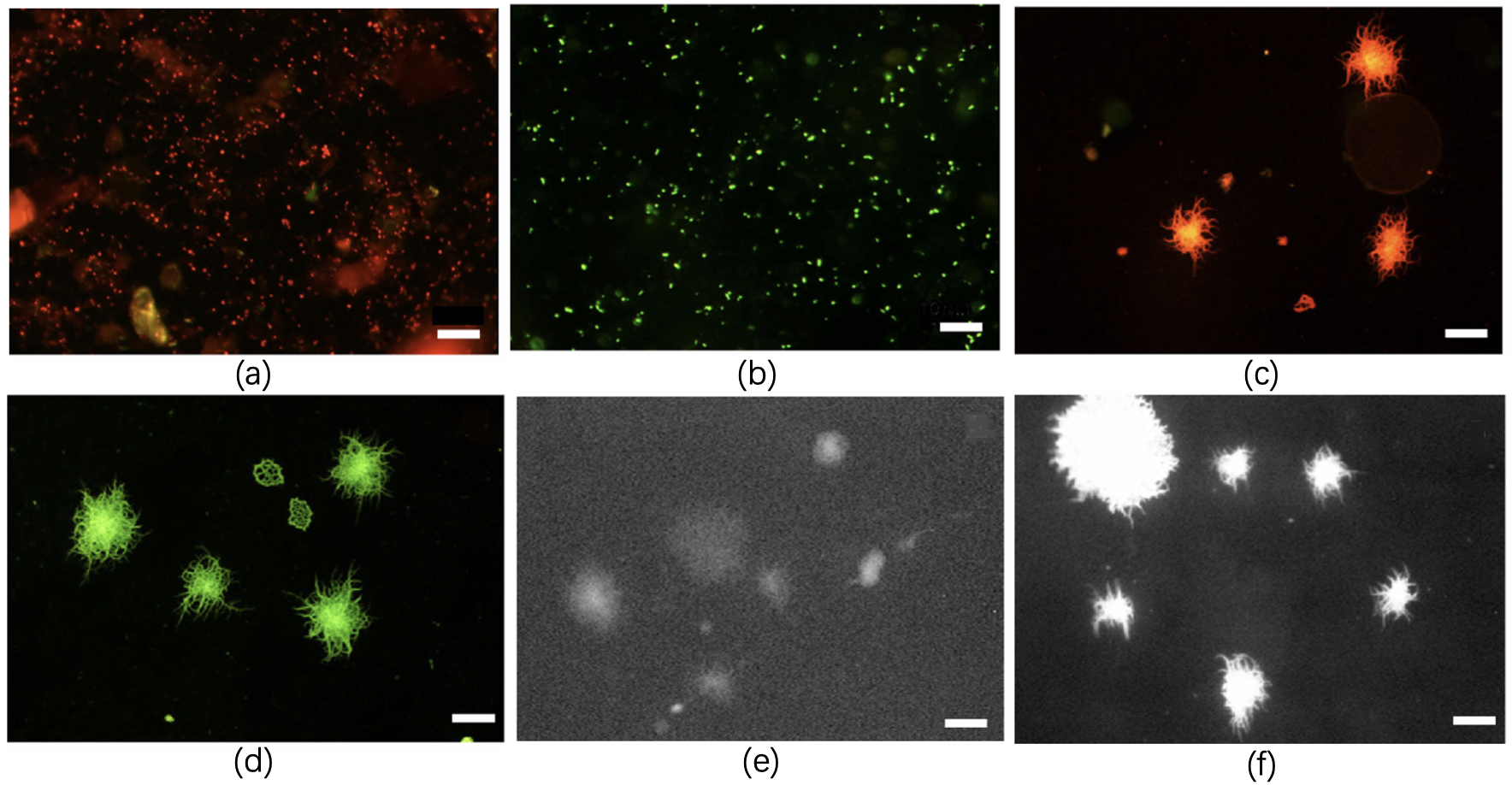}
\caption{Fluorescence images of single cells and micro-colonies of bacteria. Bacteria in compost are stained with EtBr (a, c and e) and SYBR Green II (b, d and f). Single bacterial cells in compost suspension before incubation (a and b) and micro-colonies developed after incubation on LB medium (c and d) observed under blue excitation by epifluorescence microscopy. Micro-colonies are also observed using micro-colony auto counting system (e and f) (in~\cite{Wang-2007-RAAE} fig.3).}
\label{fig:Wang-2007-RAAE}
\end{figure}

In~\cite{Wang-2007-RAAE}, a micro-colony auto counting system `MACS' (Chuo Electric Works, Osaka, Japan) is used for bacteria colony counting. The `MACS' has an automatic scanning stage and blue light emitting diode (LED) as a light source.  Micro-colonies are captured using a CCD camera and analyzed using `Micro-colony V' software (version 1.504; Chuo Electric Works). SYBR Green II is used to stain the bacteria images, and the green fluorescence is detected clearly. The stained images and counting results are shown in Fig.~\ref{fig:Wang-2007-RAAE}.

In~\cite{Hua-2009-RBFC}, `Davinci' technology is used for bacteria counting and area calculating. Image enhancement and median filter are applied to remove noises and local binary fitting (LBF) is used for image segmentation. Finally, the connected region is detected as the number of bacteria and the area of the connected region is measured. The average error between the proposed method and the manual counting method is no more than 1.6\%.

In~\cite{Freitas-2014-OAAC}, the automated enumeration software `SigmaScan Pro 5.0' (Systat Software Inc) is used for the quantification of cells in the biofilm. The intensity thresholding is used for image segmentation. There are no significant differences found using the software thresholding and the manual counting (r>0.05), indicating that the Live/Dead staining is strongly discriminative between bacteria and background, and there is no significant fluorophore bleach effect that could impair the automatic counts. 

In~\cite{Song-2018-DRAA}, `ImageJ' software version 1.52 (NIH, Bethesda, MD, USA) is applied for automatic bacteria counting. First, the threshold is adjusted to enhance the contrast of the objects of interest. Second, the image is binarized to remove the noise by rendering micro-colony regions with clear boundaries as black and the surrounding background as white. After the binarization, filling-holes processing is conducted to ensure each closed region represents one intact micro-colony. Finally, micro-colony regions above the desired size are outlined, and the number of these regions is automatically counted by `ImageJ' software.

\subsection{Summary of image analysis based counting for bacteria}
By reviewing the related work of image analysis for bacteria counting and referring to Table~\ref{tab:counting}, we find that:
\paragraph - Development trend The bacteria counting using image analysis approaches began in the 1980s and developed quickly in the 2010s. This development trend is due to the government and people attach importance to the bacteria problems in recent years, which play essential roles in the food industry and social hygiene. With the development of computer based image analysis technologies, more explorations and higher accuracies will be achieved in the future. 
\paragraph - Counting techniques  The most frequently used pre-processing methods are the median filter and Gaussian filter, image segmentation methods are thresholding, distance transform and watershed, classifier algorithms are SVMs and ANNs.

\begin{landscape}
\begin{table}
\scriptsize
\caption{\label{tab:counting}Summary of image analysis based counting for bacteria} 
\begin{tabular}{p{3cm}p{3cm}p{3cm}p{4cm}p{2cm}p{2cm}}
\hline
Related work        & Microorganism type & Pre-processing & Segmentation   & Classification  & Evaluation \\ \hline
\cite{Pettipher-1982-SACB} & Bacteria           &                & Gray-level contrast  &                       &            \\
\cite{Masuko-1991-ANMD}  & Bacteria           &                & Thresholding         &                       &            \\
\cite{Moller-1995-BGSA}     & Bacteria           &                & Gray-level intensity &                       &            \\
\cite{Massana-1997-MBSV} & Bacteria           & Edge detection & Thresholding         &                       &            \\
\cite{Trujillo-2001-AMVS}    & Bacteria           & Flatten filter & Thresholding                               &            &            \\
\cite{Ogawa-2003-DMDI}    & Bacteria           & Color and square measuring      & Thresholding       &          &            \\
\cite{Yamaguchi-2004-MEDC}    & Bacteria           & Intensity expanding                                            & Hue segmentation                           &                       &            \\
\cite{Ogawa-2005-MDIA}        & Bacteria           &  &  Multicolour segmentation    &        & 95\% confidence intervals       \\
\cite{Shenglang-2005-TJBR}    & Bacteria     & Histogram equalization,  convolutional filter and median filter & Thresholding                               &                       &            \\
\cite{Niyazi-2007-CCCA}       & Bacteria           &                                                                & Thresholding                               &                       & 95\% accuracy      \\
\cite{Gupta-2012-MABC}        & Bacteria           & Adaptive median filter                                         & Thresholding                               &                       & 95\%-100\% accuracy \\
\cite{Sethi-2012-BCCM}        & Bacteria           & Adaptive median filter and thinning        & Thresholding                               &                       & 95\%-100\% accuracy \\
\cite{Kaur-2012-ANMA}         & Bacteria           & RGB adjusting and adaptive median filter     & Morphological dilation for edge extraction                      &            \\
\cite{Chunhachart-2016-CAVE}  & Bacteria           & Morphological operations for image enhancement       & Thresholding                               &      & 97.87\% accuracy \\
\cite{Ishii-1987-TICA}       & Bacteria           &                                                                                                              & PCA                                                                                   &                       &                           \\
\cite{Mukherjee-1995-BCCD}   & Bacteria           & Thresholding and distance transform   & Region growing and connected component labelling &                       &                           \\
\cite{Pernthaler-1997-SCAI}  & Bacteria           &             & Top-hat and contrast retransformation           &                       &                           \\
\cite{Blackburn-1998-RDBA}   & Bacteria           & Marr-Hildreth operator                                                                                       & Thresholding                                                                                & ANN                   &  95\% accuracy          \\
\cite{Schonholzer-2002-AIAA} & Bacteria           & RGB processing          & Local grey value maxima       &                       &   \\
\cite{Marotz-2001-EORA}      & Bacteria           &                                                                                                              & Local adaptive threshold                                                                                       &                       & 0.968 correlation coefficient with manual count \\
\cite{Selinummi-2005-SQLB}   & Bacteria           &     & Marker-controlled watershed segmentation and global thresholding &                       & 99\% accuracy     \\
\hline
\end{tabular}
\end{table}

\newpage
\begin{table}
\scriptsize
\begin{tabular}{p{3cm}p{3cm}p{3cm}p{4cm}p{2cm}p{2cm}}
\hline
Related work        & Microorganism type & Pre-processing & Segmentation   & Classification  & Evaluation \\ \hline

\cite{Zhang-2007-AEAR}       & Bacteria           & Contrast-limited adaptive histogram equalization     & Otsu thresholding and watershed    &           &               \\
\cite{Zhang-2008-AABC}       & Bacteria           & Hypothesis testing for denoising     & Otsu thresholding and frequency distribution segmentation        &                       &                                            \\
\cite{Chen-2008-BCEA}        & Bacteria           &                                                                                                              & Watershed                                                                                                      &                       & 61\% precision                                                                            \\
\cite{Chen-2009-AABC}        & Bacteria           &    & Watershed         & SVM                   & 80\% accuracy                                                                             \\
\cite{Shenglang-2008-RDTN}   & Bacteria           & Median filter and adaptive filter                                                                            & Iterative thresholding                                                                                                  & BP neural network     &                             \\
\cite{Hong-2008-SHBC}        & Bacteria           & Gray-scale weighted thresholding     & Distance transform and watershed                                   &        & 98.2\% accuracy                                     \\
\cite{Yujie-2009-DIMB}       & Bacteria           & Gray-scale weighted thresholding         & Distance transform and watershed                               & SAGA and SVM          & 99.67\% accuracy                 \\
\cite{Kan-2008-QQSB}         & Bacteria           & Median filter          & Distance transform and watershed        &                       & 13.5\% average relative error with manual count \\
\cite{Fang-2008-ESAC}        & Bacteria           & Median filter and contrast enhancement  & Iterative thresholding                                                                                         &                       & 2.5\% average relative error with manual count  \\
\cite{Sotaquira-2009-DAQB}   & Bacteria           &         & Thresholding            &                       & 96.3\% accuracy                    \\
\cite{Ates-2009-AIPB}        & Bacteria           & Median filter           & Watershed                                          &                       &   \\
\cite{Jun-2010-RDRM}         & Bacteria           & Median filter       & Adaptive thresholding and morphological operations & BP neural network     &   \\
\cite{Mukti-2010-DACT}        & Bacteria       &         & Color thresholding                     &                       &                                  \\
\cite{Peitz-2010-SCBG}       & Bacteria           & Gaussian filter  & Otsu thresholding and erosion operation &    &               \\
\cite{Buzalewicz-2010-IPGA}   & Bacteria           & Fourier transform          & Mellin transform                 &                       & 4.51 standard deviation with manual counting  \\
\cite{Zhang-2010-ASTF}        & Bacteria           & Median filter and gray-level histogram equalization         & Otsu thresholding                                                                                                       & SVM                   & 97\% accuracy                     \\
\cite{Nayak-2010-ANAA}        & Bacteria       &      & HSI color segmentation and thresholding &     &                          \\
\cite{Shen-2010-ESAC}         & Bacteria           & Median filter       & Iterative local thresholding       &                       & 97.5\% accuracy          \\
\cite{Clarke-2010-LCHT}       & Bacteria           & Gaussian filter   & Adaptive thresholding   &                       & 97\% accuracy                                                                             \\

\hline
\end{tabular}
\end{table}

\newpage
\begin{table}
\scriptsize
\begin{tabular}{p{3cm}p{3cm}p{3cm}p{4cm}p{2cm}p{2cm}}
\hline
Related work        & Microorganism type & Pre-processing & Segmentation   & Classification  & Evaluation \\ \hline

\cite{Hongwei-2012-TTMI}     & Bacteria           & Background marking and wavelet method    & Otsu thresholding and square and  circle filter        & BP neural network     & 95\% accuracy                                                                             \\
\cite{Ferrari-2015-BCCC}      & Bacteria           &     & Thresholding                                                                                                            & CNN                   & 92.8\% accuracy                                                                           \\
\cite{Brugger-2012-ACBC}      & Bacteria           & Top-hat transform  & Otsu thresholding, adaptive \\ thresholding, distance transform and watershed & Bayes classifier      &        \\
\cite{Masschelein-2012-TACC} & Bacteria           &       & Distance transform and watershed & SVM                   &                                                                                           \\
\cite{Feng-2013-ACAI}         & Bacteria     &  &  Adaptive thresholding, distance transform and erosion &                       &                                                                                           \\
\cite{Barbedo-2013-AACM}      & Bacteria         & Gaussian filter and Canny filter                                                                             & Region growing and thresholding  &                       & 99\% accuracy  \\
\cite{Ferrari-2017-BCCC}      & Bacteria           & Contrast limited adaptive histogram equalization     & Watershed                                                                                                               & CNN                   & 92.1\% accuracy                                                                           \\
\cite{Yoon-2015-ACAC}         & Bacteria           & PCA and wavelength band selection        & Thresholding                                                                                                            &                       & 99\% accuracy                                                                            \\
\cite{Andreini-2015-AIAA}     & Bacteria           & Sobel operator and chromatic information selection    & Mean shift segmentation and thresholding         & SVM                   & 99.2\% accuracy                                                                           \\
\cite{Andreini-2016-AICT}    & Bacteria           & Morphological filter    & Random Hough circle transform and thresholding  & SVM                   & 92.1\% accuracy                                                                           \\
\cite{Chiang-2015-ACBC}      & Bacteria           & PCA and Sobel masks                              & Otsu thresholding, watershed and distance transoform       &       & 96.63\% accuracy      \\
\cite{Martinez-2016-NTBC}    & Bacteria           & Edge detection    & Distance transoform         &         & 0.994 correlation coefficient with manual count  \\
\cite{Minoi-2016-MVBA}       & Bacteria           & Edge detection           & Otsu thresholding, watershed and distance transform                       &           &         \\
\cite{Wong-2016-ACCA}        & Bacteria           & Contrast enhancement     & Watershed and distance transform    &           & 90.3\% accuracy                                                              \\
\cite{Choudhry-2016-HTMA}    & Bacteria           & Sobel filter, Gaussian blur and holes filling & Thresholding                                                                                                            & &      \\
\cite{Alves-2016-CCVA}       & Bacteria           & Laplacian filter and circular Hough transform  & Otsu thresholding                                                                                                       &       & 90\% accuracy                                                                                                                   \\
\cite{Jung-2016-RTBM}        & Bacteria           & Image reconstruction                                                                                                             & Thresholding                   &                     &       \\
\cite{Matic-2016-SAPS}       & Bacteria           & Gamma correction and Gaussian filter  & Thresholding, Canny operator and Hough transform   &        & 97\% precision and  82\% recall   \\

\hline
\end{tabular}
\end{table}

\newpage
\begin{table}
\scriptsize
\begin{tabular}{p{3cm}p{3cm}p{3cm}p{4cm}p{2cm}p{2cm}}
\hline
Related work        & Microorganism type & Pre-processing & Segmentation   & Classification  & Evaluation \\ \hline

\cite{Siqueira-2017-MFSA}    & Bacteria           & Median filter          & Thresholding, Canny operator and Hough transform         &        & 92.31\% accuracy                                          \\
\cite{Sanchez-2016-MAAC}     & Bacteria           & Histogram liner expansion                                                                                                        & Otsu thresholding         &                & 98\% accuracy                    \\
\cite{Maretic-2017-ACCH}     & Bacteria           & Gaussian filter, average filter and median filter & Thresholding                                                                                                            &    &                     \\
\cite{Austerjost-2017-ASDA}  & Bacteria           &     & Iterative thresholding and Hough transform     &       & 86.76 $\pm$ 9.76\% accuracy                 \\
\cite{Payasi-2017-DACT}      & Bacteria           & HSI color space processing                                                                                                       & Thresholding           &              & 90\% accuracy                       \\
\cite{Boukouvalas-2018-ACCS} & Bacteria           &  Median filter and Hough transform & Gaussian adaptive thresholding                                                                                          &       &          \\
\cite{Boukouvalas-2019-ASMC} & Bacteria           & Multidirectional Sobel operator                                                                                                  & Otsu thresholding, circularity filter and inertia filter                    &          & 99.8\% accuracy for high definition dataset and 95.9\% for low definition dataset \\
\cite{Zhu-2018-ACBC}         & Bacteria           & Nonliner gray transform      & Image subtraction and watershed      &                                                                                              & 99.8\% accuracy         \\
\cite{Tamiev-2020-ACBC}      & Bacteria           &      & Adaptive thresholding    & Classification-type convolutional neural network  &      \\
\cite{David-1989-EASA}       & Bacteria                  & Image enhancement                                                                                            & Model 2000 (Image Technology Corporation, Deer Park, New York) image analysis system        &                                                      &                                                    \\
\cite{Kildeso-1997-EAAM}     & Bacteria                  & Gaussian filter and Laplacian filter                                                                         & Kontron Vidas Plus (Kontron Elektronik GmbH, Germany) image analyser system                 &                                                      &                                                    \\
\cite{Shopov-2000-IIAA}      & Bacteria                  & Marr-Hildreth Gaussian-smoothed Laplacian edge-detection protocol, flat-fielding and edge-strength operators & Skidaway Tools (Skidaway Institute of Oceanography, 10 Ocean Science Circle, Savannah, USA) &                                                      &                                                    \\
\cite{Gmur-2000-AIES}        & Bacteria                  & Gradient convolution filter                                                                                  & IBAS 2.0 (Kontron Inc., Eching, West Germany)                                               &                                                      &                                                    \\
\cite{Singleton-2001-AFAM}   & Bacteria                  & Thresholding and edge-effect rule                                                                            & IBAS 2.0 (Kontron Inc., Eching, West Germany)                                               &                                                      &                                                    \\
\cite{Nunan-2001-QTSD}       & Bacteria                  & Sigma smoothing, top hat transform, high pass filter and morphological opening                               & Zeiss KS300 Imaging System 3.0                                                              &                                                      &                                                    \\
\hline
\end{tabular}
\end{table}

\newpage
\begin{table}
\scriptsize
\begin{tabular}{p{3cm}p{3cm}p{3cm}p{4cm}p{2cm}p{2cm}}
\hline
Related work        & Microorganism type & Pre-processing & Segmentation   & Classification  & Evaluation \\ \hline
\cite{Stoderegger-2005-DBSP} & Bacteria                  & Thresholding                                                                                                 & Zeiss KS300 Imaging System 3.0                                                              &                                                      &                                                    \\
\cite{Pena-2002-CAVA}        & Bacteria                  & Contour extraction, average equivalent diameter, high-Gauss filter and Gaussian filter                       & Image-Pro Plus image analysis software (Media Cybernetics, USA)                             &                                                      &                                                    \\
\cite{O-2003-MQPS}           & Bacteria                  & Morphological erosions and dilations                                                                         & Optimas 6.5 (Media Cybernetics Inc., Silver Spring, MD) image processing system             &                                                      & 96.8\% accuracy                                    \\
\cite{Putman-2005-SMAC}      & Bacteria                  &                                                                                                              & ProtoCOL (Version 4.04 from Synoptics Ltd., Cambridge, UK)                                  &                                                      &                                                    \\
\cite{Thiel-2005-AIMT}       & Bacteria                  &                                                                                                              & KS400 (Carl Zeiss Vision, Hallbergmoos, Germany)                                            &                                                      & 0.984 correlation coefficient with manual counting \\
\cite{Wang-2007-RAAE}        & Bacteria                  &                                                                                                              & Micro-colony auto counting system MACS (Chuo Electric Works, Osaka, Japan)                  &                                                      &                                                    \\
\cite{Hua-2009-RBFC}         & Bacteria                  & Image enhancement and median filter                                                                          & Davinci technology and local binary fitting                                                 &                                                      & 98.4\% accuracy                                    \\
\cite{Freitas-2014-OAAC}     & Bacteria                  &                                                                                                              & SigmaScan Pro 5.0 (Systat Software Inc) and thresholding                                    &                                                      &                                                    \\
\cite{Song-2018-DRAA}        & Bacteria                  & Thresholding and filling-holes                                                                               & ImageJ software version 1.52 (NIH, Bethesda, MD, USA)                                       &                                                      &                                                    \\
\hline
\end{tabular}
\end{table}
\end{landscape}

\section{Other microorganism counting methods}

\subsection{Classic counting methods}
\subsubsection{Counting methods based on image enhancement}
In~\cite{Zalewski-1996-MAYC}, color filtering and contour enhancement are used to separate the yeast cells and compare the automatically detected cell concentration and the traditional cell counting using a counting chamber.

In~\cite{Barbedo-2012-MCMA}, the histogram equalization is applied for microorganism counting. First, the images are converted to gray-scale images, and a median smoothing filter is applied for noise removal. Then, the histogram equalization is used for image enhancement, and the enhanced images are submitted to top-hat morphological filtering. Finally, the images are converted to binary images, and the connected regions are counted. The accuracy in correctly identifying the objects is more than 90\% and the overall deviation is 8\%. 

\begin{figure}[ht]
\centering
\includegraphics[trim={0cm 0cm 0cm 0cm},clip,width=0.6\textwidth]{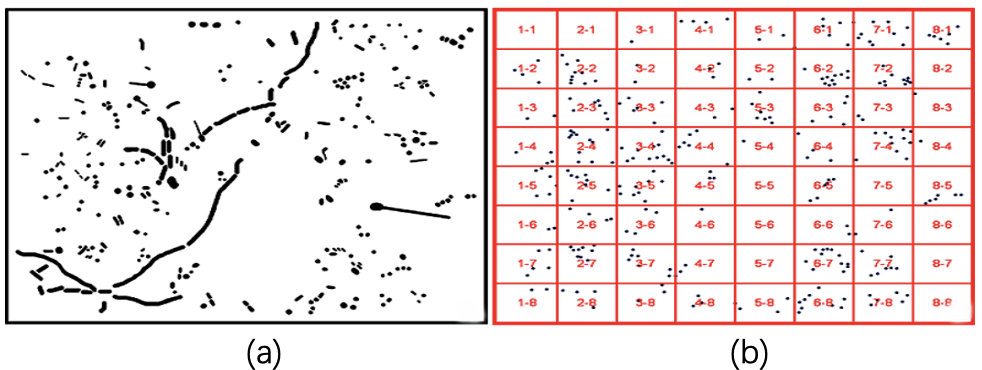}
\caption{(a) The original image. (b) All of the microoragnisms in dot map(in~\cite{Dazzo-2013-CMEI} fig.5).}
\label{fig:Dazzo-2013-CMEI}
\end{figure}

In~\cite{Dazzo-2013-CMEI}, the center for microbial ecology image analysis system (CMEIAS) is developed for understanding microbial ecology at single-cell resolution and spatial scales relevant to the individual microbes and their ecological niches in situ. An optimization method of quadrat size is proposed to reduce the complexity of calculating. Four types of quadrat sizes are proposed, that is 4$\times$4, 6$\times$6, 8$\times$8 and 10$\times$10 grid. After that, the construction of 2-dimensional scatter plots based on Cartesian coordinates of object centroids can solve serious edge effects when the best possible grid-lattice on the landscape index image still significantly overlaps foreground objects. The index image of the dot map representation derived from the original biofilm landscape image with the optimized grid raster overlay is shown in Fig.~\ref{fig:Dazzo-2013-CMEI}.

\subsubsection{Counting methods based on thresholding}
\begin{figure}[ht]
\centering
\includegraphics[trim={0cm 0cm 0cm 0cm},clip,width=0.75\textwidth]{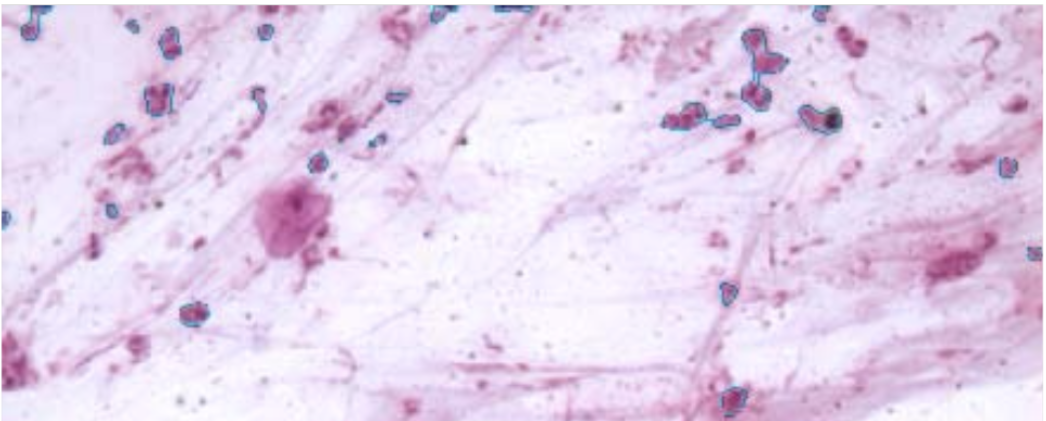}
\caption{The original image of pus cells (in~\cite{Hamid-2013-FEPC} fig.8).}
\label{fig:Hamid-2013-FEPC}
\end{figure}

In~\cite{Costello-1985-IAMT,Brown-1989-CBIA,Dias-2003-AIAI,Study-2012-SCAC,Mazzei-2014-AVIS,Cross-2004-MGTV,Packer-1990-MMFM,Tucker-1992-FAMM,Zeder-2010-AQAS}, thresholding is applied for fungi (\cite{Cross-2004-MGTV,Packer-1990-MMFM}), mycelia (\cite{Tucker-1992-FAMM}), yeast (\cite{Costello-1985-IAMT}) and protozoan (\cite{Brown-1989-CBIA,Dias-2003-AIAI}) counting.
In~\cite{Costello-1985-IAMT}, the yeast strains are counted by adjusting the gray-level of the images with the presence of high cell destinies.  In~\cite{Brown-1989-CBIA,Dias-2003-AIAI}, the minimum size of picoplankton and protozoan are set as the filters to remove the noises. Finally, the connected regions are counted as the number of chlorella in~\cite{Study-2012-SCAC}, and the flood fill algorithm is applied for labeling and tracking in~\cite{Mazzei-2014-AVIS}.
In~\cite{Zeder-2010-AQAS}, the gray-level intensities of pixels determine the threshold. `Fixels' are defined as the primary component of filaments that are can cover small parts of a filament. Once the orientation of the fixel is established, a rectangular field with the same orientation is moved along a line perpendicular to the orientation of the fixel, and the precise placement of the fixel is locally optimized by maximizing the sum of the gray-level intensities in the rectangle.  Finally, the filaments are annotated and the lengths of filaments are measured. The accuracy of the method is more than 85\%.
In~\cite{Hamid-2013-FEPC}, thresholding is applied for pus cell counting and feature extraction. 
The objects that are less than 20 pixels are eliminated after thresholding.  Then, the morphological and shape features are extracted for criteria selection. Finally, the single cells and overlap cells are classified and counted, respectively. The identified pus cells are marked on the initial image that is shown in Fig.~\ref{fig:Hamid-2013-FEPC}. It is shown that the reliability of the proposed system is above 80\% from the validation results.

In~\cite{Jones-1992-TUIA},  local threshold and morphological transformation are used to calculate spore numbers accurately. Red and blue signals are sampled for pixel segmentation. In all cases, the counts are elevated compared with those obtained by the manual method.

\begin{figure}[ht]
\centering
\includegraphics[trim={0cm 0cm 0cm 0cm},clip,width=0.6\textwidth]{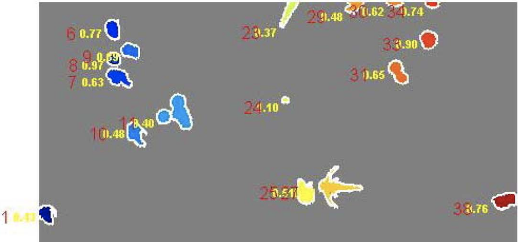}
\caption{The shape features and boundary tracks of the detected objects (in~\cite{Kim-2013-AEAH} fig.3).}
\label{fig:Kim-2013-AEAH}
\end{figure}

\begin{figure}[ht]
\centering
\includegraphics[trim={0cm 0cm 0cm 0cm},clip,width=1.0\textwidth]{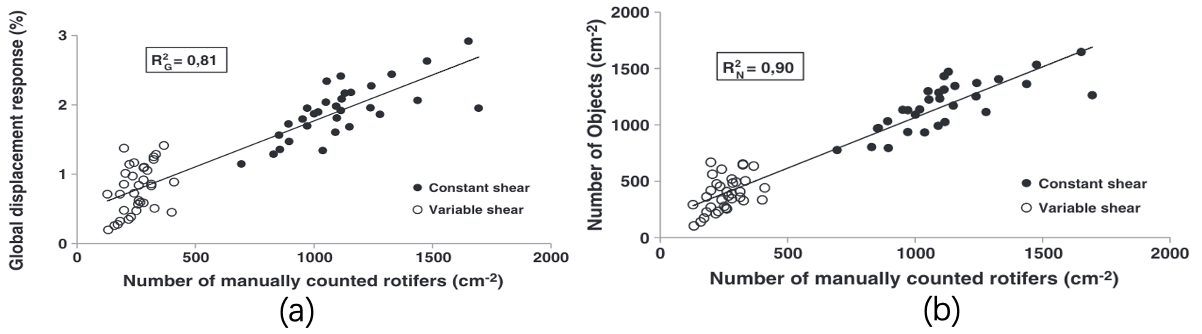}
\caption{The counting result. (a) The correlations between the manual counting result and the automatically calculated global displacement response. (b) The number of objects per c$m^{2}$ (in~\cite{Saur-2014-AAMT} fig.3).}
\label{fig:Saur-2014-AAMT}
\end{figure}

In~\cite{Robinson-1998-MCYC,Xianjiu-2012-AMAC,Kim-2013-AEAH,Saur-2014-AAMT}, Otsu thresholding is applied for the counting of microorganisms. First, the hue, lightness and saturation (HLS) model is applied to transform RGB image into the gray image. Then, the global smoothing (\cite{Robinson-1998-MCYC}) and wavelet shrinkage method (\cite{Xianjiu-2012-AMAC}) are used for noise removal. The wavelet transform is applied in the signal, and the wavelet coefficients are shrunk by thresholding. After that, the inverse wavelet transform is applied. 
Then, in~\cite{Robinson-1998-MCYC}, the binary image consisting of the regional maxima is used as the marker image for the watershed algorithm, and the Sobel filter is used to detect the edge. Finally, Otsu thresholding is used for the accurate estimation of each cell colony area. Total cell number is achieved following identification of cells by application of the shape-independent watershed algorithm.
In~\cite{Xianjiu-2012-AMAC}, the image dilation is used for image enhancement, and the combination method of Otsu thresholding and the morphological opening is used for image segmentation. Finally, the number of algae is counted based on eight neighborhood regions. The mean accuracy of the proposed method is more than 94\% comparing with the manual counting method.
In~\cite{Kim-2013-AEAH}, the morphological features are extracted after Otsu thresholding for classification and the number of the object is used for hatching rate measurement. It is shown that the maximum difference is about 19.7\%, and the average root-mean squared difference is about 10.9\% as the difference between the results using automatic counting (this study) and manual counting is compared. The result of object detection is shown in Fig.~\ref{fig:Kim-2013-AEAH}. 
In~\cite{Saur-2014-AAMT}, Otsu thresholding is used to quantify moving predators in biofilm. The Otsu thresholding is applied for image binarization, and the noises are removed by filtering. Then the global displacement response and the number of moving objects corresponding to the number of detected individual objects on the processed image are calculated. The calculating results of the two parameters above are shown in Fig.~\ref{fig:Saur-2014-AAMT}.

In~\cite{Song-2006-AARA}, the HSI thresholding is used to count the number of algae. A median filter is used to remove noises and reduce the fuzzy edges. Hue-Saturation-Intensity (HSI) threshold is used for image segmentation, and the flood fill method is used to fill the connected region. Area threshold is used to remove debris, and then the images are thinned for central point searching. Finally, the number of algae cells is counted. The accuracy of the method for identifying and counting is more than 90\%.

In~\cite{Zhonglei-2012-ADMI}, the histogram thresholding is applied for automatic fungi counting. First, the median filter and linear gray-scale transformation are applied to reduce uneven illumination and noise. Then, the image is segmented based on histogram thresholding. After that, an adaptive smooth filter is used for image enhancement, and morphological operations are applied for smoothing and hole filling. Finally, the connected regions are labeled and counted as the number of fungi to be counted.  The average relative error between the proposed method and the manual counting method is 2.56\%.

\begin{figure}[ht]
\centering
\includegraphics[trim={0cm 0cm 0cm 0cm},clip,width=1.0\textwidth]{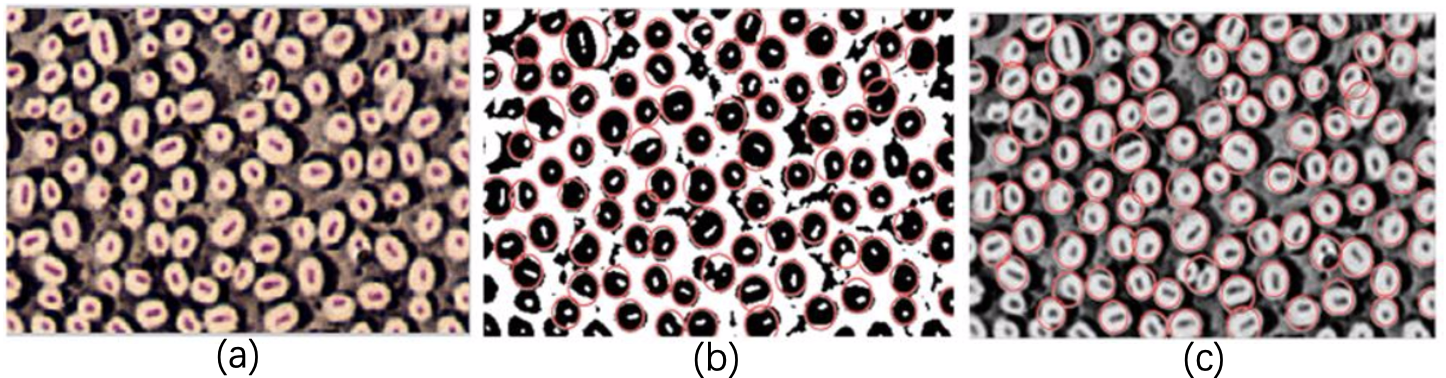}
\caption{The results after thresholding and morphological operations. (a) The original image. (b) Image after thresholding and opening operation. (c) The identified microorganisms (in~\cite{Sharma-2015-CMMD} fig.8).}
\label{fig:Sharma-2015-CMMD}
\end{figure}

In~\cite{Sharma-2015-CMMD}, two methods based on image processing are proposed for microorganisms counting in medical. The first one is based on object recognition, which means the microorganisms' shape is considered to find out the total number of microbes in the image sample. The histogram equalization is applied for image enhancement that can help to separate the objects from the background. Then the circular hough transformation technique is used to determine the circular objects in the image. The second one is based on thresholding. First, the image is enhanced with histogram equalization and converted to a binary image. Then the Moore neighbor tracing algorithm is applied to detect objects that have close boundaries. Finally, the combination method of thresholding, object recognition and morphological operation is used for counting, and the accuracy of 93\% is obtained. The segmentation result is shown in Fig.~\ref{fig:Sharma-2015-CMMD}.

\begin{figure}[ht]
\centering
\includegraphics[trim={0cm 0cm 0cm 0cm},clip,width=0.9\textwidth]{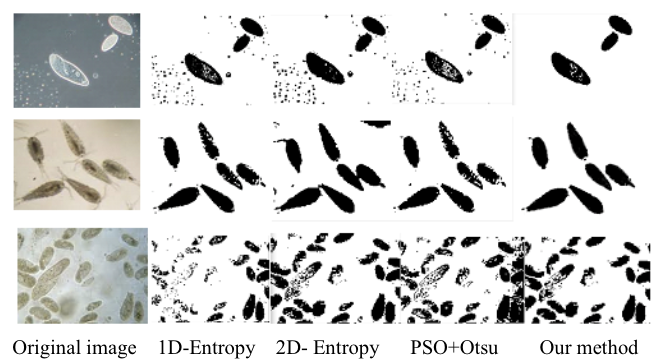}
\caption{Comparisons of target segmentation effects (in~\cite{Fang-2019-MICM} fig.1).}
\label{fig:Fang-2019-MICM}
\end{figure}

In~\cite{Fang-2019-MICM}, a multi-threshold image counting method based on improved particle swarm optimization (PSO) is proposed for automatic microbial counting. The two-dimensional maximum entropy algorithm is extended to design the objective function using exponential entropy and an improved PSO algorithm to acquire its maximum value and the best image segmentation effect. Furthermore, the breadth-first search (BFS) algorithm is applied to complete the microorganism marker and counting in the segmented images. Finally, the number of image target segmentation is determined according to the histogram peak searching method. The comparisons of target segmentation results are shown in Fig.~\ref{fig:Fang-2019-MICM}.

\subsubsection{Counting methods based on edge detection}

In~\cite{Viles-1992-MMPC,Sieracki-1995-OHBT}, the Marr-Hildreth method is used for edge detection and image segmentation of picoplankton (in~\cite{Viles-1992-MMPC}) and heterotrophic bacteria (in~\cite{Sieracki-1995-OHBT}). Combinations of edge strength and minimum and maximum cell sizes allow the user to count specific cell populations. The images are captured using a charge-coupled device (CCD) imaging system, and the result shows an accurate performance. 

\begin{figure}[ht]
\centering
\includegraphics[trim={0cm 0cm 0cm 0cm},clip,width=1\textwidth]{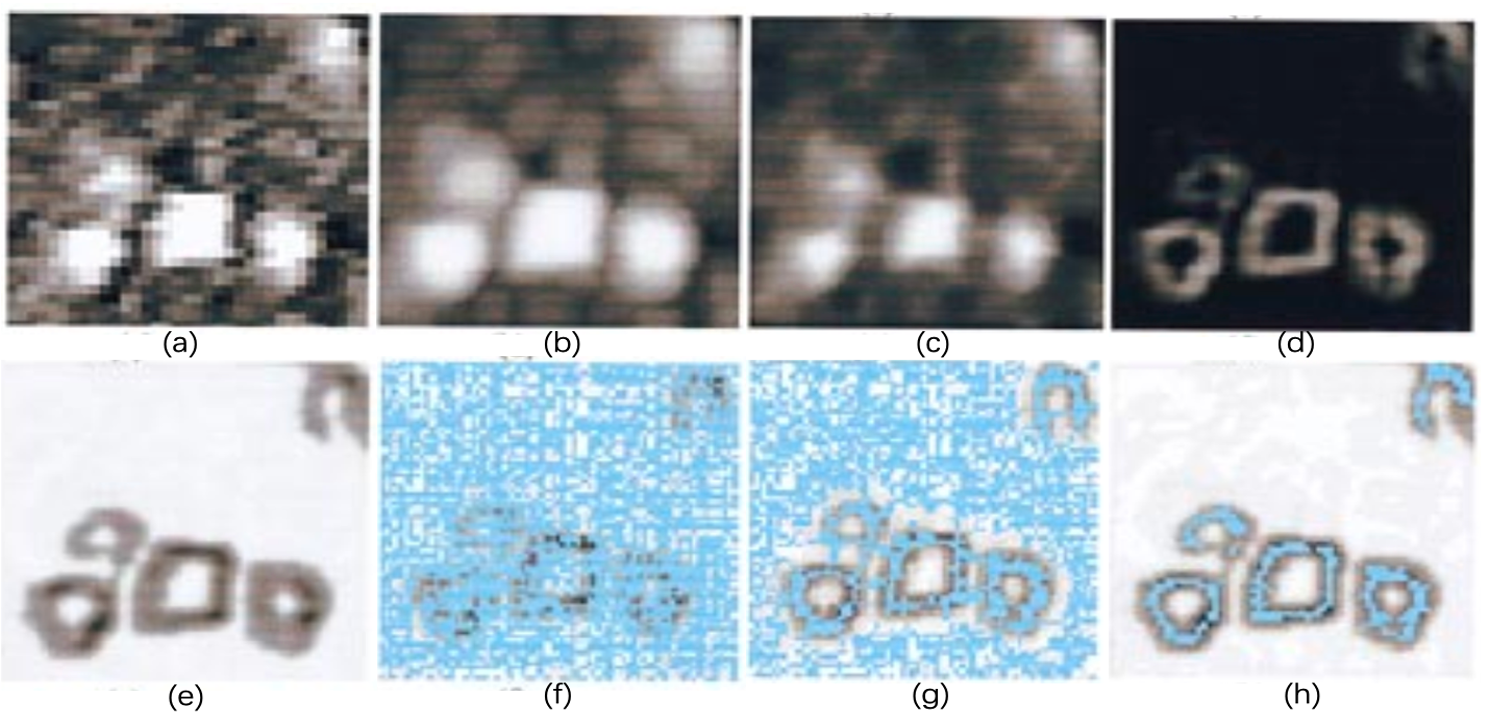}
\caption{Example of steps in segmentation (In~\cite{Kocak-1999-CVTQ} fig.12).}
\label{fig:Kocak-1999-CVTQ}
\end{figure}

In~\cite{Kocak-1999-CVTQ}, the Snake model is used for low-level interaction, which is a deformable parametric curve with its corresponding energy function, the closed curve with the minimum energy is the target contour. The edge image is produced by subtracting the eroded image from a dilated image.  An extermination algorithm is used to examine the image gradient on both sides of the snakelet by computing the directional derivative orthogonal to each of its nodes. The Snake model is used for plankton counting in this research, and the accuracy of 94.12\% is achieved. The example of segmentation is shown in Fig.~\ref{fig:Kocak-1999-CVTQ}.

\begin{figure}[ht]
\centering
\includegraphics[trim={0cm 0cm 0cm 0cm},clip,width=0.75\textwidth]{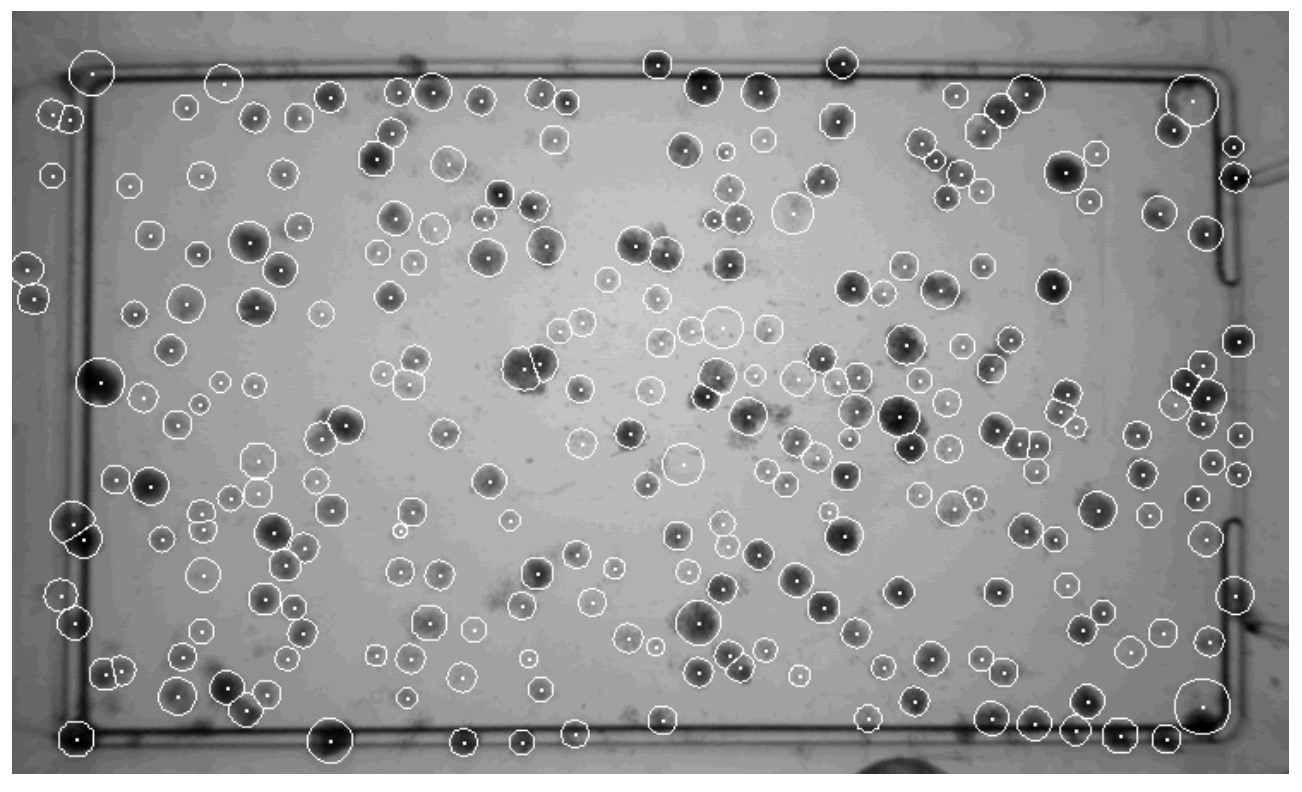}
\caption{Example of steps in segmentation (In~\cite{Barber-2000-AACC} fig.5).}
\label{fig:Barber-2000-AACC}
\end{figure}

In~\cite{Barber-2000-AACC}, the Sobel operator is used to find the edges, and the binary images are obtained using thresholding.  A compact Hough transform is used to highlight the centers of circular objects. The local area can be processed to determine a colony boundary, and so the colony area and the colony number can be calculated.  An example of colony boundary determination is shown in Fig.~\ref{fig:Barber-2000-AACC}.

In~\cite{Pernthaler-2003-AEGM}, the gradient transformation is applied for microorganism edge detection. First, a mean background gray level is determined for remapping the image gray values to the total gray value range. Edge detection is performed by using gradient transformation, and a neighborhood median filter is applied to smooth the resulting image. The stained plankton cells are calculated and counted. 

In~\cite{Tsechpenakis-2008-ANCA}, the probabilities are applied for image segmentation. First, a nonlinear morphological filter, that is, an alternating sequential filter (ASF) is used to preserve the line-type image structures in predefined orientations while filtering random noise. Then the region of animals is segmented based on the probabilities. The isolated pixels or small groups of pixels with probabilities higher than the threshold are eliminated.  After that, a nonrigid is used to recover a global transformation that brings the pose of a source shape as close as possible to that of a target shape. The shape estimation of animals is based on the maximum likelihood (ML) approach, and the animal region is extracted based on a probability map. Finally, the number of animals is calculated and compared with the manual counting result. 

In~\cite{Barbedo-2012-UFCA}, a unified framework for counting agriculture microorganisms is proposed.  There are five methods listed for object delineating. In the first method, the Laplacian of Gaussian method is applied for edge detection, then the inner regions are filled, and all connected objects are identified. The only difference between the second method and the first one is that the Canny method is used to detect the edges. In the third method, three different thresholds are applied for image binarization. In the fourth method, contrast is modified by the technique of histogram equalization. In the last method, region growing is applied for segmentation. After object delineating, a decision tree is used for classification. Then the contrast limited adaptive histogram equalization is applied, and the image is morphologically opened using as kernel a disk with a radius of 1\% of the image width. Finally,  the estimate for the number of objects is calculated based on the number of local maxima.

\subsection{Machine learning and deep learning counting methods}
In~\cite{Shabtai-1996-MMMC,Embleton-2003-ACPP}, a neural network is applied for fungus (\cite{Shabtai-1996-MMMC}) and phytoplankton (\cite{Embleton-2003-ACPP}) counting. 
In~\cite{Embleton-2003-ACPP}, the gray-level is used to separate the regions of interest,  and then a median filter is used to smooth images. A skeletonize operator is used to separating the filaments and objects. Then the final binary image is used as a mask to take measurements from the original image.  Each region in the binary image is given an identifying number, and the size, shape, color and grey level distribution are measured for that region. A neural network is used for the classification and counting of phytoplankton. The automated imaging system takes 75 images for each sample in seven minutes, and the image processing and classification take thirty to forty minutes.

In~\cite{Benyon-1999-DAFS}, seven basic features and 17 more complex features are extracted from fungal spores for image analysis. Linear and quadratic discriminant analysis are used for image classification, and then the number of every species of spores is counted based on the results of classification. Genus comparisons using only seven basic features resulted in 98\% accuracy.

\begin{figure}[ht]
\centering
\includegraphics[trim={0cm 0cm 0cm 0cm},clip,width=0.7\textwidth]{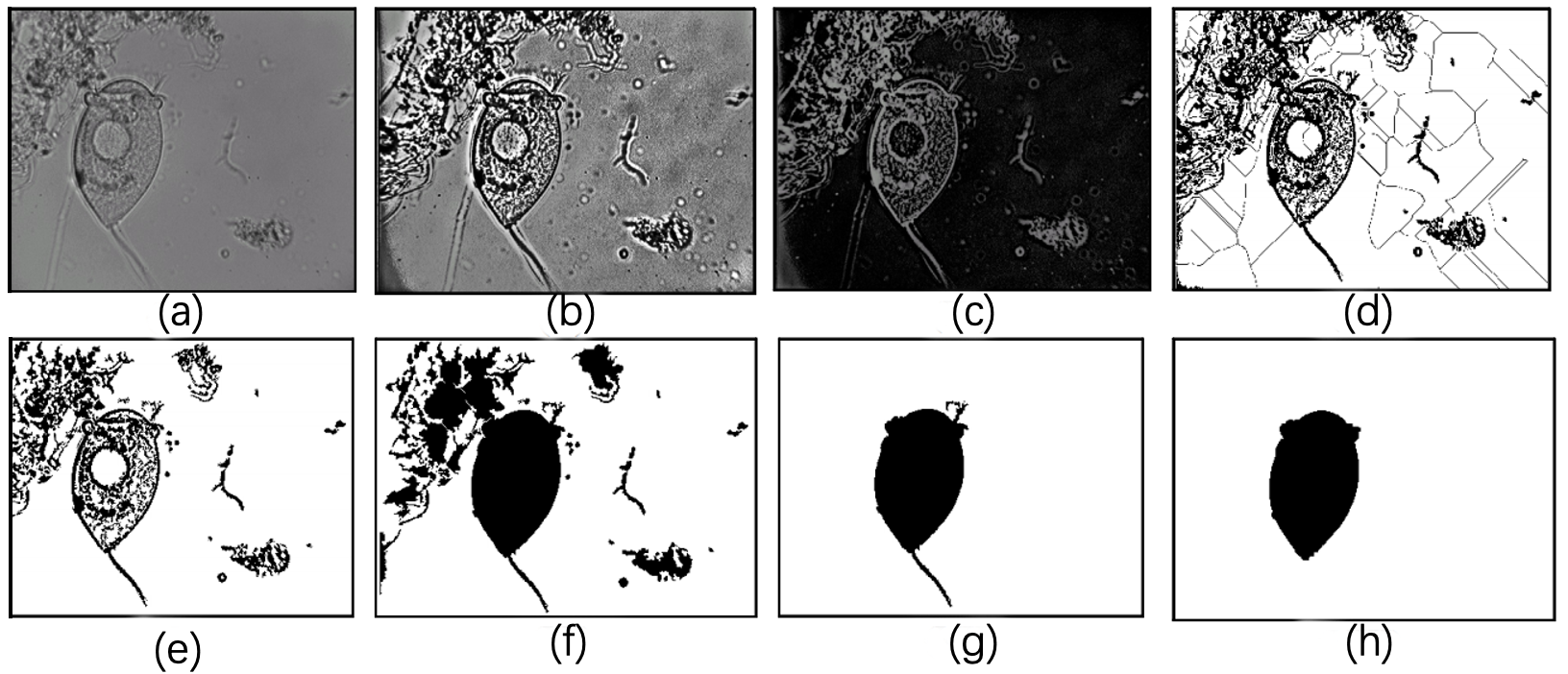}
\caption{Main steps of segmentation (In~\cite{Motta-2001-TSPP} fig.1).}
\label{fig:Motta-2001-TSPP}
\end{figure}

In~\cite{Motta-2001-TSPP,Akiba-1997-DASZ}, PCA is applied for protozoa (\cite{Motta-2001-TSPP}) and plankton (\cite{Akiba-1997-DASZ}) counting and classification. In~\cite{Motta-2001-TSPP}, the histogram local equalization is used to enhance the contours of protozoa images, the opening operation and closing operation are used to remove halo. Euclidian Distance Map is used for semi-automated segmentation. A series of erosion and reconstruction are used to eliminate the flocs of the protozoa silhouette.  Finally, PCA is used to classify different protozoa, and the number of the various species of protozoa is counted. The main steps of segmentation are shown in Fig.~\ref{fig:Motta-2001-TSPP}, (a) shows the initial image of protozoa, (b) shows the contour enhancement by histogram local equalization, (c) shows the background suppression by opening (2 iterations) and closing (55 iterations) to remove the halo, (d) shows the semi-automated segmentation based on the Euclidian Distance Map, (e) shows part of the flocs is eliminated by a border-killing routine, (f) shows the hole-filling of the silhouette and semi-automated segmentation based on the Euclidian Distance Map, (g) shows the elimination of flocs by a series of erosion and reconstruction of the protozoa silhouette, (h) shows the localization of flagella and stalk.

\begin{figure}[ht]
\centering
\includegraphics[trim={0cm 0cm 0cm 0cm},clip,width=0.5\textwidth]{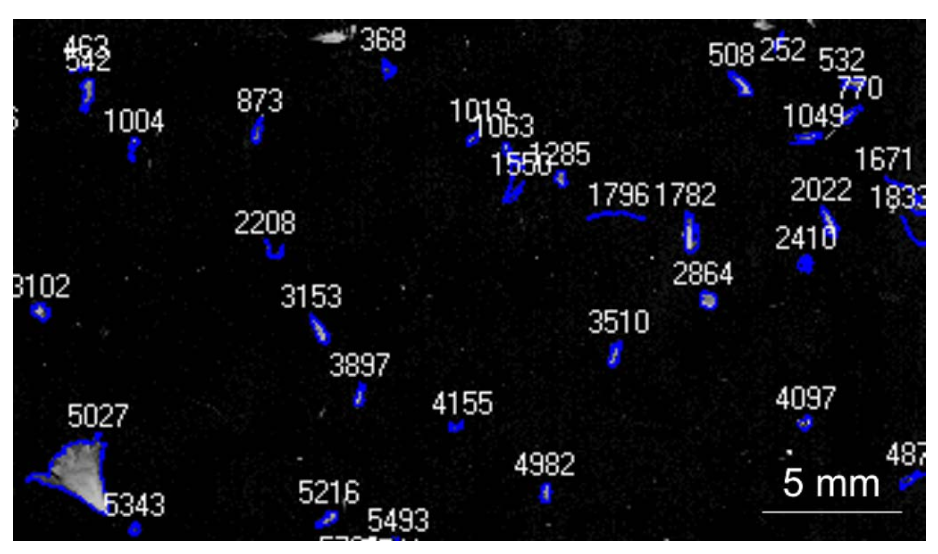}
\caption{The objects are detected, contoured, and labelled by the image analysis (In~\cite{Grosjean-2004-EMAI} fig.3).}
\label{fig:Grosjean-2004-EMAI}
\end{figure}

In~\cite{Grosjean-2004-EMAI}, the random forest and discriminant vector forest are applied for zooplankton image processing.  A threshold is used to eliminate the background and enhance the contrast, then the objects are detected, contoured, and labeled by the image analysis system. The combination method of random forest and discriminant vector forest is used for classification, and the number of each species is counted. The result of object detection is shown in Fig.~\ref{fig:Grosjean-2004-EMAI}.

In~\cite{Rong-2006-ACZB}, BP neural network and image processing are used for classification and counting of zooplankton.  Otsu thresholding is used for initial image segmentation, and region growing is used to fill holes. The noises of debris are removed by detecting the particles that are smaller than the set threshold area. Then the features such as gray level co-ocurence matrices (GLCM) and some shape features are measured and used for classification and counting by using a back propagation (BP) neural network. There are 5 nodes in the input layer, 20 nodes in the hidden layer, and 1 node in the output layer, and the Sigmoid function is applied for activation. 

\begin{figure}[ht]
\centering
\includegraphics[trim={0cm 0cm 0cm 0cm},clip,width=0.7\textwidth]{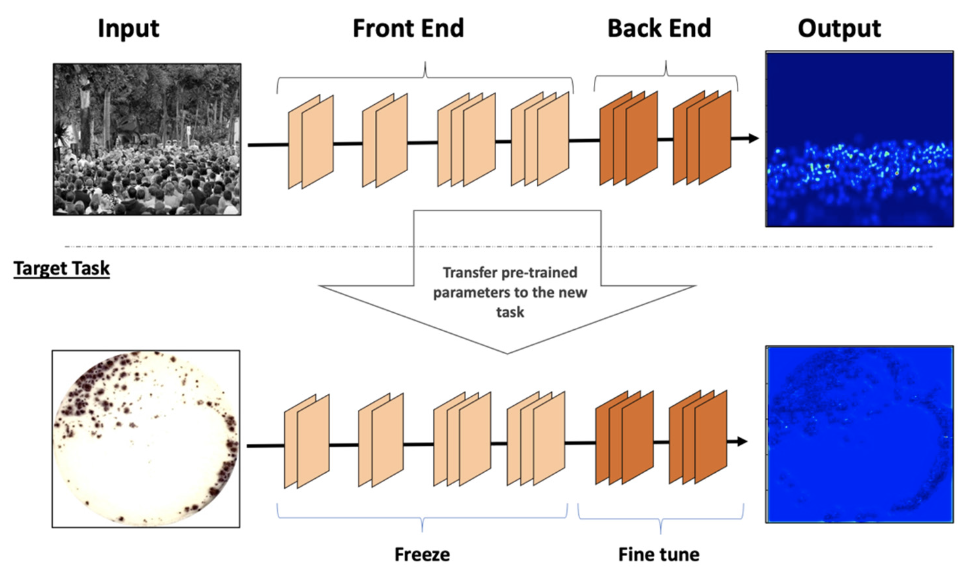}
\caption{Processing of transfer learning for counting (in~\cite{Albaradei-2020-ACCF} fig.2).}
\label{fig:Albaradei-2020-ACCF}
\end{figure}

In~\cite{Albaradei-2020-ACCF}, deep transfer learning is applied for automatic pluripotent stem cell colony counting. First, the RGB image is converted to a binary image by using thresholding. Then, some augmentation techniques are applied to expand the training dataset. The augmentation techniques include color jitter to randomly alter brightness, contrast, saturation, and hue of each image, horizontal/vertical flip, and random rotation. Moreover, the trained SRNetDL model is applied for training. The first 10 layers are frozen to remain the pre-trained network. Then the last 6 layers are fine-tuned. The stochastic gradient descent is applied for optimization. Finally, the number of cell colonies is counted. The overview of transfer learning is shown in Fig.~\ref{fig:Albaradei-2020-ACCF}.

\subsection{Third-party tools}
\begin{figure}[ht]
\centering
\includegraphics[trim={0cm 0cm 0cm 0cm},clip,width=1.0\textwidth]{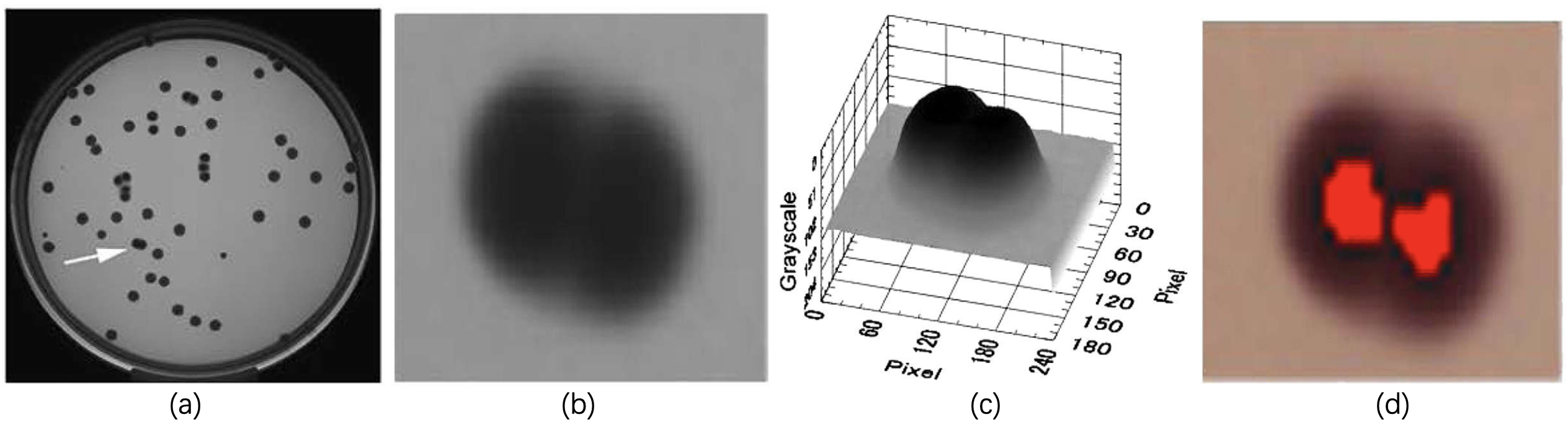}
\caption{Detection of 2 overlapping colonies of E.coli.(a) 2D shadow image of agar plate.  The arrow points to the cluster of overlapping colonies. (b) Magnified image. (c) 3D shadow image. (d) 2 colonies recognized in the cluster (in~\cite{Ogawa-2012-NFAC} fig.6).}
\label{fig:Ogawa-2012-NFAC}
\end{figure}

In~\cite{Ogawa-2012-NFAC}, a time-lapse shadow image analysis system is designed for microbial colony counting. First, an agar plate containing many clusters of microbial colonies is trans-illuminated to project their 2-dimensional (2D) shadow images on a color CCD camera. Then the 2D shadow images of every cluster distributed within a 3-mm thick agar layer are captured in focus simultaneously through a multiple focusing system and then converted to 3-dimensional (3D) shadow images. It is possible to determine whether each cluster comprised single or multiple colonies by time-lapse analysis of the 3D shadow images. Finally, the recognized colonies are counted, and the result is compared with the manual counting method, and an excellent value of correlation efficiency is obtained (r = 0.999). The colony detection method is shown in Fig.~\ref{fig:Ogawa-2012-NFAC}.

\begin{figure}[ht]
\centering
\includegraphics[trim={0cm 0cm 0cm 0cm},clip,width=0.6\textwidth]{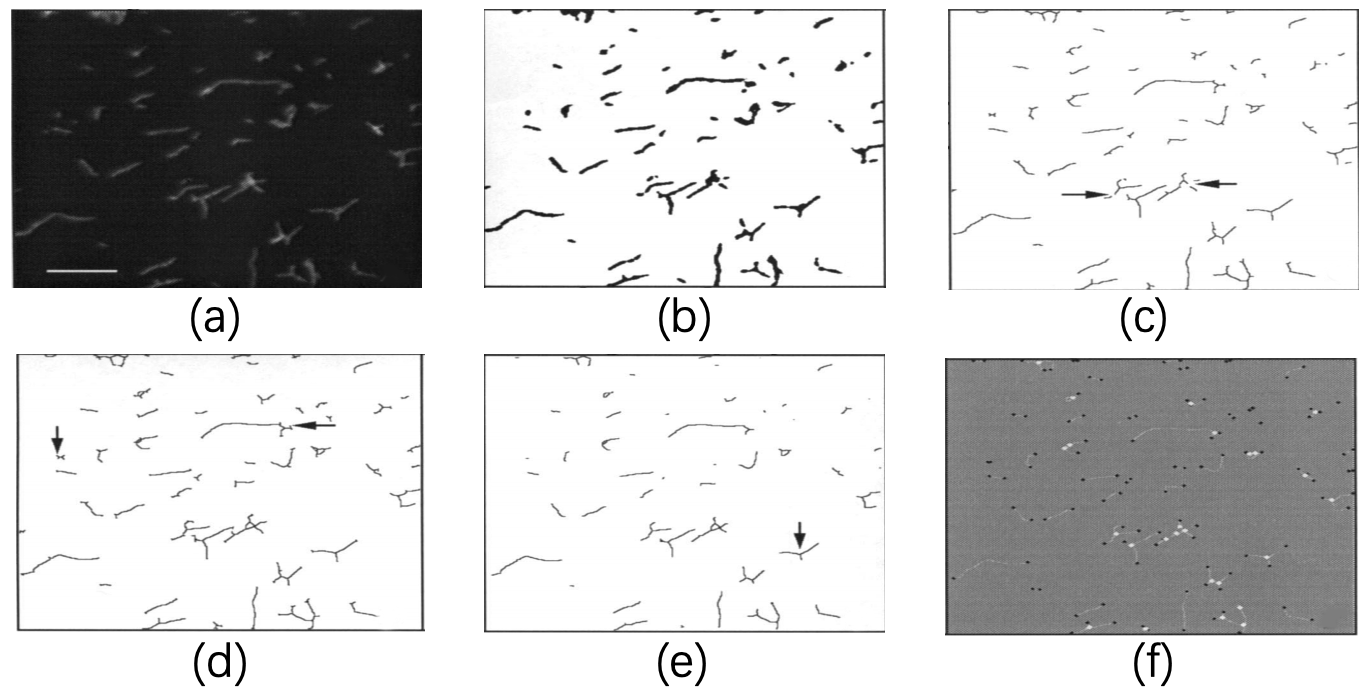}
\caption{The processing of the actinomyces image. (a) Original image. (b) The binary image after a two-step thresholding. (c) The skeleton selection. (d) The whole skeleton with branches. (e) The whole pruned skeleton. (f) The triple points and end points on the skeletons (in~\cite{Grivet-1999-AEAS} fig.3).}
\label{fig:Grivet-1999-AEAS}
\end{figure}

In~\cite{Rolke-1984-SSAZ},  `Quantimet 720' image analysis system (Leica Cambridge Ltd., Cambridge, United Kingdom) is used to detect the zooplankton, and then the total number of objects is measured.  The detector automatically selects the mean grey-level between image and background within the present range to ensure the optimal detection of the image contours. `Quantimet 570' is developed in 1990 that is used in~\cite{Bloem-1995-FADS}, the images are sharpened with maxima and minimum filter, and then all local maxima values are detected to determine the number of particles. A 5 by 5 convolution filter is used to remove noises, and a skeleton operation is used to separate the particles precisely. The mean differences between the visual and automated method are not significantly different from zero. The first method in~\cite{Grivet-1999-AEAS} is the usage of the `Quantimet 570' image analysis system for scanning and counting the adherent microorganisms. The second method shows the use of thresholding and image skeleton for enumeration. The correlation between the two enumeration methods is highly significant. The example of the second method for counting is shown in Fig.~\ref{fig:Grivet-1999-AEAS}.

In~\cite{Sieracki-1985-DEAS}, `Artek 810' image analyzer (Artek Systems Corp., Farmingdale, N.Y.) is used to detect, count and size the picoplankton. The threshold destiny level of detection can be set at 1 to 256 gray-level. Comparisons between visual and image analyzed counts show that none of the mean counts are significantly different at the 95\% significance level by the paired $t$ test.

In~\cite{Estep-1989-CSAI}, the `Zeus' image analysis system (Institute of Marine Research, Bergen) is used for algae counting, sizing and identification.  Images are enhanced, smoothed, shadowed, and then `Zeus' system is used for automated counting.  Comparison counts show no significant difference between the manual method and automatic identification and counting with the image-analysis system.

In~\cite{Wright-1991-CIPA}, a constant threshold is used to segment the images instead of an adaptive threshold, and then the Sobel operator is applied to detect the edge of micro plants and animals.  Finally, `ImageMeasure 5100' (Microscience, Div., Phoenix Trade, Inc. Seattle, Washington) is used for counting. The developed script reduces the time required to count and measure marine fouling tube worms by at least one order of magnitude over manual counts, with an error of five percent or less.

\begin{figure}[ht]
\centering
\includegraphics[trim={0cm 0cm 0cm 0cm},clip,width=0.6\textwidth]{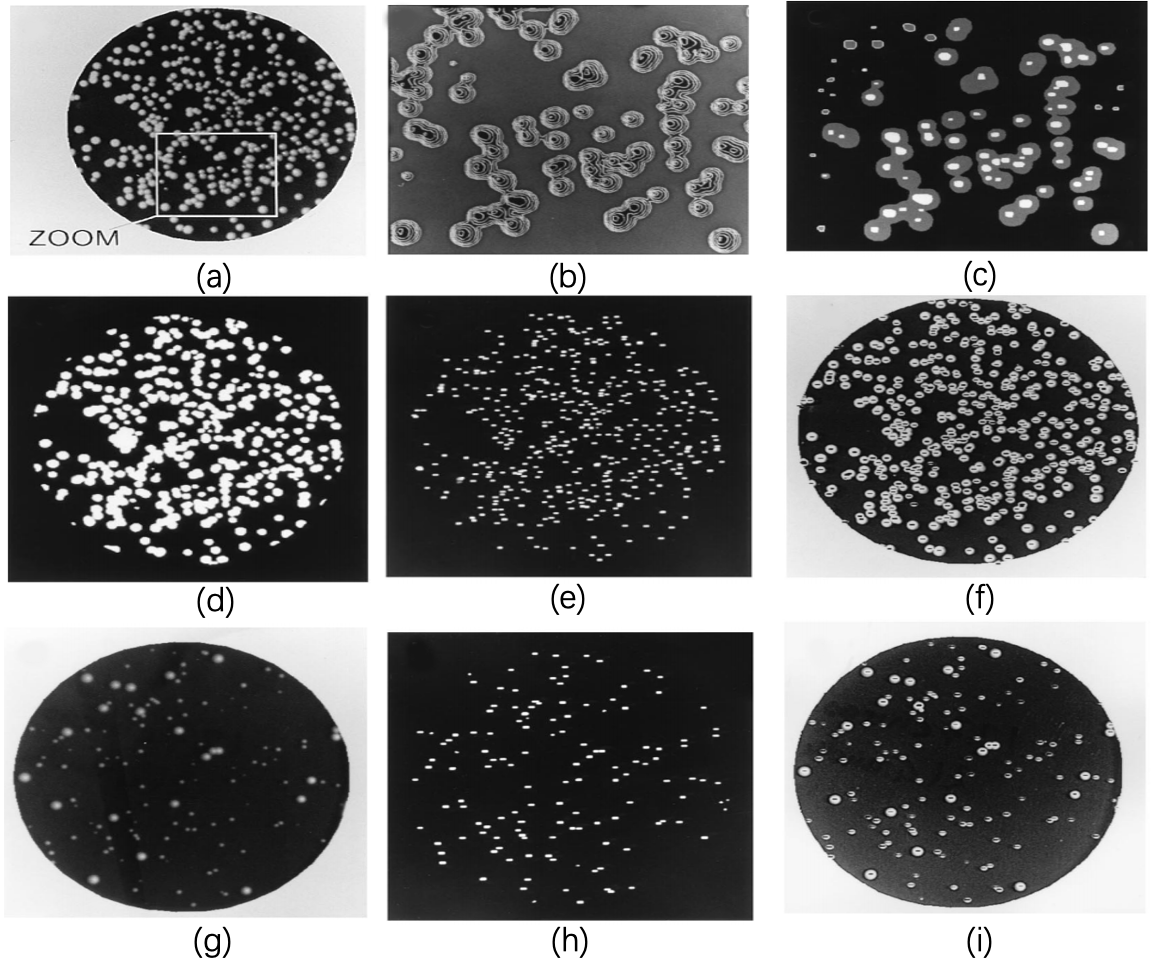}
\caption{Detection of colony. (a) The image is digitized and filtered through a binary mask. (b) Enhanced image. (c) Resulting image after adding images at T1 and T2 thresholds. (d) Colony segmentation at a single threshold level of data. (e) Colony segmentation by using COVASIAM. (f) Overlap of Fig.~\ref{fig:Corkidi-1998-CAIA}a and e. (g) Digitized image of colony in various size. (h) Colony segmentation result. (i) Overlap of Fig.~\ref{fig:Corkidi-1998-CAIA}g and h. (In~\cite{Corkidi-1998-CAIA} fig.2).}
\label{fig:Corkidi-1998-CAIA}
\end{figure}

In~\cite{Corkidi-1998-CAIA}, a commercial program `IMAGENIA 2000' (Biocom, Les Ulis, France) is used for image processing and object counting based on the multi-level threshold, and the total number of bright spots over the dark background is counted. The confluent and various sizes image analysis method (COVASIAM) is proposed, which estimates an average of 95.47\% ($\sigma$ = 8.55\%) of the manually counted colonies, while an automated method based on a single-threshold segmentation procedure estimates an average of 76\% ($\sigma$ = 16.27\%) of the manually counted colonies. Fig.~\ref{fig:Corkidi-1998-CAIA}(h) shows the segmentation result of Fig.~\ref{fig:Corkidi-1998-CAIA}(g), and  Fig.~\ref{fig:Corkidi-1998-CAIA}(i) shows the superimposed images of  Fig.~\ref{fig:Corkidi-1998-CAIA}(g) and (h). COVASIAM gave 135 CFU, representing 97.8\% of the manual counts.

\begin{figure}[ht]
\centering
\includegraphics[trim={0cm 0cm 0cm 0cm},clip,width=0.75\textwidth]{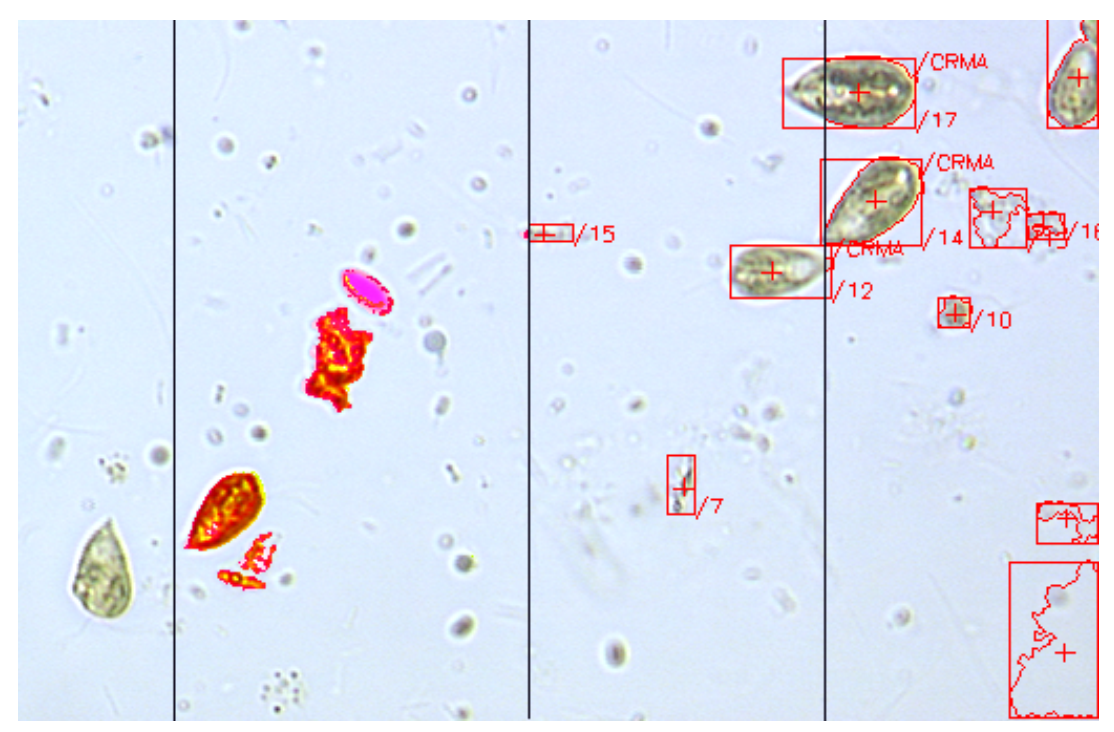}
\caption{Segmentation of one image, central field with masks, right field with marked featured objects (In~\cite{Rodenacker-2002-IAQP} fig.2).}
\label{fig:Rodenacker-2002-IAQP}
\end{figure}

In~\cite{Rodenacker-2001-SARM}, a program `IDL' (Research System Inc., Boulder, USA) is used for image analysis. Thresholding is used to segment the images. The opening operation and closing operation are used to clean the masks and fill holes. A neural network is designed to classify the microorganisms based on shape features that can help count each microorganism species in water.  The `IDL' is also used for identification and quantification of phytoplankton in~\cite{Rodenacker-2002-IAQP}, and the threshold is used for image segmentation. The segmentation result is shown in Fig.~\ref{fig:Rodenacker-2002-IAQP}. The morphological and some intensity features are used for identification, and a neural network is designed for classification, then the number of each species is counted.

In~\cite{Sandor-2001-ATRB},  `Quanitmet 570' computer system (Leica, Cambridge, United Kingdom) is used for image processing. The proportion of the clumps is determined as the mean value of their projected areas, and the mean total hyphal length and the mean number of tips are determined for the freely dispersed mycelia.

In~\cite{Nishimura-2006-UAAC}, an automatic cell counting system, `Bioplorer' (BP) (Matsushita Ecology Systems Co. Ltd, Kasugai, Aichi-ken, Japan), is used for the enumeration of yeast cells. BP system is used to exclude yeast cells measuring 5 $\mu$m in diameter from the count when cultured yeast. The intensity is used for image segmentation and bright points are counted.  In~\cite{Nishimura-2008-ACFM}, BP is used for the quantification of eukaryotic and prokaryotic cells. The cells are stained and captured using a CCD camera that can capture photons emitted from bacteria or yeast cells. Bright points are visualized on display and enumerated automatically. The threshold brightness value is optimized for each set of measurements by using BP.

In~\cite{Eickhorst-2008-IDSM}, an image analysis software, `AnalySIS' (Soft Imaging) is used for soil microorganisms counting and detection. First, the images are optimized, such as the contrast of enhancement and gradation. Then the threshold value is set for the color of fluorescent probes. After that, the parameters are set to detect the microorganism pixels and the cell number is automated counted. The ratio of automated to manual counting is 97.7\% ($\pm$ 1.0) for bacteria and 92.2\% ($\pm$ 2.4) for archaea in the investigated paddy soils.

\begin{figure}[ht]
\centering
\includegraphics[trim={0cm 0cm 0cm 0cm},clip,width=1.0\textwidth]{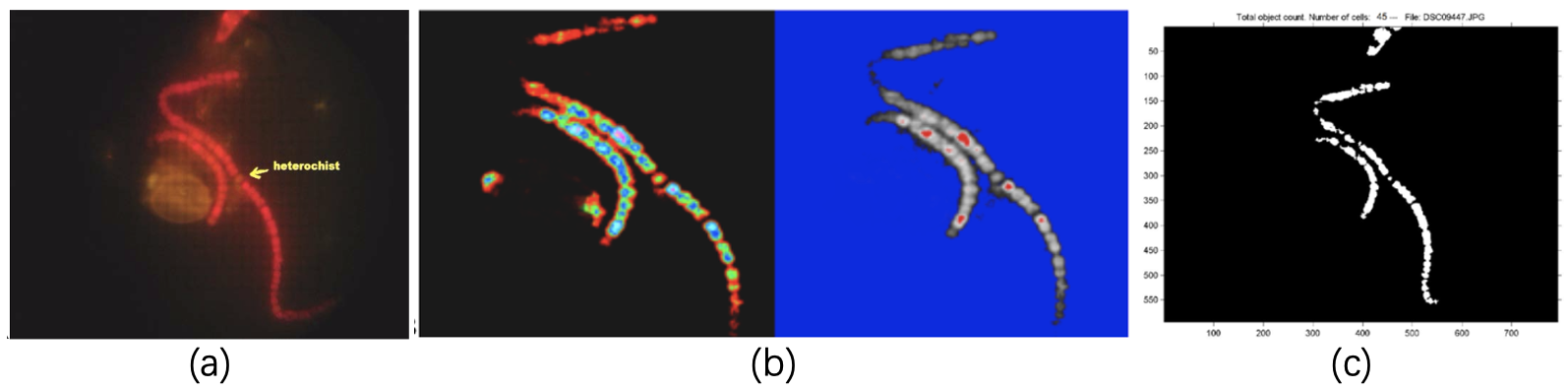}
\caption{(a)Digital image of cyanobacteria. (b) Delimitation A panel edges using Image J software. (c) The total number of cells in A panel using CellC software (In~\cite{Ghitua-2013-MIAA} fig.9).}
\label{fig:Ghitua-2013-MIAA}
\end{figure}

\begin{figure}[ht]
\centering
\includegraphics[trim={0cm 0cm 0cm 0cm},clip,width=0.7\textwidth]{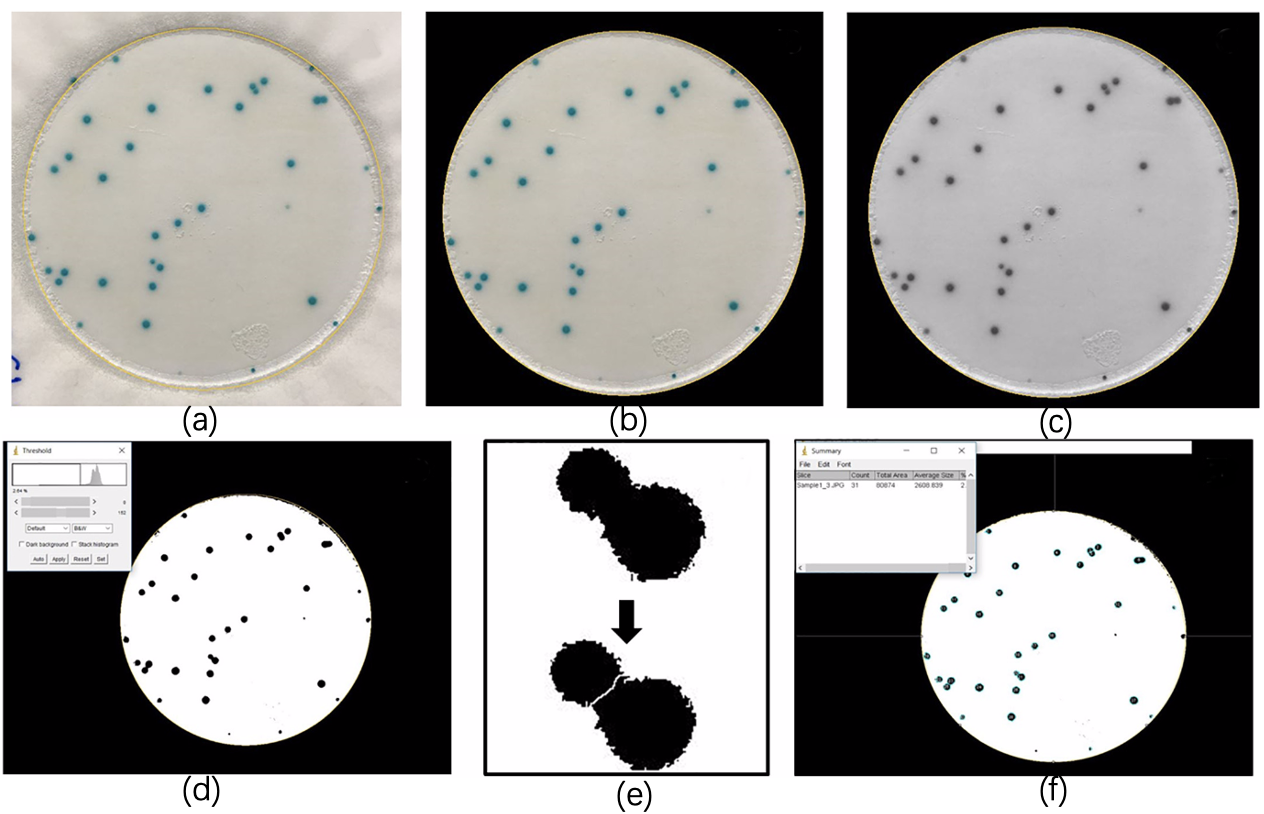}
\caption{ImageJ automated cell analysis of a Petrifilm image. (a) Petrifilm colony forming area outlined with the oval ROI tool. (b) The area outside the ROI is cleared. (c) Image is converted to 8-bit. (d) Threshold is set to highlight colonies as black particles. (e) Merging colonies split by a single pixel line via the Watershed tool. (f) Particles included in the total count, highlighted and numbered in an overlay on the image (In~\cite{Stolze-2019-AIAI} fig.1).}
\label{fig:Stolze-2019-AIAI}
\end{figure}

In~\cite{Ghitua-2013-MIAA}, `ImageJ' is applied for analysis of cyanobacteria from the marine sample. First, the background is separated from the objects based on the intra-class variance threshold method. Then the mathematical morphology operations are used to remove noises produced by specks of staining color in the image. Finally,  the clustered objects are separated and counted. The same image is then analyzed with the program `CellC' to count the cells in the filament of cyanobacteria. The result of the proposed method is shown in Fig.~\ref{fig:Ghitua-2013-MIAA}. In~\cite{Stolze-2019-AIAI}, `ImageJ' is applied for yeast colony counting based on automatic image analysis. First, the RGB image is converted to an 8-bit image, and the thresholding value is adjusted for image binarization. After that, the watershed is applied for splitting merging colonies, and finally, the number of colonies is counted. The proposed method is shown in Fig.~\ref{fig:Stolze-2019-AIAI}.

In~\cite{Bennke-2016-MAHT}, `Automated Cell Measuring and Enumeration tool 2.0' program is applied for the enumeration of the microbial cell. First, the sample is stained using DAPI and captured using a CCD camera. Then the program is applied for cell determination and enumeration. The parameter values like object area, circularity, mean gray value and signal-to-background ratio are measured.  Automated enumeration result is highly correlated with manual counts ($r^2$>0.9).

\subsection{Summary of image analysis based counting for other microorganisms}
By reviewing the related work of image analysis for other microorganisms counting and from Table~\ref{tab:othercounting}, we find that:
\paragraph - Development trend The counting for other microorganisms using image analysis approaches began in the 1980s and developed in the 2000s.  By comparing with the related research on bacteria, the development speed of research about other microorganisms counting is relatively slow, and the research is relatively limited. There are two main reasons to cause this situation, firstly, the structures of bacteria are relatively simple, and most of them are circular, which are more visualized and accessible to evaluated for segmentation results. By contrast, some other microorganisms, such as alga and fungi, are relatively complex in structure and with plenty of hyphae, challenging to be segmented precisely. Secondly, microorganism counting systems are designed but not for one specific type. A dataset is necessary when evaluating systems’ performances, but the number of bacteria datasets is relatively abundant, so the researchers tend to test their systems with bacteria datasets.
\paragraph - Counting techniques  The most frequently used pre-processing methods are the medial filter and Gaussian filter, image segmentation methods are thresholding and Otsu thresholding, classifier algorithms are PCA and neural networks.

\begin{landscape}
\begin{table}
\scriptsize
\caption{\label{tab:othercounting}Summary of image analysis based counting for other microorganisms} 
\begin{tabular}{p{3cm}p{3cm}p{3cm}p{4cm}p{2cm}p{2cm}}
\hline
Related work        & Microorganism type & Pre-processing & Segmentation   & Classification  & Evaluation \\ \hline
\cite{Costello-1985-IAMT}     & Yeast                           &                                                                                                                                                                        & Gray-level adjusting                                                                            &                                                      &                                                     \\
\cite{Brown-1989-CBIA}         & Picoplankton                    & Minimize size filter                                                                                                                                                   & Gray-level adjusting                                                                            &                                                      & 7 times faster than manual counting                 \\
\cite{Dias-2003-AIAI}          & Protozoan                       & Wiener filter                                                                                                                                                          & Thresholding and size filter                                                                    &                                                      &                                                     \\
\cite{Cross-2004-MGTV}         & Fungi                           & Square root transformation                                                                                                                                             & Line detection, thinning and thresholding                                                       &                                                      &                                                     \\
\cite{Study-2012-SCAC}         & Chlorella                       &                                                                                                                                                                        & Thresholding and connected region detection                                                     &                                                      &                                                     \\
\cite{Mazzei-2014-AVIS}        & Bioluminescence organisms       &                                                                                                                                                                        & Thresholding and flood filling                                                                  &                                                      & 92.24\% accuracy                                    \\
\cite{Packer-1990-MMFM}        & Fungi                           & Circularity test                                                                                                                                                       & Thresholding and skeletonization                                                                &                                                      &                                                     \\
\cite{Tucker-1992-FAMM}        & Fungi                           & Circularity test                                                                                                                                                       & Thresholding and skeletonization                                                                &                                                      &                                                     \\
\cite{Viles-1992-MMPC}         & Picoplankton                    &                                                                                                                                                                        & Edge strength filter, minimum and maximum cell size filter and Marr-Hildeth method              &                                                      &                                                     \\
\cite{Sieracki-1995-OHBT}      & Heterotrophic bacteria          &                                                                                                                                                                        & Edge strength filter, minimum and maximum cell size filter and Marr-Hildeth method              &                                                      &                                                     \\
\cite{Jones-1992-TUIA}         & Spores                          & RGB processing                                                                                                                                                         & Local thresholding and morphological transform                                                  &                                                      &                                                     \\
\cite{Zalewski-1996-MAYC}      & Yeast cells                     & Color filtering and contour enhancement                                                                                                                                & Thresholding                                                                                    &                                                      &                                                     \\
\cite{Shabtai-1996-MMMC}       & Fungi                           &                                                                                                                                                                        & Self-organizing multilayer neural network                                                       &                                                      &                                                     \\
\cite{Akiba-1997-DASZ}         & Plankton                        & Local auto-correlational masks                                                                                                                                         & PCA                                                                                             & Auto-correlational masks and discrimination analysis & 90\% accuracy                                       \\
\cite{Robinson-1998-MCYC}      & Cell                            & Global smoothing and Sobel operator                                                                                                                                    & Otsu thresholding and watershed                                                                 &                                                      &                                                     \\
\cite{Kocak-1999-CVTQ}         & Plankton                        & Morphological erosion and dilation                                                                                                                                     & Snake model                                                                                     &                                                      & 94.12\% accuracy                                    \\
\cite{Shabtai-1996-MMMC}       & Fungi                           &                                                                                                                                                                        & Self-organizing multilayer neural network                                                       &                                                      &                                                     \\
\hline
\end{tabular}
\end{table}

\newpage
\begin{table}
\scriptsize
\begin{tabular}{p{3cm}p{3cm}p{3cm}p{4cm}p{2cm}p{2cm}}
\hline
Related work        & Microorganism type & Pre-processing & Segmentation   & Classification  & Evaluation \\ \hline

\cite{Akiba-1997-DASZ}         & Plankton                        & Local auto-correlational masks                                                                                                                                         & PCA                                                                                             & Auto-correlational masks and discrimination analysis & 90\% accuracy                                       \\
\cite{Robinson-1998-MCYC}      & Cell                            & Global smoothing and Sobel operator                                                                                                                                    & Otsu thresholding and watershed                                                                 &                                                      &                                                     \\
\cite{Kocak-1999-CVTQ}         & Plankton                        & Morphological erosion and dilation                                                                                                                                     & Snake model                                                                                     &                                                      & 94.12\% accuracy                                    \\
\cite{Benyon-1999-DAFS}        & Fungi                           & Feature extraction                                                                                                                                                     &                                                                                                 & Linear and quadratic discriminant analysis           & 98\% accuracy                                       \\
\cite{Barber-2000-AACC}        & Mammalian cellular              & Sobel operator and Hough transform                                                                                                                                     & Thresholding                                                                                    &                                                      &                                                     \\
\cite{Motta-2001-TSPP}         & Protozoa                        & Histogram local equalization and morphological operations                                                                                                              & Euclidian distance map                                                                          & PCA                                                  &                                                     \\
\cite{Embleton-2003-ACPP}      & Phytoplankton                   & Median filter                                                                                                                                                          & Gray level and skeletonize operator                                                             & Neural network                                       &                                                     \\
\cite{Pernthaler-2003-AEGM}    & Plankton                        & Gradient transformation and neighbourhood median filter                                                                                                                & Thresholding                                                                                    &                                                      &                                                     \\
\cite{Grosjean-2004-EMAI}      & Zooplankton                     &                                                                                                                                                                        & Thresholding                                                                                    & Random forest and discriminant vector forest         &                                                     \\
\cite{Rong-2006-ACZB}          & Zooplankton                     &                                                                                                                                                                        & Otsu thresholding and region growing                                                            & BP neural network                                    &                                                     \\
\cite{Song-2006-AARA}          & Alga                            & Median filter                                                                                                                                                          & Hue-Saturation-Intensity thresholding, flood fill method and area thresholding                  &                                                      & 90\% accuracy                                       \\
\cite{Tsechpenakis-2008-ANCA} & Animal                          & Alternating sequential filter                                                                                                                                          & Nonrigid local registration and thresholding                                                    & Maximum likelihood                                   &                                                     \\
\cite{Zeder-2010-AQAS}         & Cyanobacteria                   &                                                                                                                                                                        & Thresholding and gray-level intensities maximizing                                              &                                                      & 85\% accuracy                                       \\
\cite{Ogawa-2012-NFAC}        & Colony                          &                                                                                                                                                                        & Time-lapse shadow approach                                                                      &                                                      & 0.999 correlation efficiency with manual counting   \\

\hline
\end{tabular}
\end{table}

\newpage
\begin{table}
\scriptsize
\begin{tabular}{p{3cm}p{3cm}p{3cm}p{4cm}p{2cm}p{2cm}}
\hline
Related work        & Microorganism type & Pre-processing & Segmentation   & Classification  & Evaluation \\ \hline

\cite{Xianjiu-2012-AMAC}       & Alga                            & Wavelet shrinkage image dilation                                                                                                                                       & Otsu thresholding and morphological opening                                                     &                                                      & 94\% accuracy                                       \\
\cite{Zhonglei-2012-ADMI}      & Fungi                           & Median filter, linear gray-scale transformation and adaptive smooth filter                                                                                             & Histogram thresholding                                                                          &                                                      & 97.44\% accuracy                                    \\
\cite{Barbedo-2012-MCMA}       & Colony                          & Median filter, histogram equalization and top-hat filter                                                                                                               & Thresholding                                                                                    &                                                      & 90\% accuracy                                       \\
\cite{Barbedo-2012-UFCA}       & Agriculture microorganism       & Laplacian filter, Canny filter and histogram equalization                                                                                                              & Thresholding and region growing                                                                 & Decision tree                                        &                                                     \\
\cite{Kim-2013-AEAH}          & Cysts                           & Morphological operation                                                                                                                                                & Otsu thresholding                                                                               &                                                      & 10.9\% average root-mean squared difference         \\
\cite{Hamid-2013-FEPC}        & Pus cell                        & Image enhancement                                                                                                                                                      & Thresholding                                                                                    &                                                      & 80\% reliability                                    \\
\cite{Dazzo-2013-CMEI}        & Colony                          &                                                                                                                                                                        & Quadrat size segmentation                                                                       &                                                      &                                                     \\
\cite{Saur-2014-AAMT}         & Moving predator                 & Thresholding                                                                                                                                                           & Global displacement response                                                                    &                                                      &                                                     \\
\cite{Sharma-2015-CMMD}       & Medical microorganism           & Histogram equalization and circular hough transformation                                                                                                               & Thresholding and Moore neighbour tracing                                                        &                                                      & 93\% accuracy                                       \\
\cite{Fang-2019-MICM}         & Microorganism                   & Histogram peak searching                                                                                                                                               & Particle swarm optimization, breadth-first search and exponential entropy                       &                                                      &                                                     \\
\cite{Albaradei-2020-ACCF}    & Stem cell                       & Augmentation techniques include color jitter to randomly alter brightness, contrast, saturation, and hue of each image, horizontal/vertical flip, and random rotation. & Thresholding                                                                                    & SRNetDL                                              &                                                     \\
\cite{Rolke-1984-SSAZ}        & Zooplankton                     & Mean grey-level selection                                                                                                                                              & Quantimet 720 image analysis system (Leica Cambridge Ltd., Cambridge, United Kingdom)           &                                                      &                                                     \\
\cite{Bloem-1995-FADS}        & Microorganism                   & Image sharpen and convolutional filter                                                                                                                                 & Quantimet 570                                                                                   &                                                      &                                                     \\
\cite{Grivet-1999-AEAS}       & Microorganism                   &                                                                                                                                                                        & Quantimet 570, thresholding and image skeleton                                                  &                                                      &                                                     \\

\hline
\end{tabular}
\end{table}

\newpage
\begin{table}
\scriptsize
\begin{tabular}{p{3cm}p{3cm}p{3cm}p{4cm}p{2cm}p{2cm}}
\hline
Related work        & Microorganism type & Pre-processing & Segmentation   & Classification  & Evaluation \\ \hline
\cite{Sieracki-1985-DEAS}     & Picoplankton                    & Thresholding                                                                                                                                                           & Artek 810 image analyzer (Artek Systems Corp., Farmingdale, N.Y.)                               &                                                      &                                                     \\
\cite{Estep-1989-CSAI}        & Alga                            &                                                                                                                                                                        & Zeus image analysis system (Institute of Marine Research, Bergen)                               &                                                      &                                                     \\
\cite{Wright-1991-CIPA}       & Micro plant and animal          & Sobel operator and thresholding                                                                                                                                        & ImageMeasure 5100 (Microscience, Div., Phoenix Trade, Inc. Seattle, Washington)                 &                                                      &                                                     \\
\cite{Corkidi-1998-CAIA}      & Microorganisms                  &                                                                                                                                                                        & IMAGENIA 2000 (Biocom, Les Ulis, France) and multilevel thresholding                            &                                                      & 95.47\% accuracy                                    \\
\cite{Rodenacker-2001-SARM}   & Microorganisms in water         & Morphological operation and hole filling                                                                                                                               & IDL (Research System Inc., Boulder, USA) and thresholding                                       & Neural network                                       &                                                     \\
\cite{Rodenacker-2002-IAQP}   & Phytoplankton                   &                                                                                                                                                                        & IDL (Research System Inc., Boulder, USA) and thresholding                                       & Neural network                                       &                                                     \\
\cite{Sandor-2001-ATRB}       & Fungi                           &                                                                                                                                                                        & Quanitmet 570 computer system (Leica, Cambridge, United Kingdom)                                &                                                      &                                                     \\
\cite{Nishimura-2006-UAAC}    & Yeast                           &                                                                                                                                                                        & Bioplorer (BP) (Matsushita Ecology Systems Co. Ltd, Kasugai, Aichi-ken, Japan) and intensity    &                                                      &                                                     \\
\cite{Nishimura-2008-ACFM}    & Eukaryotic and prokaryotic cell &                                                                                                                                                                        & Bioplorer (BP) (Matsushita Ecology Systems Co. Ltd, Kasugai, Aichi-ken, Japan) and thresholding &                                                      &                                                     \\
\cite{Eickhorst-2008-IDSM}    & Microorganisms in soil          & Contrast enhancement and gradation                                                                                                                                     & AnalySIS (Soft Imaging) and thresholding                                                        &                                                      & 97.7\% accuracy for bacteria and 92.2\% for archaea \\
\cite{Ghitua-2013-MIAA}       & Cyanobacteria                    & Morphological operation                                                                                                                                                & ImageJ                                                                                          &                                                      &                                                     \\
\cite{Stolze-2019-AIAI}       & Yeast                           & Thresholding                                                                                                                                                           & ImageJ and watershed                                                                            &                                                      &                                                     \\
\cite{Bennke-2016-MAHT}       & Cell                            &                                                                                                                                                                        & Automated Cell Measuring and Enumeration tool 2.0                                               &                                                      &                                                     \\

\hline
\end{tabular}
\end{table}

\end{landscape}

\section{Analysis of image processing based counting methods}
The image processing methods based microorganism counting are summarized from Sects. 3 to 4.  It can be seen that the most effective approaches for microorganism image counting are image pre-processing, image segmentation, image classification, connected region detection, and feature extraction. In order to find out the reasons why they are widely used and reveal the potential future direction, in this section, the properties of these methods with their application domains are analyzed and summarized. In order to illustrate the correlation between methods and their applications, some representative works are selected as examples.

\subsection{Image pre-processing methods}
Because many kinds of microorganisms are colorless, staining methods are necessary to apply before image capture. Different staining methods lead to different color images, so the color feature is not appropriate for automatic microorganism counting. Moreover, due to the illumination and image noise's inhomogeneity, pre-processing methods should be applied to solve the problems and prepare for image segmentation.  

Firstly, to reduce the effect of different colors for image segmentation, the RGB images are usually converted to gray-scale images by adjusting the proportions of red, green, and blue channels. The RGB can also be converted to HSI (Hue-Saturation-Intensity) color space to assist the colony boundaries detection, such as the works in~\cite{Song-2006-AARA},~\cite{Nayak-2010-ANAA},~\cite{Payasi-2017-DACT}. HSI color space can adjust the Intensity but does not change the color type of the original image when processing colored images. Furthermore, it can ultimately reflect the primary attribute of color perception and corresponds to the result of color perception, which is helpful for the following segmentation.

\begin{figure}[ht]
\centering
\includegraphics[trim={0cm 0cm 0cm 0cm},clip,width=0.8\textwidth]{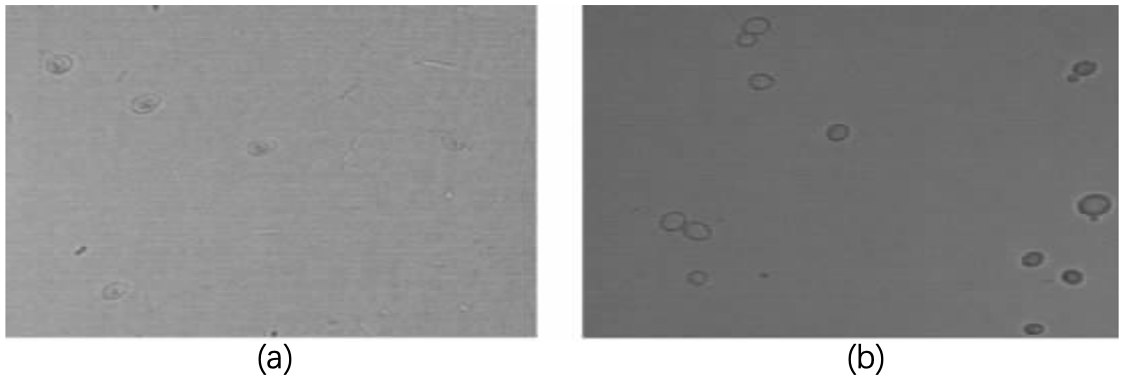}
\caption{Contrast of the yeast preprocessing image. (a) Original image. (b) Filtered image (In~\cite{Zhonglei-2012-ADMI} fig.1).}
\label{fig:Zhonglei-2012-ADMI2}
\end{figure}

Secondly, the uneven illumination can result in shading and a nonuniform background, which can usually be corrected using background subtraction, linear gray-scale transformation and low pass filtering, such as the works in~\cite{Choudhry-2016-HTMA} and~\cite{Zhonglei-2012-ADMI}(see Fig.~\ref{fig:Zhonglei-2012-ADMI2}).

Thirdly, noise removal is also one of the most necessary parts of pre-processing. Median filter and Gaussian filter are applied for denoising that are easy to approach and perform well in this part. The morphological open and close operations can be used to remove halos that appear while imaging, such as the work in~\cite{Motta-2001-TSPP}. 

\begin{figure}[ht]
\centering
\includegraphics[trim={0cm 0cm 0cm 0cm},clip,width=0.6\textwidth]{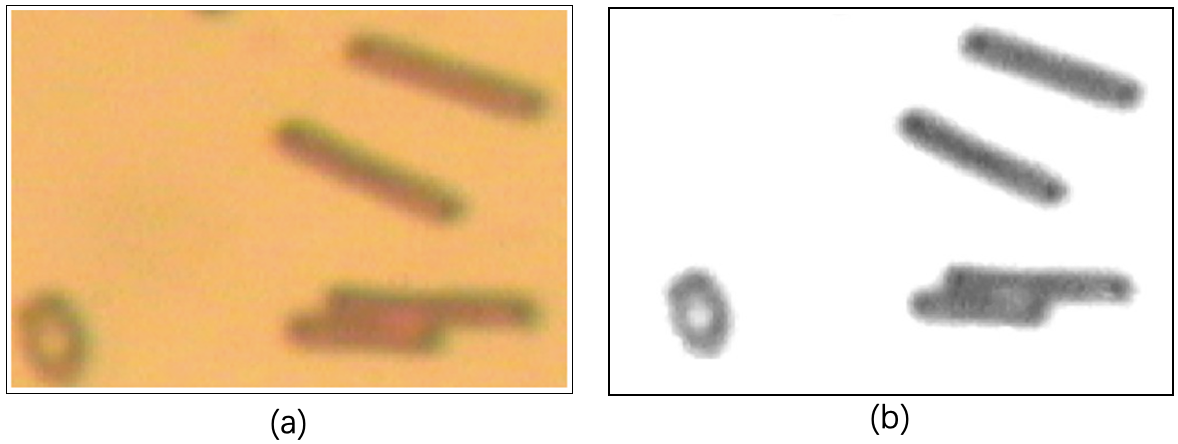}
\caption{Comparison of original image and enhanced image. (a) Original image. (b) Image obtained by histogram equalization (In~\cite{Zhang-2010-ASTF} fig.2, fig. 3).}
\label{fig:Zhang-2010-ASTF2}
\end{figure}

\begin{figure}[ht]
\centering
\includegraphics[trim={0cm 0cm 0cm 0cm},clip,width=1.0\textwidth]{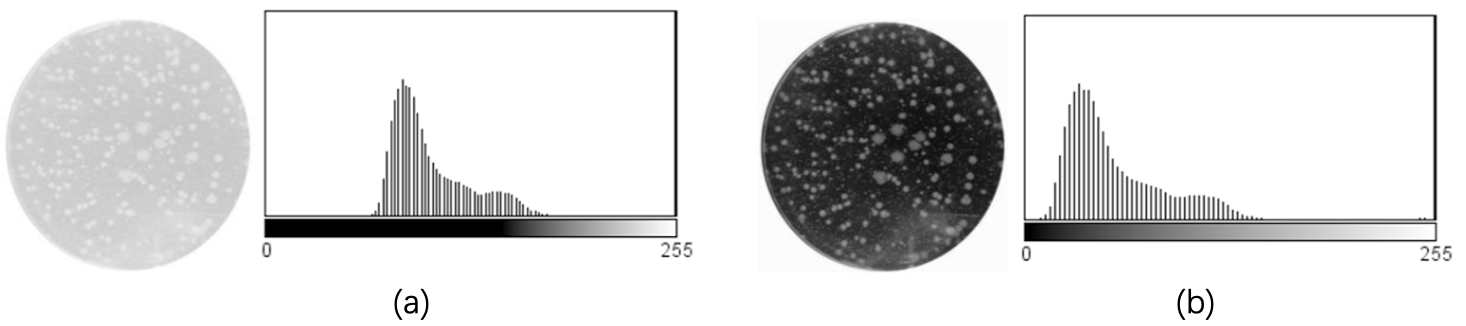}
\caption{Comparison of original image and enhanced image. (a) Original image with its gray-level histogram. (b) Resulting image with its gray-level histogram (In~\cite{Sanchez-2016-MAAC} fig.7, fig. 8).}
\label{fig:Sanchez-2016-MAAC2}
\end{figure}

Finally, the contrast of images may not be striking and need to be enhanced for image segmentation. The gray-level histogram equalization is the most acclaimed method that is easy to operate and can enhance the contrast in a global field, such as the work in~\cite{Zhang-2010-ASTF} and the image after enhancement is shown in Fig.~\ref{fig:Zhang-2010-ASTF2}.  A contrast limited adaptive histogram equalization (CLAHE) is proposed in~\cite{Ferrari-2017-BCCC} for local contrast enhancement, and a linear histogram expansion method is applied in~\cite{Sanchez-2016-MAAC} that is based on a transformation of the gray levels, a linear distribution of the values that are within the range of 0 to 255 is performed. The performance is shown in Fig.~\ref{fig:Sanchez-2016-MAAC2}.

\subsection{Image segmentation methods based on thresholding}
Image segmentation is the most significant part of microorganism counting methods. The extraction of a region of interest can be regarded as the segmentation of the colony part. Segmentation based on thresholding is the basic technique widely applied for microorganism counting, while many new segmentation methods are proposed to segment the area for counting accurately. 

Firstly, the segmentation technique based on thresholding is applied in many works for microorganism counting. Global thresholding is the easiest method for image segmentation when it has strong contrast, and an excellent result can be obtained, such as the works in~\cite{Gupta-2012-MABC} and~\cite{Chunhachart-2016-CAVE}.  Most of the segmentation methods are developed from the thresholding method and can improve performance in complex environments. 

\begin{figure}[ht]
\centering
\includegraphics[trim={0cm 0cm 0cm 0cm},clip,width=0.7\textwidth]{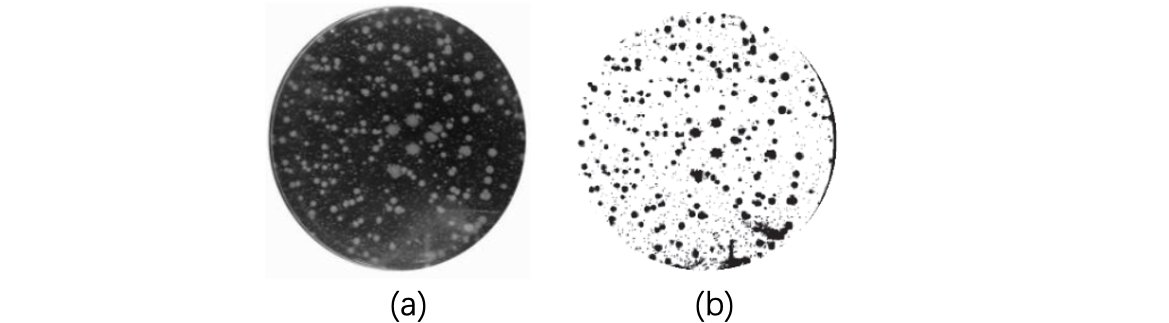}
\caption{Comparison of gray-scale image and binary image. (a) Gray-scale image. (b) Binary image based on Otsu thresholding. (In~\cite{Sanchez-2016-MAAC} fig. 8, fig. 9).}
\label{fig:Sanchez-2016-MAAC3}
\end{figure}

Secondly, Otsu thresholding is applied in many works such as~\cite{Zhang-2007-AEAR},~\cite{Zhang-2008-AABC} and~\cite{Peitz-2010-SCBG}. The Otsu thresholding is simple and easy to calculate. It can be used to segment the image effectively when the area difference between the target and the background is negligible. Nevertheless, the target and background can not be separated accurately when the gray-scale of the target and the background have a large overlap because the gray-scale distribution is used as the basis of image segmentation. It is also sensitive to noises, so denoising processing is usually applied first. The example of Otsu thresholding based image binarization is shown in Fig.~\ref{fig:Sanchez-2016-MAAC3}.

\begin{figure}[ht]
\centering
\includegraphics[trim={0cm 0cm 0cm 0cm},clip,width=0.6\textwidth]{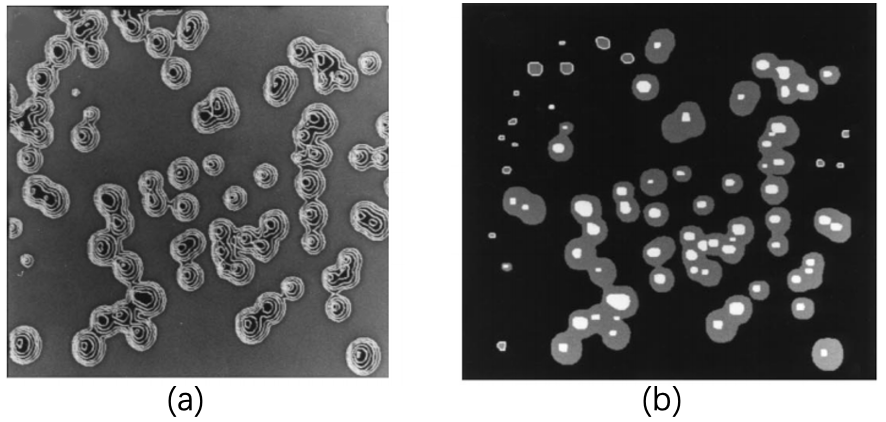}
\caption{Comparison of gray-scale image and binary image. (a) Enhanced image. (b) Multi-level thresholding image (In~\cite{Corkidi-1998-CAIA} fig. 2).}
\label{fig:Corkidi-1998-CAIA23}
\end{figure}

Finally, the segmentation result of the single threshold method is not satisfied when the gray-level of the image may be unevenly distributed, resulting in the influence of illumination. An iterative local threshold method is applied in~\cite{Shen-2010-ESAC}, the point with a local maximum threshold is obtained using a Laplacian operator, that is, the initial local thresholds. The microorganism images captured by microscope can be affected by lighting distribution, so the idea of the algorithm is not to calculate the global image threshold, but to calculate the local threshold according to the brightness distribution of different regions of the image, which means different thresholds can be calculated adaptively for different regions of the image. Another method to improve the segmentation performance when traditional thresholding does not work well is the multi-level threshold that is applied in~\cite{Corkidi-1998-CAIA}. The performance of multi-level thresholding for segmentation is shown in Fig.~\ref{fig:Corkidi-1998-CAIA23}. Thresholding based on multi-level method is divided into multi-spatial-level and multi-threshold-level, it can help segment more detailed pieces of information that may be lost using global thresholding.

\subsection{Other image segmentation methods}

First of all, edge detection is applied in a mass of works that performs well in image segmentation.  Sobel operator is a classical method that is easy to be applied, such as the works in~\cite{Andreini-2015-AIAA}, \cite{Chiang-2015-ACBC} and \cite{Choudhry-2016-HTMA}. Laplacian operator is a second order differential operator that is isotropic but sensitive to noises, such as the works in~\cite{Barbedo-2013-AACM}, and Laplacian operator can be combined with Sobel operator that can obtain a better detection result, such as the work in~\cite{Ogawa-2003-DMDI}. Canny operator is a multi-stage optimization operator with filtering, enhancement and detection that performs best but is relatively complex to use, such as the works in~\cite{Matic-2016-SAPS} and \cite{Barbedo-2012-UFCA}. Another popular method is Marr-Hildreth operator, that is, a Gaussian filter is applied first for smoothing and a Laplacian filter is applied for image enhancement, such as the works in~\cite{Viles-1992-MMPC} and \cite{Blackburn-1998-RDBA}, that performs well for images with the low signal-to-noise ratio.

\begin{figure}[ht]
\centering
\includegraphics[trim={0cm 0cm 0cm 0cm},clip,width=0.8\textwidth]{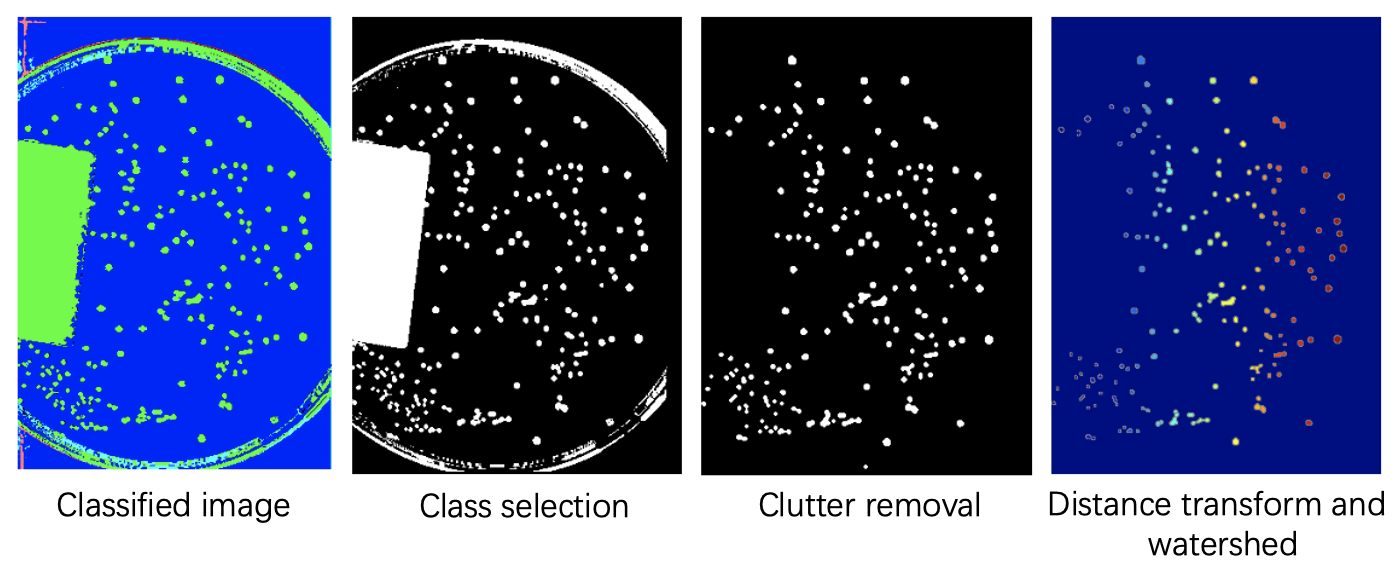}
\caption{Sample step-sequence for an automated colony counting system (In~\cite{Masschelein-2012-TACC} fig. 15).}
\label{fig:Masschelein-2012-TACC15}
\end{figure}

Secondly, the method combined with distance transform and watershed is applied in colony segmentation. Distance transform can extract the distance between a non-zero pixel and the nearest zero pixel, that is, the gray-scale value of each pixel in the image is the distance between the pixel and the nearest background pixel. The distance transform is usually applied for image segmentation with watershed, such as the works in~\cite{Hong-2008-SHBC}, \cite{Yujie-2009-DIMB} and \cite{Masschelein-2012-TACC}. The performance of distance transform and watershed is shown in Fig.~\ref{fig:Masschelein-2012-TACC15}.

\begin{figure}[ht]
\centering
\includegraphics[trim={0cm 0cm 0cm 0cm},clip,width=0.5\textwidth]{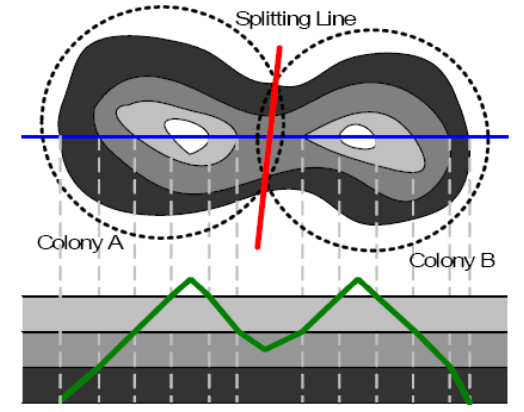}
\caption{The concept of watershed algorithm (In~\cite{Zhang-2008-AABC} fig. 5).}
\label{fig:Zhang-2008-AABC5}
\end{figure}

Thirdly, watershed segmentation is always applied for the separation of connected colonies. The original watershed algorithm performs well in the segmentation process for adherent colonies, such as the works in~\cite{Zhang-2008-AABC}, \cite{Ates-2009-AIPB} and \cite{Stolze-2019-AIAI}. The segmentation method is shown in Fig.~\ref{fig:Zhang-2008-AABC5} and Fig.~\ref{fig:Ates-2009-AIPB}. However, the original watershed may get the results of over-segmentation because of the noise and local discontinuity of the images, the marker-controlled watershed algorithm is applied in~\cite{Selinummi-2005-SQLB}. The Hough transformation is applied to extract the object's marker, and the background is marked and eliminated separately.

Finally, top-hat transform and bottom-hat transform are morphological operations applied for image segmentation when the illumination is uneven. The top-hat transformation is the difference between the image and the image after the open operation, such as the works in~\cite{Pernthaler-1997-SCAI}, \cite{Brugger-2012-ACBC} and \cite{Barbedo-2012-MCMA}, while the bottom-hat transform is the difference between the image after the close operation and the image, such as the work in~\cite{Chiang-2015-ACBC}. The top hat transformation is used for light objects on a dark background, while the bottom hat transformation is used for the opposite.

\subsection{Image classification methods}

Classification is a necessary operation when the microorganisms need to be counted respectively. Firstly, a decision tree is a supervisor learning that is widely used based on probability analysis. In~\cite{Barbedo-2012-UFCA}, a decision tree is applied for the classification of agriculture microorganisms.  The decision tree is easy to understand and explain, and can make possible and practical results for large data sources in a relatively short period, but the overfitting problem while classification needs to be solved.

Secondly, support vector machine (SVM) is a kind of linear classifier that classifies data in a binary way according to supervised learning, such as the works in~\cite{Chen-2009-AABC} and \cite{Masschelein-2012-TACC}.  In~\cite{Yujie-2009-DIMB}, the shape invariant moment and gray level co-ocurence matrices (GLCM) are extracted for SVM training, and the classification accuracy of bacteria is 99.67\%. SVM performs well with small sample and can be trained to solve the problem of high dimensional, but it is sensitive to missing data, and the choice of features has enormous implications for classification results.

\begin{figure}[ht]
\centering
\includegraphics[trim={0cm 0cm 0cm 0cm},clip,width=1.0\textwidth]{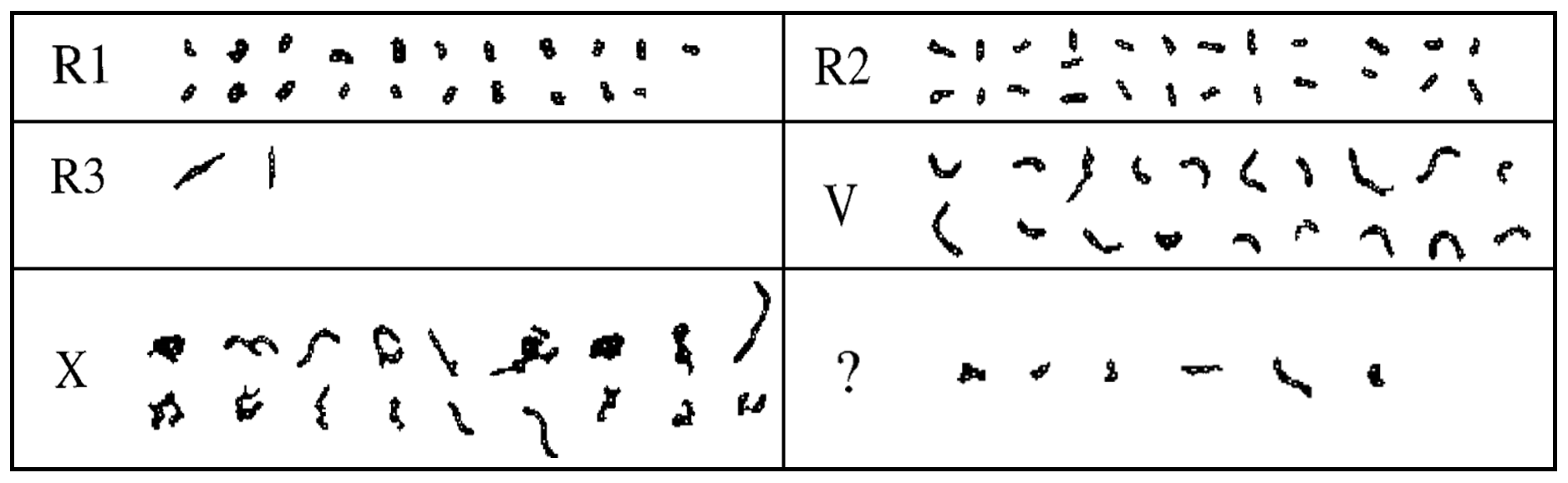}
\caption{The classification result (In~\cite{Blackburn-1998-RDBA} fig. 4).}
\label{fig:Blackburn-1998-RDBA4}
\end{figure}

Thirdly, artificial neural network (ANN) is a network with self-learning, self-organization, self-adaptation and strong nonlinear function approximation ability, that has strong fault tolerance. In~\cite{Blackburn-1998-RDBA}, an ANN is trained for classification and quantification of bacteria, and about 95\% of all objects are classified in each image. The classification result is shown in Fig.~\ref{fig:Blackburn-1998-RDBA4}. ANN has high classification accuracy and strong robustness to noise nerves, but the learning process is unobservable and the output is hard to interpret. Moreover,  back propagation (BP) neural network is a multi-layer feedforward network trained by error back propagation, that is the most widely used ANN. In~\cite{Jun-2010-RDRM}, BP neural network is applied for bacteria classification, and in~\cite{Rong-2006-ACZB}, BP neural network is used for zooplankton classification and counting. BP neural network has strong nonlinear mapping ability and flexible network structure, but the convergence rate is slow. Moreover, it is easy to fall into local minima.  Furthermore, the convolutional neural network (CNN) is a feedforward neural network with deep structure and convolution computation representing learning. In~\cite{Ferrari-2015-BCCC}, CNN is applied for bacteria colony counting, and the accuracy of 92.8\% is obtained. In~\cite{Tamiev-2020-ACBC}, a classification-type convolutional neural network (cCNN) is designed for bacteria classification and counting.CNN can automatically extract the features of images and process high-dimensional data quickly, but the pooling layer may lose much valuable information while training.

By reviewing all the existing deep learning based microorganism counting methods, the classification can be achieved automatically, but the segmentation part still needs to be adjusted manually.
In order to show the development of deep learning and the time for microorganism counting, the training time and counting time of deep learning methods in this review are summarized in Table~\ref{tab:dltime}.

\begin{table}
\scriptsize
\caption{\label{tab:dltime}The training time and counting time of deep learning methods.} 
\begin{tabular}{p{2cm}p{1cm}p{2cm}p{1.5cm}p{2cm}p{1cm}}
\hline
Related work        & Model & GPU & Training time   & Counting time  & Accuracy \\ \hline
\cite{Blackburn-1998-RDBA}     & ANN     &   -      & -  &  100 images per hour    &     95\% \\
\cite{Shabtai-1996-MMMC} & ANN & PC-Vision Plus & - & - & - \\
\cite{Embleton-2003-ACPP} & ANN & Pentium processor & - & 75 images per minutes & >90\% \\
\cite{Hongwei-2012-TTMI}      & BPNN   &- & -     &< 1 hour   &  95\%            \\
\cite{Shenglang-2008-RDTN} & BPNN  & SDK-2000 & - &   1 image in 3 seconds  &  Student's t test $p$>0.05\\
\cite{Rong-2006-ACZB} & BPNN & - & - & - & - \\
\cite{Ferrari-2015-BCCC} & CNN & Nvidia Titan Black & 50000 iterations in 3 hours & - & 92.8\% \\
\cite{Ferrari-2017-BCCC} & CNN & Nvidia Titan X &  50000 iterations in 1 hours & - & 92.1\% \\
\cite{Tamiev-2020-ACBC} & cCNN & NVIDIA Quadro K620& - & 3.8 times faster & 86\% \\
\cite{Albaradei-2020-ACCF} & SRNet & - & - & - & RMSE = 22.38 \\

\hline
\end{tabular}
\end{table}

\subsection{Analysis of potential methods} 
Through summarizing the work of image analysis based microorganism counting, it can be found that the accuracy of microorganism counting is continuously improving with the development of computer vision and deep learning technologies, which indicates that computer vision based microorganism counting methods will completely replace the traditional manual counting methods. 
However, the deep learning methods are mainly applied for microorganism classification, while the microorganism segmentation methods are still adopted the traditional techniques, such as thresholding or watershed, resulting in a huge gap with the state-of-the-art technology. The application of the latest semantic segmentation technology can classify microorganisms at the same time of segmentation, which will be the trend of future development.

\begin{figure}[ht]
\centering
\includegraphics[trim={0cm 0cm 0cm 0cm},clip,width=0.7\textwidth]{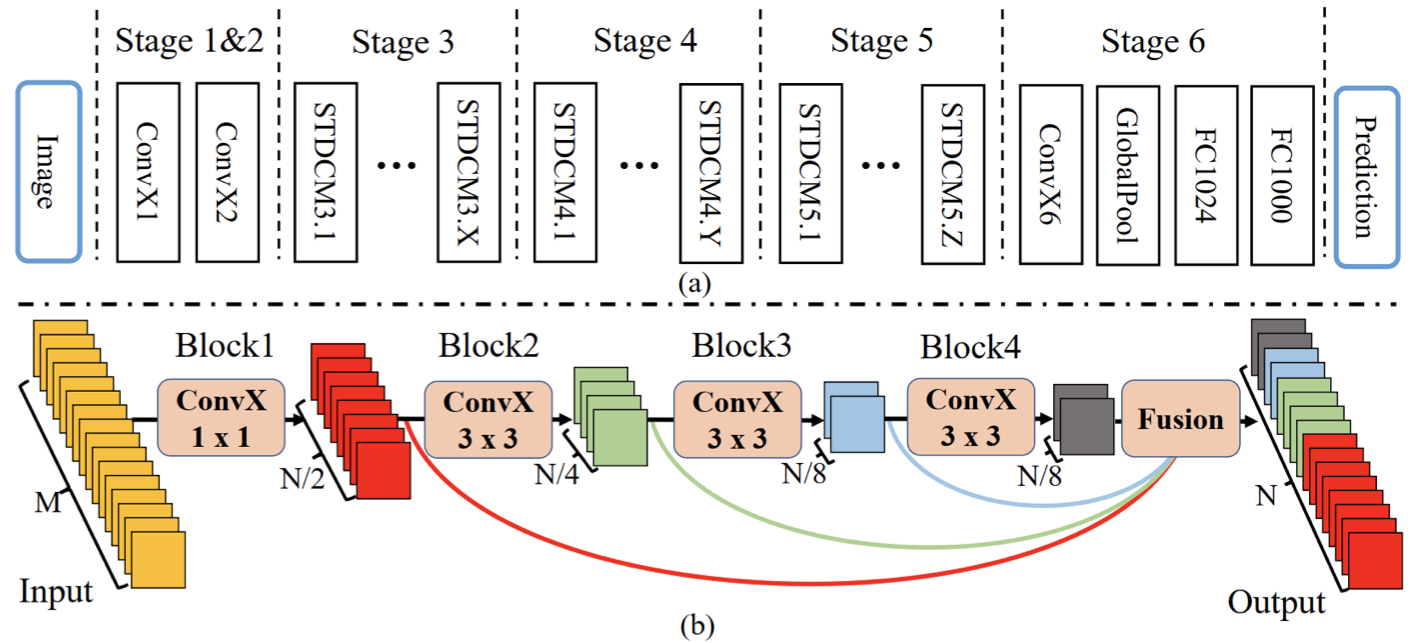}
\caption{(a) The original architecture of STDC network. (b) The proposed architecture of STDC network (In~\cite{Fan-2021-RBRS} fig. 3).}
\label{fig:BiseNet}
\end{figure}

According to the existing microorganism counting work, the work of imaging and image analysis are often separated. Therefore, it is difficult to obtain real-time microorganism counting information, leading to time-consuming and workforce waste. 
BiSeNet is one of the real-time semantic segmentation networks~\citep{Yu-2018-BBSN}. However, an extra encoding path is applied for spatial information, which is time-consuming. In~\cite{Fan-2021-RBRS},  a novel architecture is designed as a Short-Term Dense Concatenate network (STDC network), which is shown in Fig.~\ref{fig:BiseNet}. Multiple contiguous layers of response maps are connected, and each layer encodes the input image at different scales and in its own field to achieve multi-scale feature representation. Then the Detail Guidance is applied for decoding, which can guide the low-level layers to learn spatial details. Finally, the spatial information and segmentation of deep layers are combined to show the final results. The 71.9\% of mIoU is obtained, and the computing speed is 45.2\% faster than the original method.

\begin{figure}[ht]
\centering
\includegraphics[trim={0cm 0cm 0cm 0cm},clip,width=1.0\textwidth]{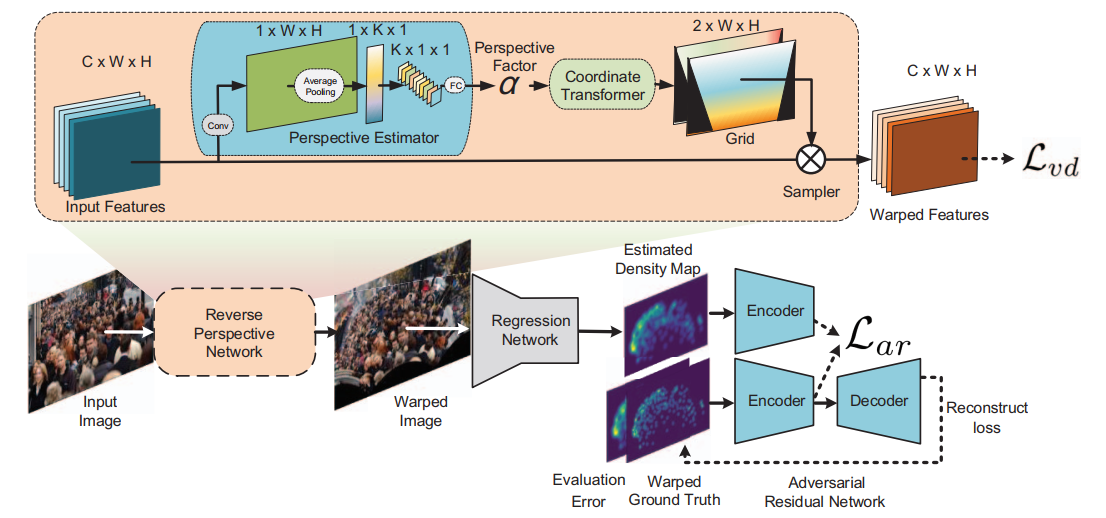}
\caption{The architecture of the reverse perspective network(In~\cite{Yang-2020-PRNP} fig. 4).}
\label{fig:PAOC}
\end{figure}

The work of deep learning based counting methods also provides a new direction for the field. 
In~\cite{Yang-2020-PRNP}, the scale variations of images are solved based on a reverse perspective network, which is shown in Fig.~\ref{fig:PAOC}. The reverse perspective networks can reduce the scale variances of images before regression, reducing the complexity of the network. The original is sampled firstly, and then the number of objects can be evaluated by a regression network. 
The reverse perspective networks can evaluate perspective distortion precisely, which can be correct by uniformly distorting the image. 
Finally, the images with similar scales are transmitted to the regressor, and 61.2 of mean average error (MAE) is obtained.

\begin{figure}[ht]
\centering
\includegraphics[trim={0cm 0cm 0cm 0cm},clip,width=1.0\textwidth]{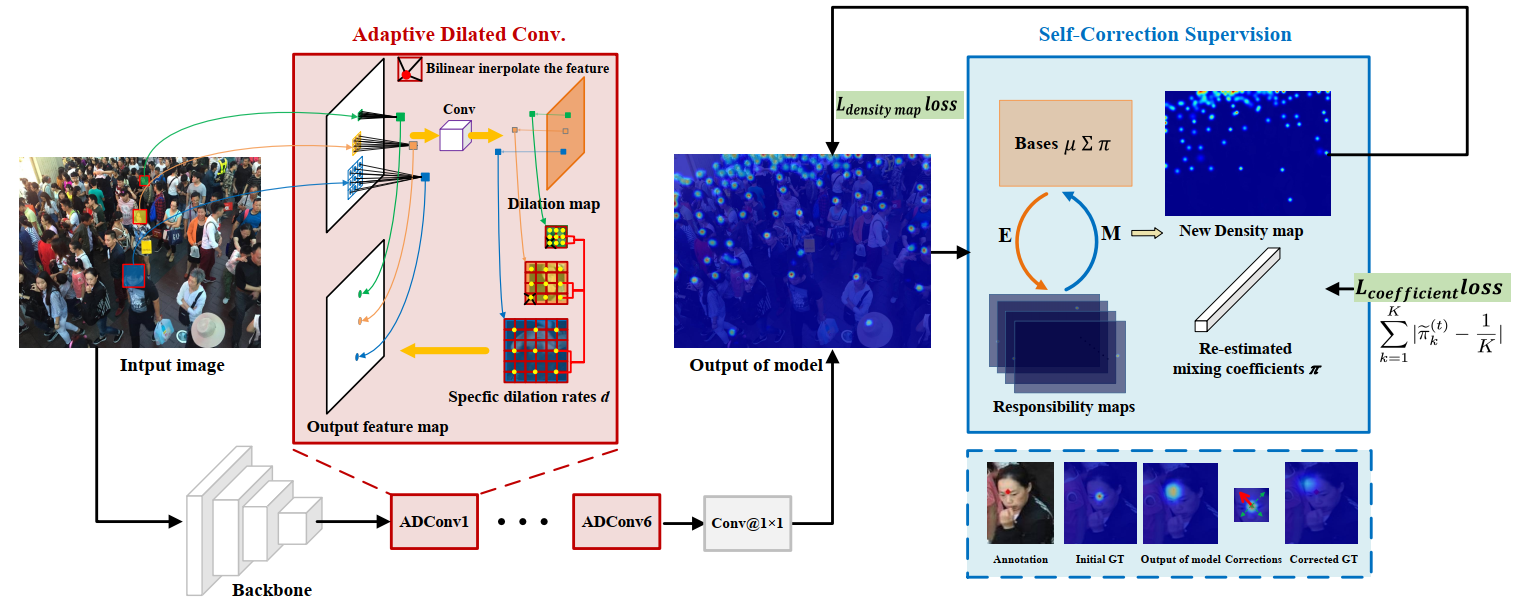}
\caption{The overview of the adaptive dilated convolution network (In~\cite{Bai-2020-ADNS} fig. 2).}
\label{fig:Bai-2020-ADNS}
\end{figure}

In~\cite{Bai-2020-ADNS}, an adaptive dilated convolution and a novel supervised learning framework is proposed for self-correlation counting works, which is shown in Fig.~\ref{fig:Bai-2020-ADNS}. In classical counting methods, the models are optimized by comparing the ground truth and predicted image, and the density map is not precise because of the labeling deviation. 
First, the image is input into the model for feature extraction, and then the density map is output by using six adaptive convolutions. After that, the sample locations are calculated by dilation rates. The result can adapt the scale variation of the images, and the MAE of 66.5 is obtained.

Deep learning frameworks have quickly become the primary method for analyzing microscopic images. We can infer and predict machine and deep learning based methods will be applied in microorganism counting as well as in other research such as digital pathology, which is summarized in~\cite{Salvi-2020-TIPP, Madabhushi-2016-IAAM}. However, there are two main limitations during the surveying. First, the number of research based on deep learning is limited, proving the vast development potential in this field. Second, most of the deep learning methods are applied for classification but not for segmentation, and the deep learning based segmentation can carry out more precise segmentation results by comparing with classic methods. These two limitations also show the development directions and opportunities in the future.
First, traditional manual counting methods will be replaced by deep learning based microorganism counting methods, which cannot only be used for classification, but also for precise segmentation. 
Second, in the future, the microorganism counting systems will be integrated with sampling, imaging and analyzing systems using deep learning, helping researchers monitor the microorganism timely.

\section{Conclusion and future work}
In this paper, a comprehensive review of image analysis methods for microorganism counting is proposed. The counting methods are summarized and grouped based on the types of microorganisms, including bacteria counting and other microorganisms counting. 
Then the methods are separated based on segmentation approaches, such as thresholding methods, edge detection methods, third-party tools and deep learning based methods. 
By reviewing all the related works, we can find that the classic methods in Sect. 2.1 and Sect. 3.1 are developed from the 1980s to 2000s, such as the Otsu thresholding method, watershed algorithm and edge detection methods, which shows a blooming development of digital image processing for microorganism analysis. 
Since the 2010s, the development of deep learning carries out the microorganism counting results with high accuracy. Furthermore, the development of professional microorganism counting systems is summarized in Sect. 2.3 and Sect. 3.3, such as `ImageJ' and `CellC' show people pay more and more attention to microorganism counting.
In summary, the successful development of image analysis based microorganism counting methods shows vast research potential in this field. Moreover, the most frequently used microorganism counting approaches of image preprocessing, image segmentation and image classification are analyzed in Sect. 4. 

The image analysis based microorganism counting methods discussed in this paper can be referred in other digital image analysis fields. 
For example, microorganism classification is a significant application field of the microorganism analysis, referring to environmental microorganism classification~\citep{Kosov-2018-EMCC}, cervical cell classification~\citep{Mamunur-2021-DADL}, blood cell classification~\citep{Su-2014-ANNA}, classification for different types of microorganisms~\citep{Li-2019-ASTA}. 
Furthermore, the segmentation methods for microorganisms can be referred to by digital image processing workers, such as stem cell segmentation~\citep{Huang-2016-SCMI}, cancer cell segmentation~\citep{Chen-2006-ASCA}, environmental microorganism segmentation~\citep{Zhang-2021-LANL}. Moreover, microscopic image processing performs an essential role in industrial analysis, such as the monitoring for waste water~\citep{Amaral-2005-ASMA}, beef carcass evaluation~\citep{Cross-1983-BCEU}, monitoring of bacteria in milk~\citep{Pettipher-1982-SACB}, monitoring flames in an industrial boiler~\citep{Yu-2004-MFAI}, softwood lumber grading~\citep{Bharati-2003-SLGO} and so on. 
Finally,  the summarized counting methods can also be applied in small object detection, such as sperm counting~\citep{Peng-2015-CDCC}, crowd counting~\citep{Zhang-2015-CSCC} and vehicle counting~\citep{Li-2019-SDAC}.

In the future, deep learning based microorganism counting methods are promising. Since the COVID-19 broke out in 2019, people pay increasing attention to microorganism analysis. There is still a considerable limitation and opportunity in microorganism research. This review can contribute a lot to the research of microorganism counting for future researchers.

\section*{Acknowledgements}
This work is supported by the ``Natural Science Foundation
of China'' (No. 61806047). We thank Miss Zixian Li and Mr. 
Guoxian Li for their important discussion.

{\small
\bibliographystyle{spbasic}
\bibliography{Jiawei}
}

\end{document}